\documentclass[preprint,notoc]{JHEP3}
\usepackage{epsfig}

\def\eslt{E_T^{\rm miss}}

\def\to{\rightarrow}

\def\bi{\begin{itemize}}
\def\ei{\end{itemize}}

\def\ta{\tilde a}

\def\tu{\tilde u}

\def\ta{\tilde a}

\def\tb{\tilde b}
\def\tf{\tilde f}

\def\tst{\tilde t}
\def\ttau{\tilde{\tau}}
\def\tmu{\tilde{\mu}}
\def\tg{\tilde g}
\def\tnu{\tilde\nu}
\def\tell{\tilde\ell}
\def\tq{\tilde q}

\def\tw{\widetilde W}
\def\tz{\widetilde Z}
\def\pino{\tilde{\gamma}}
\def\delew{\Delta_{\rm EW}}
\def\msusy{M_{\rm SUSY}}
\def\alt{\lesssim}
\def\agt{\gtrsim}
\def\be{\begin{equation}}
\def\ee{\end{equation}}
\def\bea{\begin{eqnarray}}
\def\eea{\end{eqnarray}}

\newcommand\sjp[3]{{\it Sov.\ J.\ Nucl.\ }{\bf #1} (#2) #3}

\title{Radiative natural supersymmetry: Reconciling electroweak
fine-tuning and the Higgs boson mass } 
\author{Howard Baer$^a$, Vernon
Barger$^b$, Peisi Huang$^b$, Dan Mickelson$^a$, Azar Mustafayev$^c$ and
Xerxes Tata$^c$\\ $^a$Dept.\ of Physics and Astronomy, University of
Oklahoma, Norman, OK 73019, USA\\ $^b$Dept. of Physics, University of
Wisconsin, Madison, WI 53706, USA\\ $^c$Dept.\ of Physics and Astronomy,
University of Hawaii, Honolulu, HI 96822, USA\\ E-mail:
\email{baer@nhn.ou.edu},
\email{barger@pheno.wisc.edu}, \email{phuang7@wisc.edu},
\email{mickelso@nhn.ou.edu}, \email{azar@phys.hawaii.edu}, \email{tata@phys.hawaii.edu}
}

\preprint{\vbox{UH-511-1204-12 }}

\abstract{ Models of natural supersymmetry seek to solve the little
hierarchy problem by positing a spectrum of light higgsinos $\alt 200-300$~GeV 
and light top squarks $\alt 600$~GeV along with very heavy squarks
and~TeV-scale gluinos.  Such models have low electroweak fine-tuning and
satisfy the LHC constraints.  However, in the context of the MSSM, they
predict too low a value of $m_h$, are frequently in conflict with the
measured $b\to s\gamma$ branching fraction and the relic density of
thermally produced higgsino-like WIMPs falls well below dark matter (DM)
measurements.  We propose a framework dubbed {\it radiative natural SUSY} (RNS)
which can be realized within the MSSM (avoiding the addition of extra
exotic matter) and which maintains features such as gauge coupling
unification and radiative electroweak symmetry breaking. The RNS model
can be generated from SUSY GUT type models with non-universal Higgs
masses (NUHM).  Allowing for high scale soft SUSY breaking Higgs mass
$m_{H_u}> m_0$ leads to {\it automatic cancellations} during
renormalization group (RG) running, and to {\it radiatively-induced low
fine-tuning} at the electroweak scale.  Coupled with large mixing in the
top squark sector, RNS allows for fine-tuning at the 3-10\% level with
TeV-scale top squarks and a 125~GeV light Higgs scalar $h$.  The model
allows for at least a partial solution to the SUSY flavor, $CP$ and
gravitino problems since first/second generation scalars (and the
gravitino) may exist in the 10-30~TeV regime.  We outline some possible
signatures for RNS at the LHC such as the appearance of low invariant
mass opposite-sign isolated dileptons from gluino cascade decays. The
smoking gun signature for RNS is the appearance of light higgsinos at a
linear $e^+e^-$ collider. If the strong $CP$ problem is solved by the
Peccei-Quinn mechanism, then RNS naturally accommodates mixed
axion-higgsino cold dark matter, where the light higgsino-like WIMPS --
which in this case make up only a fraction of the measured relic
abundance -- should be detectable at upcoming WIMP detectors.}

\begin{document}

\section{Introduction}
\label{sec:intro}

The recent discovery by Atlas and CMS of a Higgs-like resonance at the
CERN LHC\cite{atlas_h,cms_h} adds credence to supersymmetric models
(SUSY) of particle physics in that the mass value $m_h\simeq 125$~GeV
falls squarely within the narrow window predicted by the Minimal
Supersymmetric Standard Model (MSSM): $m_h\sim 115-135$~GeV\cite{mhiggs}.  
At the same time, the lack of a SUSY signal at LHC7
and LHC8 implies $m_{\tg}\agt 1.4$~TeV (for $m_{\tg}\sim m_{\tq}$) and
$m_{\tg}\agt 0.9$~TeV (for $m_{\tg}\ll
m_{\tq}$)\cite{atlas_susy,cms_susy}.  While weak scale SUSY\cite{wss}
provides a solution\cite{big} to the gauge hierarchy problem via the cancellation
of quadratic divergences, the apparently multi-TeV sparticle masses
required by LHC searches seemingly exacerbates {\it the little hierarchy
problem} (LHP): 
\bi
\item how do multi-TeV values of SUSY model parameters conspire to yield a $Z$-boson
(Higgs boson) mass of just 91.2  (125)~GeV?
\ei

Models of {\it natural supersymmetry}\cite{kn} address the LHP by
positing a spectrum of light higgsinos $\alt 200$~GeV and light top- and
bottom-squarks with $m_{\tst_{1,2},\tb_1}\alt 600$~GeV along with very
heavy first/second generation squarks and~TeV-scale
gluinos\cite{ah,others,bbht}.  Such a spectrum allows for low
electroweak fine-tuning (EWFT) while at the same time keeping sparticles
safely beyond LHC search limits.  Because third generation scalars
are in the few hundred~GeV range, the radiative
corrections to $m_h$, which increase logarithmically 
with $m_{\tst_i}^2$, are never very large
and these models have great difficulty in accommodating a light SUSY Higgs
scalar with mass $m_h\sim 125$~GeV~\cite{bbht,h125}.  Thus, we are faced with
a new conundrum: 
\bi
\item how does one reconcile low EWFT with such a large value of
$m_h$ \cite{hpr}?  
\ei 
A second problem occurs in that 
\bi
\item the predicted branching fraction for $b\to s\gamma$ decay is
frequently at odds with the measured value due to the very light third
generation squarks\cite{bbht}.  
\ei 
A third issue appears in that
\bi
\item the light higgsino-like WIMP particles predicted by models of
natural SUSY lead to a thermally-generated relic density which is
typically a factor 10-15 below\cite{bbh,bbht} the WMAP measured value of
$\Omega_{CDM}h^2\simeq 0.11$.  
\ei

One solution to the fine-tuning/Higgs problem is to add extra fields to
the theory, thus moving beyond the MSSM\cite{hpr}. For example, adding
an extra singlet as in the NMSSM permits a new quartic coupling in the Higgs
potential thus allowing for an increased value of $m_h$\cite{nmssm}.
Alternatively, one may add extra vector-like matter to increase $m_h$
while maintaining light top squarks\cite{vecmatter}. In the former case
of the NMSSM, adding extra gauge singlets may lead to re-introduction of
destabilizing divergences into the theory\cite{bpr}. In the latter case,
one might wonder about the ad-hoc introduction of extra weak scale
matter multiplets and how they might have avoided detection. A third
possibility, which is presented below, is to re-examine EWFT and to
ascertain if there do indeed exist sparticle spectra {\it within the
MSSM} that lead to $m_h\sim 125$~GeV while maintaining modest levels of
electroweak fine-tuning.

\subsection{Electroweak fine-tuning}

One way to evaluate EWFT in SUSY models is to examine the minimization
condition from the Higgs sector scalar potential which determines the $Z$
boson mass. (Alternatively, one may examine the mass formula for $m_h$
and arrive at similar conclusions.)  Minimization of the one-loop
effective potential $V_{\rm tree} + \Delta V$, leads to 
\be \frac{M_Z^2}{2} =
\frac{m_{H_d}^2 + \Sigma_d^d -
(m_{H_u}^2+\Sigma_u^u)\tan^2\beta}{\tan^2\beta -1} -\mu^2 \;,
\label{eq:loopmin}
\ee where $\Sigma_u^u$ and $\Sigma_d^d$ are radiative corrections that
arise from the derivatives of $\Delta V$ evaluated at the minimum.
Eq.~(\ref{eq:loopmin}) reduces to the familiar tree-level
expression\cite{wss} for $M_Z^2$ when radiative correction terms are
ignored.  As we will discuss in detail below, $\Sigma_u^u$ and
$\Sigma_d^d$ include contributions, listed in the Appendix, from
various particles and sparticles with sizeable Yukawa and/or gauge
couplings to the Higgs sector.
To obtain a {\it natural} value of $M_Z$ on the left-hand-side, one
would like each term $C_i$ (with $i=H_d,\ H_u$, $\mu$ as well as
$\Sigma_u^u(k)$, $\Sigma_d^d(k)$, where $k$ denotes the various
contributions to the $\Sigma$s that we just mentioned) on the
right-hand-side to have an absolute value of order $M_Z^2/2$.  Noting
that all entries in (\ref{eq:loopmin}) are defined at the weak scale, we
are led to define the {\it electroweak fine-tuning
parameter}\footnote{Barbieri and Giudice\cite{bg} (and, even earlier,
Ellis {\em et al.}\cite{ellis}) define a fine tuning measure
$\Delta_{BG}=max|(a_i/M_Z^2)\partial M_Z^2/\partial a_i|$ for input
parameters $a_i$.  Our definition coincides with theirs when $M_Z^2$
depends linearly on input parameters (such as $\mu^2$, $m_{H_u}^2$ or
$m_{H_d}^2$ using electroweak scale parameters) but differs when the
parameter dependence is non-linear. For electroweak scale parameters,
the non-linear dependence only occurs in the radiative correction
terms $\Sigma_u^u$ and $\Sigma_d^d$ and in $\tan\beta$.}
by~\cite{ltr}
\be 
\delew \equiv max_i \left(C_i\right)/(M_Z^2/2)\;, 
\label{eq:Deltaew}
\ee 
where $C_{H_u}=|-m_{H_u}^2\tan^2\beta /(\tan^2\beta -1)|/$,
$C_{H_d}=|m_{H_d}^2/(\tan^2\beta -1)|/$ and $C_\mu =|-\mu^2|$, along
with analogous definitions for $C_{\Sigma_u^u(k)}$ and
$C_{\Sigma_d^d(k)}$.  Low $\delew$ means less fine-tuning.
Since $C_{H_d}$ and $C_{\Sigma_d^d(k)}$ terms are suppressed by
$\tan^2\beta -1$, for even moderate $\tan\beta$ values this expression
reduces approximately to
\be 
\frac{M_Z^2}{2} \simeq -(m_{H_u}^2+\Sigma_u^u)-\mu^2\;.
\label{eq:approx}
\ee 
We see that to get low $\delew$ we require $|-m_{H_u}^2|\sim
M_Z^2/2$ and $\mu^2\sim M_Z^2/2$.  The question then arises: what is the
model and can we find a set of model parameters such that $\delew \sim 1-30$, 
corresponding to better than $\delew^{-1}=3\%$
EWFT?   Note that $\delew$ depends only on the weak scale
parameters of the theory and hence is essentially fixed by the particle
spectrum, independent of how superpartner masses arise. 

To understand how the underlying framework for superpartner masses may be
relevant, consider a model with input
parameters defined at some high scale $\Lambda\gg \msusy$, where
$\msusy$ is the SUSY breaking scale $\sim 1$~TeV and $\Lambda$ may be
as high as $M_{\rm GUT}$ or even the reduced Planck mass $M_{P}$. Then 
\be
m_{H_u}^2(\msusy)=m_{H_u}^2(\Lambda )+\delta m_{H_u}^2 \label{eq:dmh} 
\ee 
where 
\be
\delta m_{H_u}^2\simeq
-\frac{3f_t^2}{8\pi^2}\left(m_{Q_3}^2+m_{U_3}^2+A_t^2 \right)
\log\left(\frac{\Lambda}{\msusy}\right) .  
\ee 
Requiring $\delta m_{H_u}^2 \leq \Delta \times \frac{m_h^2}{2}$
then leads for $m_h=125$~GeV to,
\be \sqrt{m_{\tst_1}^2+m_{\tst_2}^2} \alt 600 \ {\rm GeV}
\frac{\sin\beta}{\sqrt{1+R_t^2}}\left(\frac{\log\frac{\Lambda}{{\rm
TeV}}}{3}\right)^{-1/2}\left(\frac{\Delta}{5}\right)^{1/2}\;,
\label{eq:papucci}
\ee 
where $R_t= A_t/\sqrt{m_{\tst_1}^2+m_{\tst_2}^2}$.  Taking $\Delta =10$
and $\Lambda$ as low as 20~TeV corresponds to\cite{kn,ah,others}
\bi
\item $|\mu | \alt 200\ {\rm GeV}$,
\item $m_{\tst_i},\ m_{\tb_1}\alt 600\ {\rm GeV}$,
\item $m_{\tg}\alt 1.5-2\ {\rm TeV}$.
\ei
The last of these conditions arises because the squark radiative corrections
$\delta m_{\tst_i}^2\sim (2g_s^2/3\pi^2)m_{\tg}^2 \times \log\Lambda$.  Setting
the $\log$ to unity and requiring $\delta m_{\tst_i}^2<m_{\tst_i}^2$
then implies $m_{\tg}\alt 3m_{\tst_i}$, or $m_{\tg}\alt 1.5-2$~GeV for
$\Delta\alt 10$.  
Taking $\Lambda$ as high as $M_{GUT}$ leads to even tighter constraints:
$m_{\tst_{1,2}},m_{\tb_1}\alt 200$ GeV and $m_{\tg}\alt 600$ GeV, 
almost certainly in violation of LHC sparticle search constraints. 
Since (degenerate) first/second generation
squarks and sleptons enter into (\ref{eq:loopmin}) only at the two loop
level, these can be much heavier: beyond LHC reach and also possibly
heavy enough to
provide a (partial) decoupling solution to the SUSY flavor and $CP$
problems.  In gravity mediation where $m_{\tq}\sim m_{3/2}$, then one
also solves the cosmological gravitino problem\cite{linde,moroi} and in
GUTs one also suppresses proton decay.  Then we may also have
\bi
\item $m_{\tq,\tell}\sim 10-50$~TeV.  \ei 
The generic natural SUSY (NS) solution reconciles lack of a SUSY signal
at LHC with allowing for electroweak naturalness. It also predicts that
the $\tst_{1,2}$ and $\tb_1$ may soon be accessible to LHC searches. New
limits from direct top- and bottom-squark pair production searches,
interpreted within the context of simplified models, have begun to bite
into the NS parameter space\cite{lhc_stop}.  Of course, if
$m_{\tst_{1,2}},\ m_{\tb_1}\simeq m_{\tz_1}$, then the visible decay
products from stop and sbottom production will be soft and difficult to
see at the LHC.

A more worrisome problem comes from the newly discovered value of the
Higgs mass $m_h\simeq 125$~GeV.  In the MSSM, one obtains\cite{mhiggs} 
(assuming that the $t$-squarks are not very split),
\be
m_h^2\simeq M_Z^2\cos^2 2\beta +\frac{3g^2}{8\pi^2}\frac{m_t^4}{m_W^2}\left[
\ln\frac{m_{\tst}^2}{m_t^2}+\frac{X_t^2}{m_{\tst}^2}\left(1-\frac{X_t^2}{12m_{\tst}^2}\right)\right]
\ee
where $X_t=A_t-\mu\cot\beta$ and $m_{\tst}^2\simeq m_{Q_3}m_{U_3}$.  For
a given $m_{\tst}^2$, this expression is maximal for large mixing in the
top-squark sector with $X_t^{max}=\sqrt{6}m_{\tst}$.  With top-squark
masses below about 500~GeV, the radiative corrections to $m_h$ are not
large enough to yield $m_h\simeq 125$~GeV even with maximal
mixing\cite{hpr,bbht}.  This situation has been used to argue that
additional multiplets beyond those of the MSSM must be present in order
to raise up $m_h$ while maintaining very light third generation
squarks\cite{hpr}. Added to these are the two issues mentioned
earlier: 1. the very light third generation squarks\cite{bbht} endemic to
NS lead to a predicted branching fraction for $b\to s\gamma$ decay which
is frequently much lower than the measured value\cite{bbht}, and 2. that
the relic abundance of higgsino-like WIMPs inherent in NS, calculated in
the standard MSSM-only cosmology, is typically a factor 10-15 below
measured values\cite{bbht}.  These issues have led to increasing
skepticism of weak scale SUSY as realized in the natural SUSY
incarnation described above.

A possible resolution to the above issues associated with a NS spectrum
is to simply invoke a SUSY particle spectrum at the weak scale (or some other
nearby scale\cite{casas}), as in
the pMSSM model\cite{pmssm} so that large logarithms associated with a
high value of $\Lambda$ are absent.  In this case, $\Lambda\sim \msusy$ and 
$\delta m_{H_u}^2$ is not enhanced by large logarithms
and we may select parameters
$m_{H_u}^2\sim \mu^2\sim M_Z^2\sim m_h^2$. Of course,
heavy top squarks are needed to obtain the observed value of $m_h$.
While a logical possibility, this
solution loses several attractive features of models which are valid up
to scales as high as $\Lambda\sim M_{\rm GUT}$, such as gauge coupling
unification and radiative electroweak symmetry breaking driven by a
large top quark mass.

Another alternative is to use $\delew$ defined above as a
fine-tuning measure even for models defined at the high scale.  This use
of weak scale parameters to define the fine-tuning criterion is a {\em
weaker condition} since it allows for possible cancellations in
(\ref{eq:dmh}).  Indeed this is precisely what happens in what is known
as the hyperbolic branch or focus point region (HB/FP) of
mSUGRA\cite{hb_fp}: $m_{H_u}^2(\Lambda)+\delta m_{H_u}^2\sim
m_{H_u}^2(\msusy)\sim \mu^2 \sim M_Z^2$.
The HB/FP region of mSUGRA occurs, however, only for small values of
$A_0/m_0$~\cite{sugra} and yields $m_h < 120$~GeV, well below the
Atlas/CMS measured value of $m_h\simeq 125$~GeV.  Scans over parameter
space show that the HB/FP region is nearly excluded if one requires both
low $|\mu |$ and $m_h\sim 123-127$~GeV~\cite{dm125,sugra}.

To obtain a viable high scale model we see that we clearly need to go
beyond mSUGRA.  The small value of $-m_{H_u}^2(M_{SUSY})$ that we
require for low EWFT can be obtained in several ways. For instance, we
could introduce non-universality of gaugino masses and adopt very high
GUT scale values of the $SU(2)$ gaugino mass parameters\cite{him2}, or a
low value of the $SU(3)$ gaugino mass parameter\cite{lm3}. Both choices
would lead to a larger chargino to gluino mass ratio than in models with
universal gaugino masses\cite{shafi}, and since charginos couple
directly to the Higgs sector, potentially significant contributions to
the radiative corrections for gluinos that satisfy the LHC bound. The
other way of obtaining small values of $-m_{H_u}^2(M_{SUSY})$ without
undue cancellations in (\ref{eq:loopmin}) is to introduce
non-universality in the scalar Higgs sector.  To facilitate our
analysis, we use the two parameter non-universal Higgs mass (NUHM2)
extension\cite{nuhm2} of the mSUGRA model where $m_{H_u}^2(M_{\rm GUT})$
and $m_{H_d}^2(M_{\rm GUT})$, or equivalently, the weak scale parameters
$\mu$ and $m_A$, are chosen independently of matter scalar mass
parameters, and the model is completely specified by the parameter set,
\be
m_0,\ m_{1/2},\ \mu,\ m_A,\ A_0,\ \tan\beta \qquad {\rm (NUHM2)}. 
\label{eq:nuhm2}
\ee
Modest electroweak fine-tuning
is then obtained due to large cancellations between
$m_{H_u}^2(\Lambda=M_{\rm GUT})$ and $\delta m_{H_u}^2$. 

Along with $-m_{H_u}^2(M_{SUSY})\sim M_Z^2/2$, low $\Delta_{\rm EW}$ 
also requires $\mu^2\sim M_Z^2/2$. In gravity-mediated SUSY breaking models--
where the $\mu$ problem is solved by the Guidice-Masiero mechanism\cite{gm}--
one expects that $\mu\sim \lambda m_{3/2}$ where $\lambda$ is a hidden-visible 
sector coupling. For small values of $\lambda$, then we expect 
$|\mu | \ll m_{3/2}$ which is then significantly smaller than the 
typical soft-SUSY breaking masses. 

Later in our analysis, we will also allow for the possibility of split matter
generations where the third generation mass parameter $m_0(3)$ is
independent from the corresponding parameter $m_0(1,2)$ for the first
two generations. We refer to this case as the NUHM3 model. 
The NUHM3 model allows for
an improved decoupling solution to the SUSY flavour problem. Finally, we
mention that a small magnitude of $\mu$ is also possible in the NUHM1
model where we take the two GUT scale Higgs mass parameters to have a
common value $m_\phi^2$ which can be raised above $m_0$ until
$m_{H_u}^2(\msusy)$ becomes comparable to $M_Z^2$~\cite{nuhm1}. In this
case, $m_A$ is of course determined. We consider this case briefly.  In
all cases, intra-generation splitting is avoided since as noted in
Ref.~\cite{sugra}, it can lead to large fine-tuning if scalars are very
heavy. We emphasize that $\delew$ is determined by
physical sparticle masses and couplings so that {\it our results can be
applied to any model that yields a similar spectrum}, irrespective of
how sparticles acquire their SUSY breaking mass.

At this point the reader may legitimately wonder about the validity of
using $\delew$ as a measure of fine-tuning in the NUHM2 model which is,
by construction, assumed to be a description of physics up to energy
scales as large as $M_{\rm GUT}$. The introduction of the $\Delta_{\rm
HS}$ in Ref.~\cite{sugra} was precisely to include the impact of these
large logarithms -- which also appear in (\ref{eq:papucci}) -- on the
fine-tuning. The use of $\delew$ as a fine-tuning measure allows for the
possibility of large cancellation between $m_{H_{u,d}}^2(\Lambda)$ and
the term $\delta m_{H_{u,d}}^2$ (which may include large logarithms).
For instance, special regions of parameter space of some models ({\it
e.g.}  focus-point SUSY, the mixed-modulus-anomaly-mediation model for
special values of the ratio $\alpha$ \cite{nilles} or particular regions
of parameter space of non-universal Higgs mass models) enjoy nearly
complete cancellations between the terms with large logarithms and
$m_{H_{u,d}}^2(\Lambda)$. In a more encompassing framework that includes
the origin of soft SUSY breaking parameters, such cancellations might
not only be allowed, but might be automatic~\cite{nilles,antusch}.
We note, however, that $\Delta_{\rm HS}$, as we have defined it, does 
{\it not} take such a cancellation into account; under these circumstances 
$\Delta_{\rm EW}$ is the appropriate fine-tuning measure to use.

The fine-tuning measure $\delew$ introduced in Ref's. \cite{ltr} and \cite{sugra} has several attractive features that merit consideration. 
\bi
\item {\it Model independent} (within the context of models which reduce 
to the MSSM at the weak scale): 
$\delew$ is essentially determined by the sparticle
spectrum\cite{sugra}, and -- unlike $\Delta_{\rm HS}$ and other
measures of fine-tuning -- does not depend on the mechanism by which
sparticles acquire masses.
Since $\delew$ is determined only from weak scale Lagrangian parameters, the
phenomenological consequences which may be derived by requiring low
$\Delta_{EW}$ will apply not only for the NUHM2 model considered here,
but also for other possibly more complete (or less complete, such as pMSSM) 
models which lead to look-alike spectra at the weak scale.
\item {\it Conservative}: $\delew$ captures the minimal fine-tuning 
that is necessary for any given sparticle spectrum, 
and so leads to the {\em most conservative
conclusions} regarding fine-tuning considerations.
\item {\it Measureable}: $\delew$ is in principle measurable in that it 
can be evaluated   if the underlying weak scale parameters 
can be extracted from data.
\item {\it Unambiguous}: Fine-tuning measures which depend 
on high scale parameter choices, such as the
Barbieri-Guidice measure $\Delta_{\rm BG}$ discussed previously, are
highly sensitive to exactly which set of model input parameters 
one adopts: for example, it is well-known that significantly different
values of $\Delta_{\rm BG}$ result depending on whether the
high scale top-Yukawa coupling is or is not included as an 
input parameter\cite{fmm}. 
There is no such ambiguity in the fine-tuning sensitivity as
measured by both $\delew$ and $\Delta_{\rm HS}$.
\item {\it Predictive}: While $\delew$ is less restrictive than $\Delta_{HS}$, 
it still remains highly restrictive. The requirement of low $\Delta_{EW}$ 
highly disfavors models such as mSUGRA/CMSSM\cite{sugra}, 
while allowing for very distinct predictions from more general 
models such as NUHM2.
\item {\it Falsifiable}: The most important prediction from 
requiring low $\Delta_{EW}$ is that $|\mu |$ cannot be too far removed from 
$M_Z$. This implies the existence of light higgsinos $\sim 100-300$ GeV 
which are hard to see at hadron colliders, but which are easily detected at 
a linear $e^+e^-$ collider with $\sqrt{s}\agt 2|\mu |$. If no
higgsinos appear at ILC1000, then the idea of electroweak naturalness
in SUSY models is dead.
\item {\it Simple to calculate}: 
$\delew$ is extremely simple to encode in sparticle mass spectrum
programs, even if one adopts models with very large numbers of 
input parameters.
\ei 

To illustrate how a low value of $m_{H_u}^2(\msusy)$ is obtained, in
Fig.~\ref{fig:evol} we show the running of various SUSY parameters
versus the renormalization scale $Q$ for the RNS2 benchmark point from
Ref.~\cite{ltr}.  The RNS2 point has parameters $m_0=7025$~GeV,
$m_{1/2}=568.3$~GeV, $A_0=-11426.6$~GeV, $\tan\beta =8.55$ with $\mu
=150$~GeV and $m_A=1000$~GeV.  The gaugino and matter scalar mass
parameters evolve from $m_{1/2}$ and $m_0$ to their weak scale values,
resulting in a pattern of masses very similar to that in mSUGRA. The
parameter $\mu$ hardly evolves, and for such a low value of $\tan\beta$,
$m_{H_d}^2$ also suffers little evolution. Of most interest to us here
is the RG evolution of $m_{H_u}^2$. As is well known, the SUSY breaking
parameters $m_{Q_3}^2$, $m_{U_3}^2$ and $m_{H_u}^2$ of the scalar fields
that couple via the large top quark Yukawa coupling are driven down with
reducing values of the scale $Q$. The reduction is the greatest for
$m_{H_u}^2$ which, in fact, is driven negative, triggering the radiative
breakdown of electroweak symmetry \cite{rewsb}. We see from the figure
that the weak scale value of $-m_{H_u}^2$ has a magnitude $\sim M_Z^2$,
and is much smaller than the weak scale value of other mass parameters.
This is not an accident because {\it the NUHM2 model provides us the
flexibility to adjust the GUT scale value of $m_{H_u}^2$ so that it
barely runs to negative values at the weak scale}.  Since $m_{H_u}^2$ is
driven radiatively to $\sim -M_Z^2$ at the weak scale, this scenario has
been dubbed {\it Radiative Natural SUSY}, or RNS for short.
\FIGURE[tbh]{
\includegraphics[width=10cm,clip]{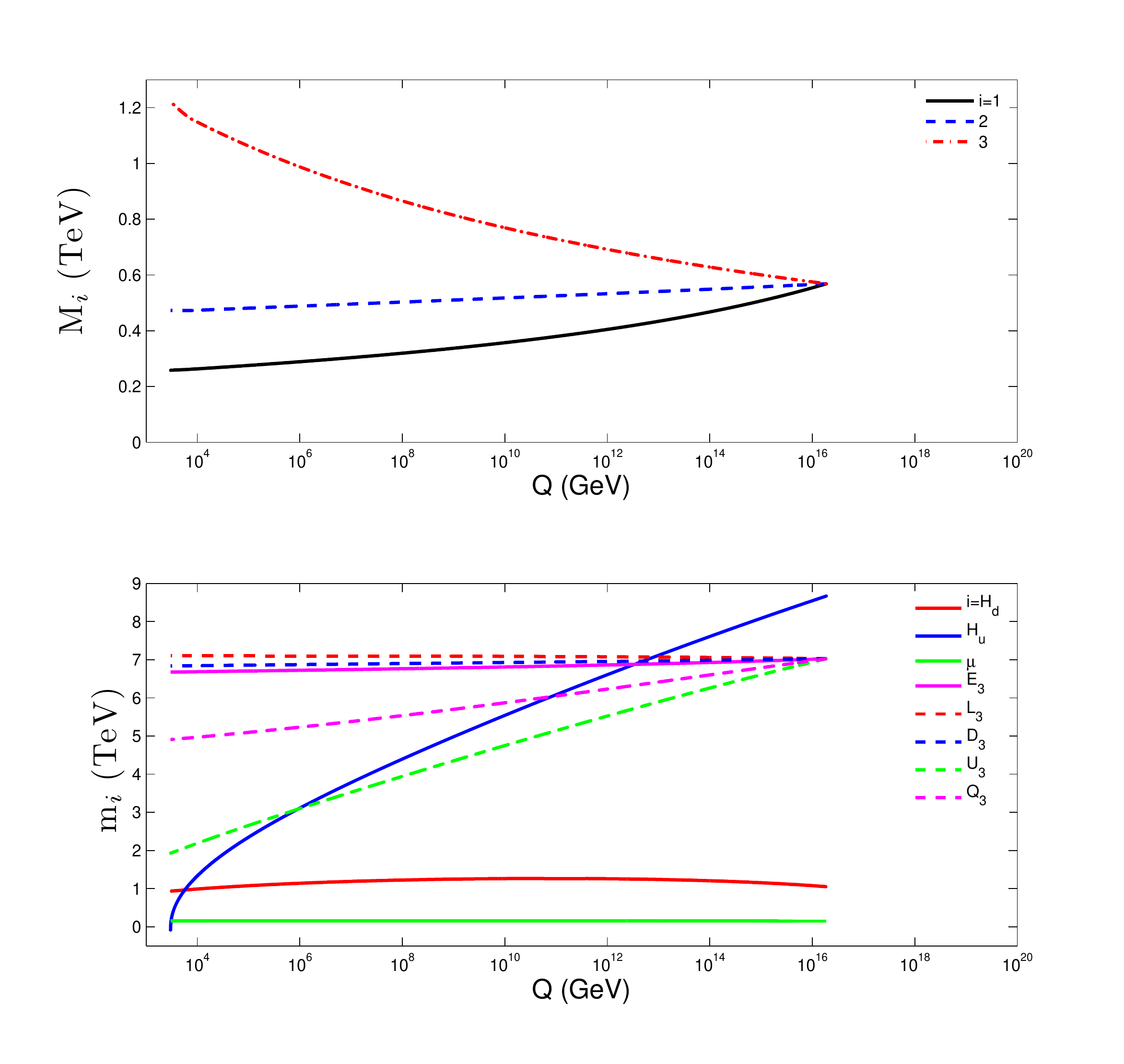}
\caption{Evolution of SSB parameters from $M_{\rm GUT}$ to $M_{weak}$ for
the RNS2 benchmark point taken from in Ref.~\cite{ltr} whose parameters
are given in the text. The graph extends to values below
$Q^2=m_{\tst_1}m_{\tst_2}$ where the Higgs mass parameters are extracted. }
\label{fig:evol}}

In Fig.~\ref{fig:DeltaHS} we scan over parameter space of the NUHM2
model -- while enforcing $m_h=125\pm 2$ GeV and LHC sparticle mass
limits -- and plot the value of $\Delta_{\rm HS}$ versus the high scale
matter scalar mass parameter $m_0$. We see that the smallest value of
$\Delta_{\rm HS}$ is $10^3$ for the lowest values of $m_0$
allowed. This is because of the large logarithms that we discussed above.
The NUHM2 would be fine-tuned to at least 0.1\%, and
usually even higher fine-tuning is necessary. We will see below
that much smaller values of $\Delta_{\rm EW}$ are possible in some parts of
parameter space. An underlying  high-scale
theory that automatically leads to an NUHM2-like spectrum in this
parameter region would then not be so fine-tuned. In contrast, there are
no analogous regions of mSUGRA/CMSSM parameters for which we have shown
that $\Delta_{\rm EW} \agt 100$\cite{sugra}.
\FIGURE[tbh]{
\includegraphics[width=10cm,clip]{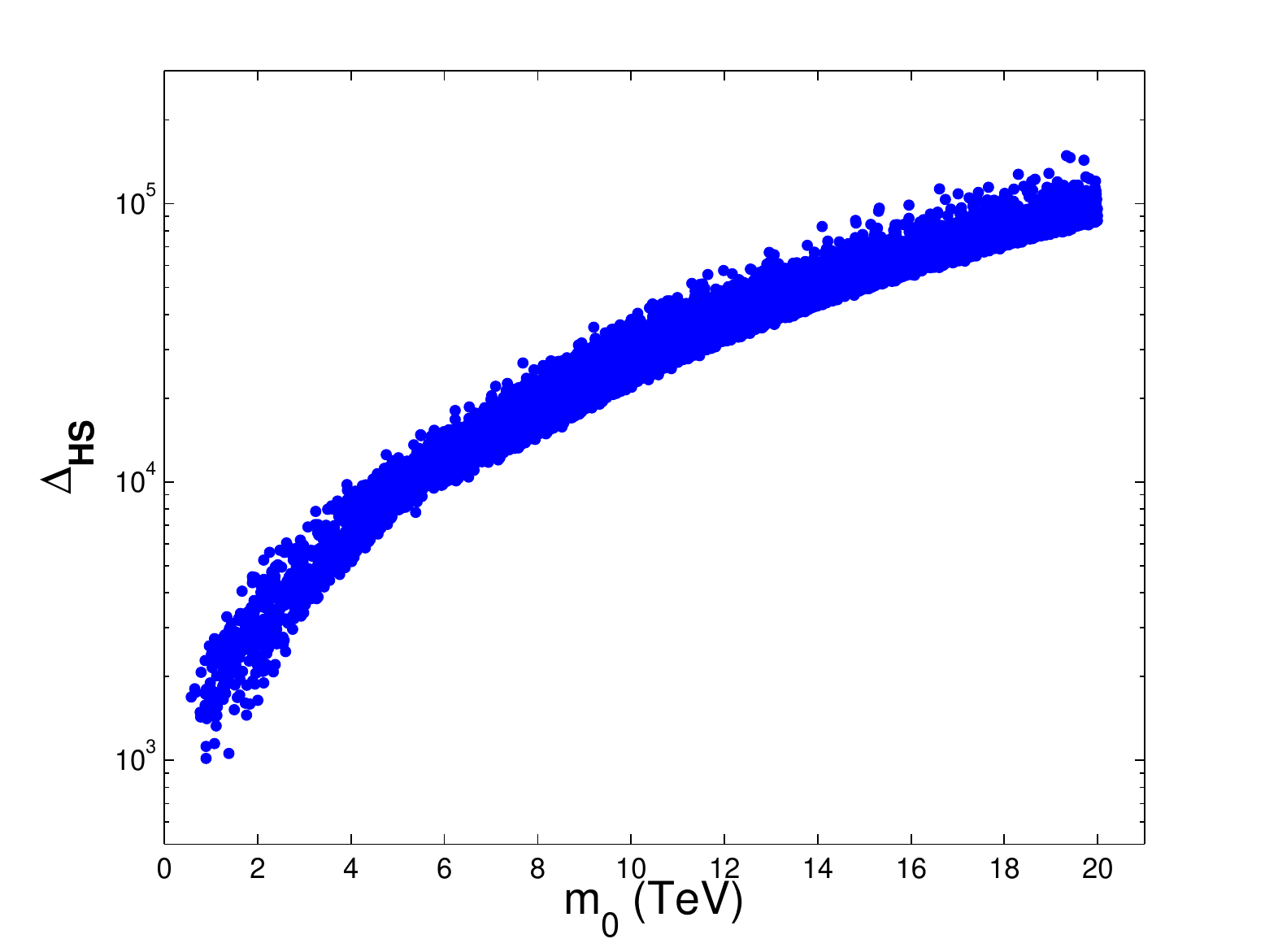}
\caption{Plot of $\Delta_{\rm HS}$ versus $m_0$ from a
scan over NUHM2 model parameters, accepting only points where
$m_h=125\pm 2$ GeV and which obey LHC sparticle mass constraints.  }
\label{fig:DeltaHS}}

In the remainder of this paper, we will explore what parameter choices
lead to low values of $\Delta_{\rm EW}$.  While $\Delta_{EW}$ seems
bounded from below by about 100 in mSUGRA/CMSSM\cite{sugra}, we will
find that $\Delta_{EW}$ as low as $\sim 10$ can be obtained in NUHM2. In
addition, requiring low $\Delta_{EW}\alt 30$ places strong restrictions
on the allowed sparticle mass spectra, leading to distinctive
predictions for collider and dark matter searches.

\subsection{Radiative natural supersymmetry}
\label{sec:rns}

Motivated by the possibility of cancellations occurring in
$m_{H_u}^2(\msusy)$, we return to the EWSB minimization condition
(\ref{eq:loopmin}) which was introduced earlier, and examine more
carefully the radiative corrections embodied in $\Sigma_u^u$ and
$\Sigma_d^d$ that we have not discussed up to now. These affect the
minimization condition in an important way when 
$m_{H_u}^2(m_{\rm SUSY})$ and $\mu^2$ are much smaller 
than the scale of other weak scale SUSY breaking parameters. 
At the one-loop level, $\Sigma_u^u$
contains the contributions\cite{an,deboer} $\Sigma_u^u(\tst_{1,2})$,
$\Sigma_u^u(\tb_{1,2})$, $\Sigma_u^u(\ttau_{1,2})$,
$\Sigma_u^u(\tw_{1,2})$, $\Sigma_u^u(\tz_{1-4})$, $\Sigma_u^u(h,H)$,
$\Sigma_u^u(H^\pm)$, $\Sigma_u^u(W^\pm)$, $\Sigma_u^u(Z)$, and
$\Sigma_u^u(t)$. $\Sigma_d^d$ contains similar terms along with
$\Sigma_d^d(b)$ and $\Sigma_d^d(\tau)$ while $\Sigma_d^d(t)=0$.  The
complete set of one-loop contributions to these is listed in the
Appendix.  There are additional contributions from first/second
generation sparticles from their $D$-term couplings to Higgs scalars.
If these squarks, and separately sleptons, are degenerate then these
contributions cancel within each generation because the sum of weak
isospins/hypercharges of squarks/sleptons total to zero\cite{sugra}.
In the parameter space region where RNS is realized,
{\it i.e.} where $-m_{H_u}^2(\msusy)\sim \mu^2\sim M_Z^2$, the radiative correction
terms from $\Sigma_u^u$ may give the largest contributions to
$\delew$.

The largest of the $\Sigma_u^u$ terms almost always come from top
squarks for which we find,
\bea
\Sigma_u^u(\tst_{1,2} )&=&\frac{3}{16\pi^2}F(m_{\tst_{1,2}}^2)\times
\left[ f_t^2-g_Z^2\mp \frac{f_t^2 A_t^2-8g_Z^2(\frac{1}{4}-\frac{2}{3}x_W)\Delta_t}{m_{\tst_2}^2-m_{\tst_1}^2}
\right]
\label{eq:Siguu}
\eea
where $\Delta_t=(m_{\tst_L}^2-m_{\tst_R}^2)/2+M_Z^2\cos
2\beta(\frac{1}{4}-\frac{2}{3}x_W)$, $g_Z^2=(g^2+g^{\prime 2})/8$,
$x_W\equiv \sin^2\theta_W$ and $F(m^2)=m^2\left(\log
(m^2/Q^2)-1\right)$, with $Q^2=m_{\tst_1}m_{\tst_2}$.  In
Ref.~\cite{ltr}, it is shown that for the case of the $\tst_1$
contribution, as $|A_t|$ gets large there is a suppression of
$\Sigma_u^u(\tst_1)$ due to a cancellation between terms in the square
brackets of Eq.~(\ref{eq:Siguu}).  The $\tst_2$ contribution is
suppressed
if there is a
sizeable splitting between $m_{\tst_2}$ and $m_{\tst_1}$ due to a large
cancellation within $F(m_{\tst_2}^2)$ because
$\log(m_{\tst_2}^2/Q^2) = \log (m_{\tst_2}/m_{\tst_1})\simeq 1$.
The large $|A_t|$ values suppress both top squark
contributions to $\Sigma_u^u$, and at the same time lift up the
value of $m_h$, which is near maximal for large negative $A_t$.
Combining all effects, one sees that the same mechanism responsible for
boosting the value of $m_h$ into accord with LHC measurements can also
suppress the $\Sigma_u^u$ contributions to EWFT, leading to a model with
low EWFT.

To display the quality of EWFT explicitly, we show in
Fig.~\ref{fig:mzcont}{\it a} the various {\it signed} contributions to
$M_Z^2/2$ that enter Eq.~(\ref{eq:loopmin}) for the RNS2 point from
Fig.~\ref{fig:evol} and Ref.~\cite{ltr}.  In this figure, we
label these signed contributions by $C_i$ where $i=H_u, H_d, \mu, \Sigma_u^u,
\Sigma_d^d$.  The largest contributions come from $C_{\Sigma_u^u}\sim
0.04$~TeV$^2$ and $C_{H_u}\sim -0.03$~TeV$^2$.  In frame {\it b}), we
show these same quantities for the mSUGRA model (where $\mu$ and $m_A$
are outputs instead of input parameters).  Here, the maximal
contributions $C_{H_u}\sim 15$~TeV$^2$ and $C_\mu\sim -15$~TeV$^2$.
Frame {\it c}) compares results from the two models using a common
scale.  Here, it is clearly seen that the mSUGRA model is enormously
fine-tuned compared to the RNS2 benchmark point.
\FIGURE[tbh]{
\includegraphics[width=13cm,clip]{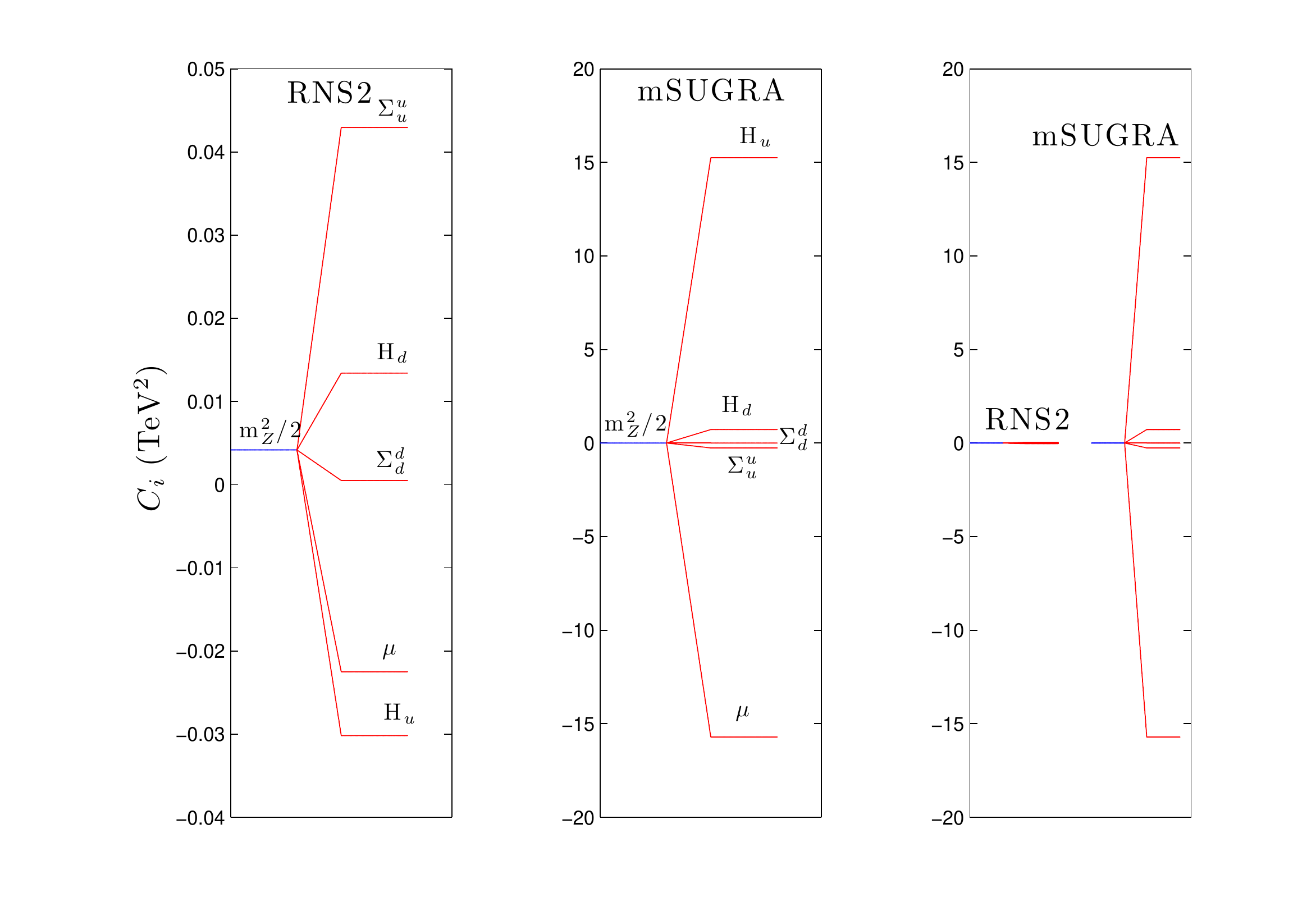}
\caption{Signed contributions to $M_Z^2/2$ from terms in the EWSB
minimization condition Eq.~\ref{eq:loopmin} from {\it a}) the RNS2
benchmark point defined in the text and {\it b}) the corresponding
mSUGRA model as RNS2 with $\mu$ and $m_A$ as outputs rather than inputs.
In frame {\it c}), the results for both models are plotted on a common
scale.  }
\label{fig:mzcont}}

Our goal in this paper is to provide a rather complete characterization
of radiative natural SUSY.  This should provide a
comprehensive picture as to where in model parameter space we can find
1)~$m_h\sim 125$~GeV along with 2)~low EWFT $\delew\alt 30$
while at the same time 3)~respecting LHC constraints on sparticle
masses.  With this goal in mind, in Sec.~\ref{sec:nuhm2} we show
parameter space regions leading to low $\delew$ from scans over
the 2-parameter non-universal Higgs model NUHM2 which allow for
radiative natural SUSY.  In Sec.~\ref{sec:nuhm3} we extend the results
to include the split generation non-universal Higgs model NUHM3, wherein
high scale third generation scalar masses $m_0(3)$ need not equal
first/second generation scalar masses $m_0(1,2)$.  While the former
implementation allows for fewer parameters, the additional freedom in
the NUHM3 model allows for a more robust decoupling solution to the SUSY
flavor and $CP$ problems because heavier multi-TeV first/second
generation sfermion masses are then possible.  In Sec.~\ref{sec:b}, we show
that constraints from $B$-physics -- especially $BF(b\to s\gamma )$ are
much more easily respected in RNS than in generic NS models.  In
Sec.~\ref{sec:lhc}, we discuss prospects for detecting RNS at the LHC.
We also show the new RNS $m_0\ vs.\ m_{1/2}$ parameter plane, which
offers a template for future searches for RNS at the LHC.
Searches for RNS at the ILC are
discussed in Sec.~\ref{sec:ilc} while direct and indirect detection of
higgsino-like WIMPs is discussed in Sec.~\ref{sec:dm}.  In an Appendix,
we present formulae needed for implementation of our measure of
electroweak fine-tuning $\delew$.

\section{Radiative natural SUSY from the NUHM models}
\label{sec:nuhm2}

The direct supersymmetrization of the Standard Model -- augmented by weak
scale soft supersymmetry breaking terms -- leads to the Minimal
Supersymmetric Standard Model (MSSM).  Since the mass scale of the
MSSM is stable to radiative corrections even when the MSSM is embedded
into a high scale framework, it is tempting to speculate that the MSSM
arises as the low energy limit of an underlying SUSY grand unified
theory with a unification scale $M_{\rm GUT}\simeq 2\times 10^{16}$~GeV.
Indeed, the MSSM (possibly with additional gauge singlets and/or
additional complete $SU(5)$ multiplets) receives some indirect support
from experiment in that 1)~the measured weak scale gauge couplings
nearly unify at $M_{\rm GUT}$ under MSSM RG evolution,
2)~radiative corrections due to the large top quark Yukawa
coupling -- consistent with $m_t\sim 173$~GeV -- dynamically breaks electroweak
symmetry, and 3)~a light SM-like Higgs boson has been discovered
to be lying squarely within the narrow mass window predicted
by the MSSM.

Motivated by these successes, the interesting question arises as to
whether a natural SUSY sparticle mass spectrum, {\it i.e.} one with a
modest value of $\delew$, can be consistently generated from a model
with parameters defined at the high scale $Q=M_{\rm GUT}$.
Naturalness requires $|\mu| \sim M_Z\sqrt{\delew/2}$, whilst the
recently measured\cite{lhcb} value of the branching fraction $BF(B_s
\to\mu^+\mu^- )$ qualitatively agrees with the predicted SM value,
which in turn requires the $CP$ odd boson $A$ to be relatively
heavy. We are thus led to adopt the 2-parameter non-universal Higgs
model (NUHM2)\cite{nuhm2}, wherein weak scale values of $\mu$ and
$m_A$ may be used as inputs in lieu of GUT scale values of $m_{H_u}^2$
and $m_{H_d}^2$.\footnote{Since the Higgs fields belong to different
  multiplets from matter fields, it is easy to envisage models with
  independent SUSY breaking mass parameters for Higgs and matter
  scalars.
} In Sec.~\ref{ssec:nuhm2}, for simplicity we take a common GUT scale
mass parameter $m_0$ for {\it all} the matter scalars. Motivated by
Grand Unification, we also assume a common GUT scale gaugino mass
parameter.  Later in Sec.~\ref{sec:nuhm3} we also explore the
possibility of split first/second versus third generation matter scalars
where we allow the third generation GUT scale mass parameter $m_0(3)$ to
differ from $m_0(1,2)$ for the first/second generation scalars.
Universality within each generation is well-motivated by $SO(10)$ GUT
symmetry, since all matter multiplets of a single generation belong to a
16-dimensional spinor rep of $SO(10)$. We can also envisage some degree
of non-universality between $m_0(1)$ and $m_0(2)$ as long as both lie in
the tens of~TeV regime: such a scenario invokes a partial
decoupling-partial degeneracy solution to the SUSY flavor and $CP$
problems (for constraints from FCNC processes \cite{silvestrini}, see
{\it e.g.}  Ref.~\cite{nmh}).  For convenience, we will take
$m_0(1)=m_0(2)$.

\subsection{RNS from the NUHM2 model}
\label{ssec:nuhm2}

The NUHM2 model is defined by the NUHM2 parameter set (\ref{eq:nuhm2})
introduced earlier.
%
%
We take $m_t=173.2$~GeV throughout this paper.
For our calculations, we use the Isajet 7.83~\cite{isajet} SUSY
spectrum generator Isasugra\cite{isasugra}.  Isasugra begins the
calculation of the sparticle mass spectrum with input $\overline{DR}$
gauge couplings and $f_b$, $f_\tau$ Yukawa couplings at the scale
$Q=M_Z$ ($f_t$ running begins at $Q=m_t$) and evolves the 6 couplings
up in energy to scale $Q=M_{\rm GUT}$ (defined as the value $Q$ where
$g_1=g_2$) using two-loop RGEs.  We do not enforce the exact
unification condition $g_3 = g_1 = g_2$ at $M_{\rm GUT}$, since a few
percent deviation from unification can be attributed to unknown
GUT-scale threshold corrections~\cite{Hisano:1992jj}.  Next, we use
the SSB boundary conditions at $Q=M_{\rm GUT}$ and evolve the set
of 26 coupled two-loop MSSM RGEs~\cite{mv,yamada} back down in
scale to $Q=M_Z$.  Full two-loop MSSM RGEs are used for soft term
evolution, and the gauge and Yukawa coupling evolution includes
threshold effects in the one-loop beta-functions, so the gauge and
Yukawa couplings transition smoothly from the MSSM to SM effective
theories as different mass thresholds are passed.  In Isasugra, the
values of SSB terms which mix are frozen out at the scale $Q =
\msusy=\sqrt{m_{\tst_L} m_{\tst_R}}$, while non-mixing SSB terms are
frozen out at their own mass scale~\cite{isasugra}.  The scalar
potential is minimized using the RG-improved one-loop MSSM effective
potential evaluated at an optimized scale $Q=\msusy$ to account
for leading two-loop effects~\cite{haber}.  Once the tree-level
sparticle mass spectrum is obtained,  one-loop radiative
corrections are calculated for all sparticle and Higgs boson masses,
including complete one-loop weak scale threshold corrections for the
top, bottom and tau masses at scale $Q=\msusy$~\cite{pbmz}.  Since
Yukawa couplings are modified by the threshold
corrections, the solution must be obtained iteratively, with
successive up-down running until a convergence at the required
level is found.  Since Isasugra uses a ``tower of effective theories''
approach to RG evolution, we expect a more accurate evaluation of the
sparticle mass spectrum for models with split spectra than with
programs such as SuSpect, SoftSUSY or Spheno, which make an
all-at-once transition from the MSSM to SM effective theories.

Our goal in this section is to find parameter ranges of the NUHM2 model
which satisfy LHC sparticle and Higgs boson mass constraints while maintaining
a low level of EWFT. We will also calculate the allowed mass range for
various sparticles in low fine-tuned/phenomenologically viable parameter space.
Toward this end, we search for regions of the NUHM2 parameter space
with $\delew\alt 30$, where  fine-tuning is better than about 3\%.
We will also require
that our calculated light Higgs scalar mass lies within the range
$m_h=125\pm 2$~GeV to allow for an estimated  uncertainty
in our calculation of $m_h$. We will also require that the
parameters $m_0$ and $m_{1/2}$ respect the
recent LHC limits on squark and gluino masses obtained within
the mSUGRA model~\cite{atlas_susy,cms_susy}.

We search for radiative Natural SUSY solutions  by
first performing a random scan over the following NUHM2 parameter ranges:
\bea
m_0 &:& \ 0-20\ {\rm TeV}, \nonumber\\
m_{1/2} &:& \  0.3-2\ {\rm TeV},\nonumber\\
-3 &<& A_0/m_0 \ <3,\nonumber\\
\mu &:& \ 0.1-1.5\ {\rm TeV}, \label{eq:param}\\
m_A &:& \ 0.15-1.5\ {\rm TeV},\nonumber\\
\tan\beta &:& 3-60 . \nonumber
\eea
We require of our solutions that:
\bi
 \item electroweak symmetry be radiatively broken (REWSB),
 \item the neutralino $\tz_1$ is the lightest MSSM particle,
 \item the light chargino mass obeys the model
independent LEP2 limit, $m_{\tw_1}>103.5$~GeV\cite{lep2ino},
\item LHC search bounds on $m_{\tg}$ and $m_{\tq}$ are respected,
\item $m_h=125\pm 2$~GeV.
\ei
%

To begin our investigation of NUHM2 model parameters leading to low
$\delew$, in Fig.~\ref{fig:nuhm2} we plot each scan point as a red ``+''
in frames of $\delew$ versus {\it a})~$m_0$, {\it b})~$m_{1/2}$, {\it
  c})~$A_0/m_0$, {\it d})~$\tan\beta$, {\it e})~$\mu$, and {\it
  f})~$m_A$.
Since low $\delew$ solutions are only
possible for low values of $\mu$, we have performed a separate narrow scan,
but this time with $\mu$ restricted between 100--300~GeV. The results of this
second scan is shown by the blue crosses in the figure.

We see from the plots that $\delew$ varies from as low as $\sim
10$ ($\Delta_{EW}^{-1}=10\%$ EWFT) to over 1000.  While the bulk of points shown are
fine-tuned with large $\delew\agt 100$, there do exist many
solutions with $\delew\alt 30$, corresponding to better than
3\% EWFT. The RNS solutions with $\delew\alt 30$ are obtained
for values of $m_0\sim 1-8$~TeV.  In the cases where $m_0$ is as high as
5-10~TeV, the top squark masses are driven to much lower values via
1)~the large top-quark Yukawa coupling $f_t$ which suppresses top-squark
soft masses during RG evolution, 2)~large mixing effects which can
suppress $m_{\tst_1}$ and yield a large $m_{\tst_1}-m_{\tst_2}$
splitting, and 3)~two-loop RGE suppression of diagonal top squark mass
terms arising from large first/second generation sfermion
masses\cite{am,graesser,imh}.  If $m_0$ is too large -- in this case
above $\sim 10$~TeV -- then these suppression mechanisms are insufficient
to drive $m_{\tst_{1,2}}$ to low enough values to allow for low EWFT.
Thus, the span of points shown in frame {\it a}) trends upward in
$\delew$ as $m_0$ increases past about 8~TeV.  We also see that
for the red pluses in frame {\it a}) $\delew$ has an upper
bound close to about 500 if $m_0\alt 10$~TeV. For still larger values of
$m_0$ then $\delew$ increases with $m_0$. This is because while
$\mu^2$ (or equivalently $-m_{H_u}^2$) is the largest of the quantities
in (\ref{eq:loopmin}) for the lower range of $m_0$, for very large
values of $m_0$ then $\Sigma_u^u$ begins to dominate.  The blue crosses
from the narrow scan with small $\mu$ have a different shape from the
red broad scan since the upper edge is mostly determined by
$\Sigma_u^u$, and so increases with $m_0$.
\FIGURE[tbh]{ \includegraphics[width=7cm,clip]{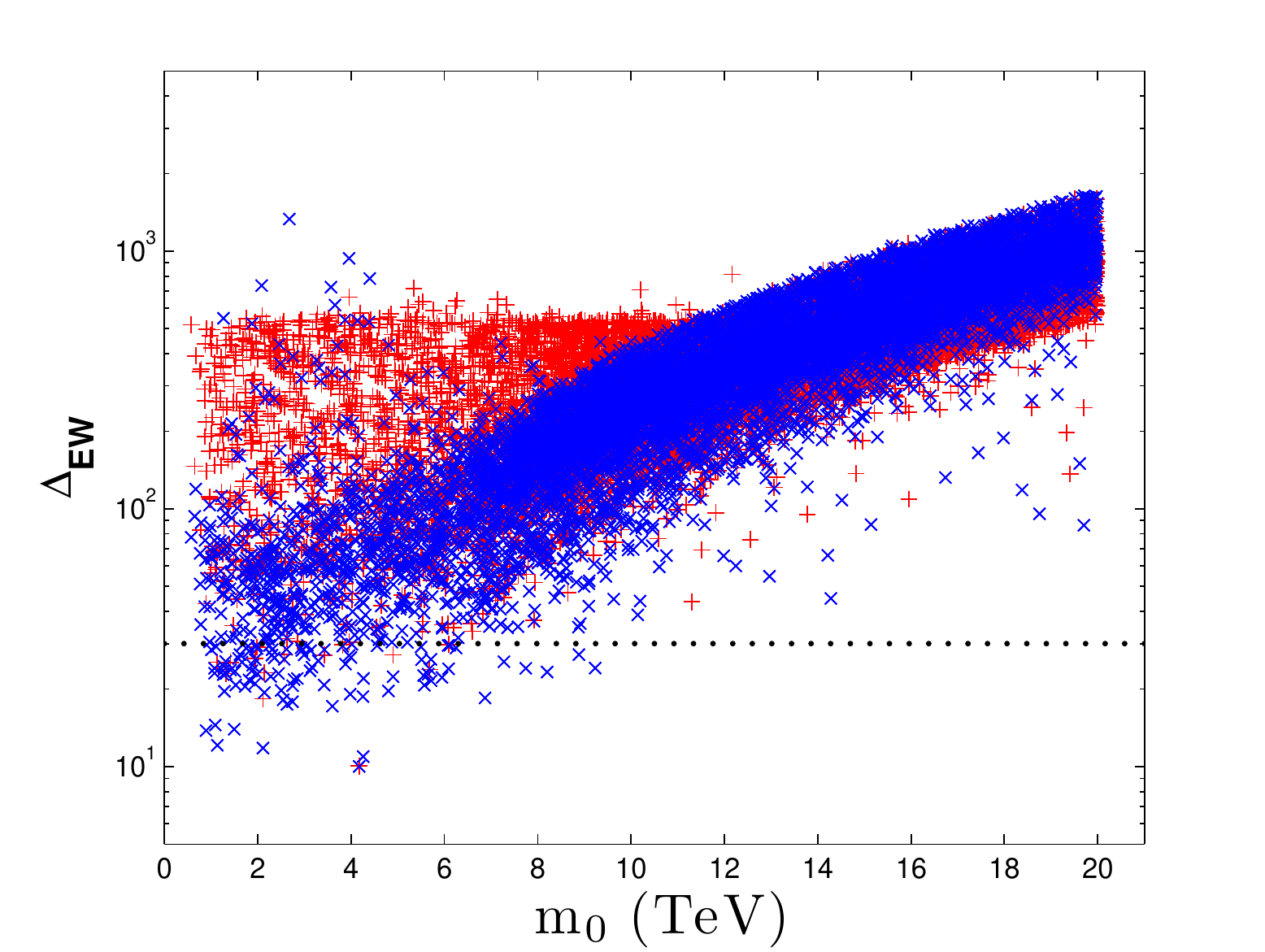}
  \includegraphics[width=7cm,clip]{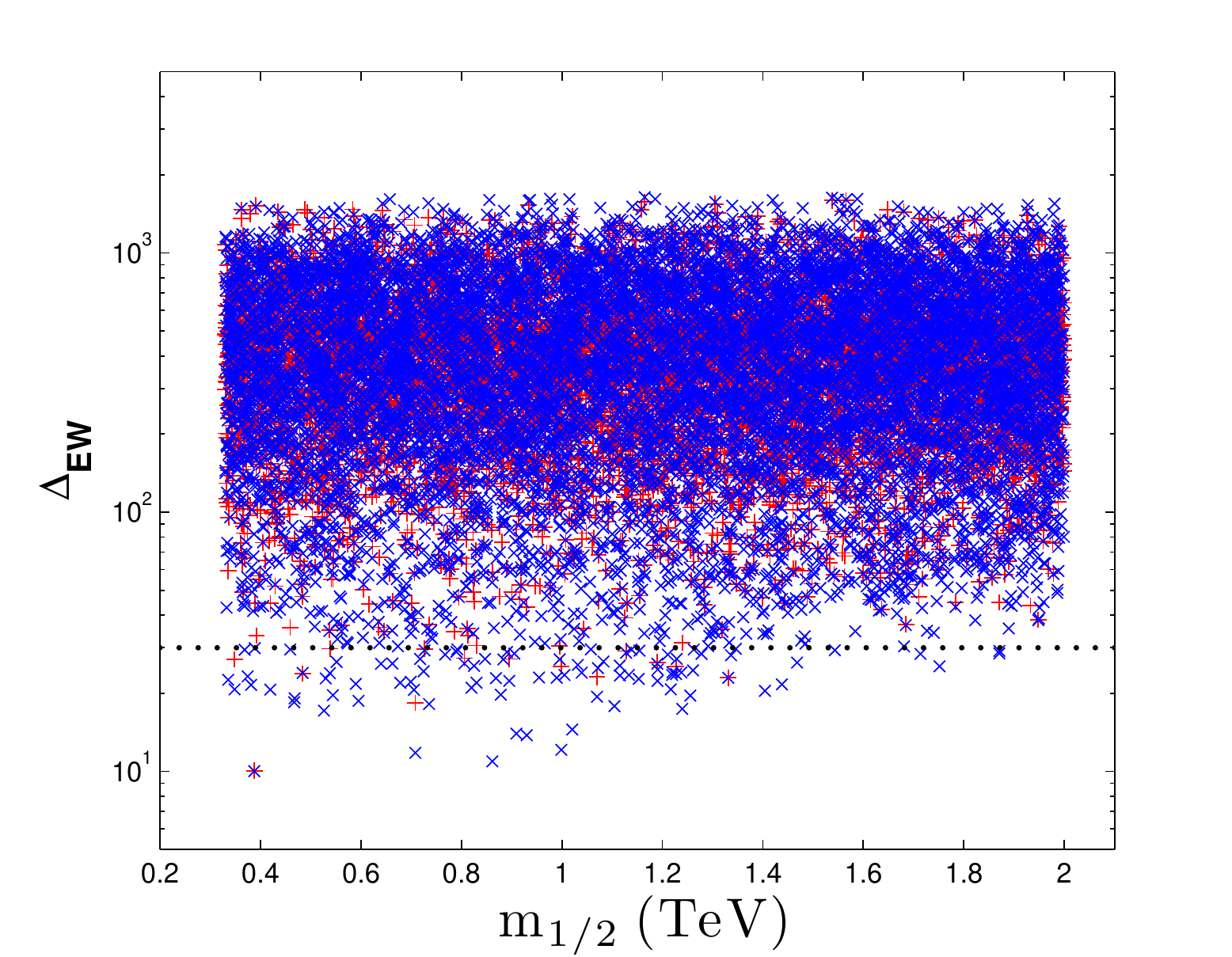}
  \includegraphics[width=7cm,clip]{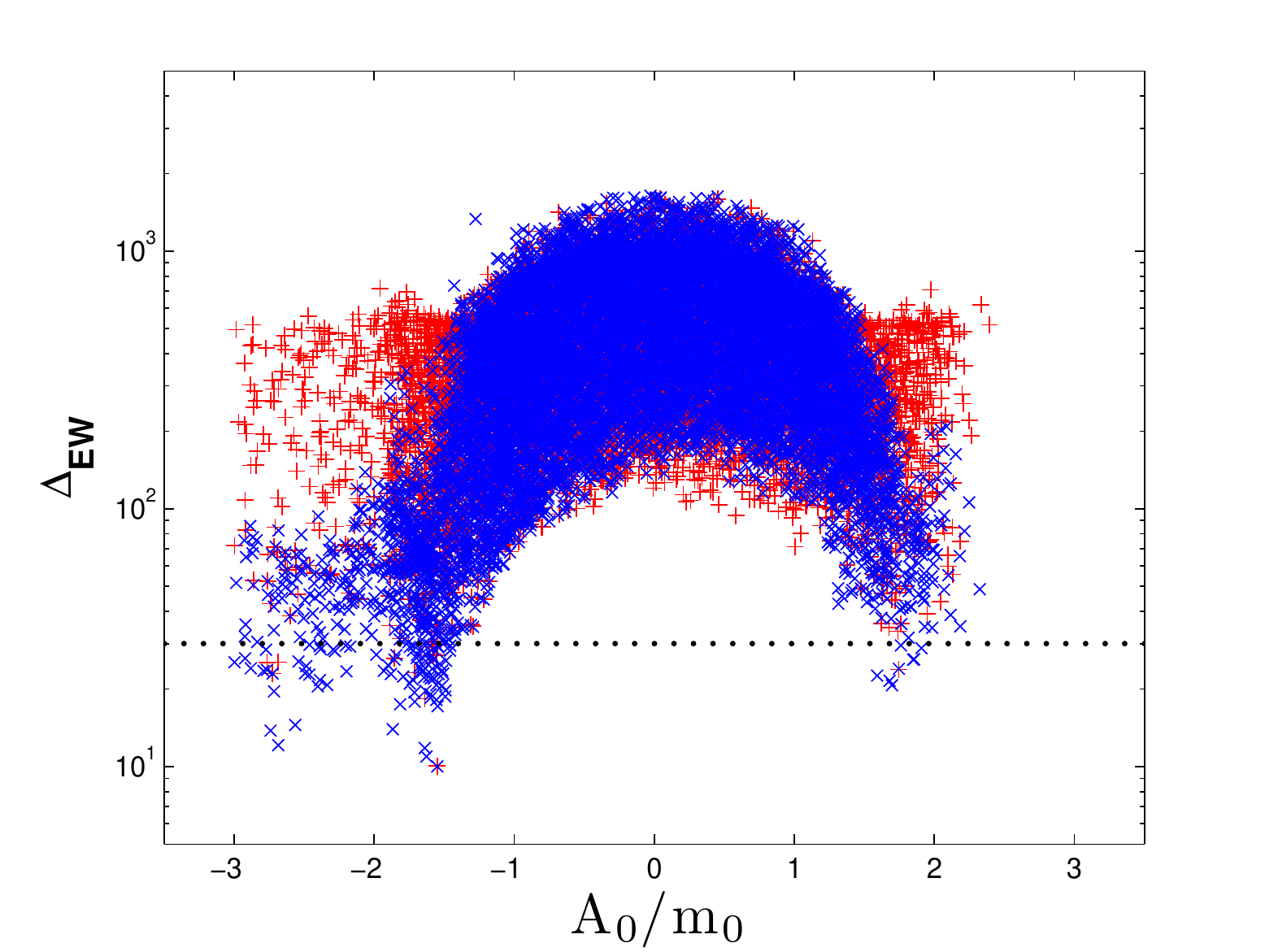}
  \includegraphics[width=7cm,clip]{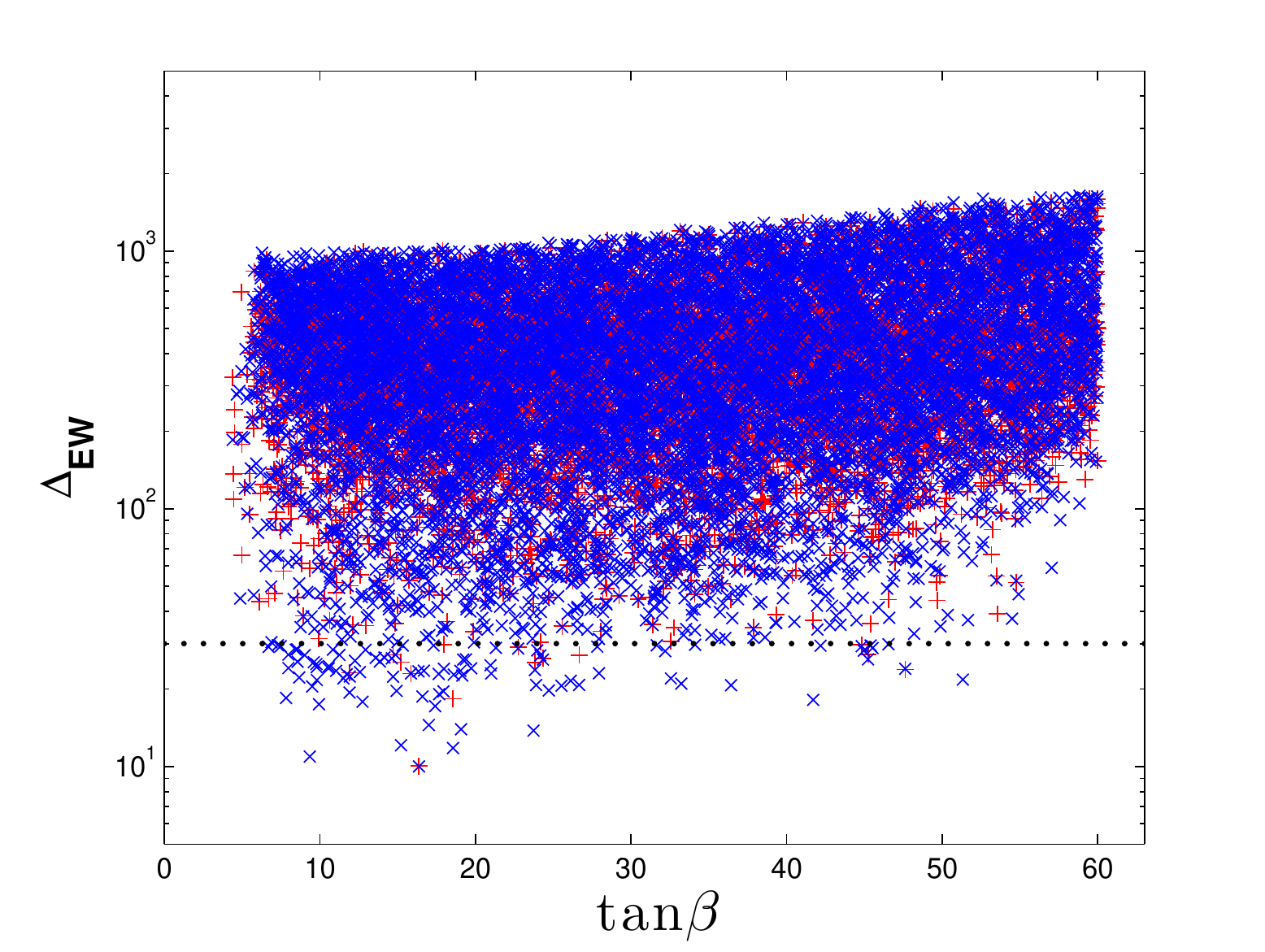}
  \includegraphics[width=7cm,clip]{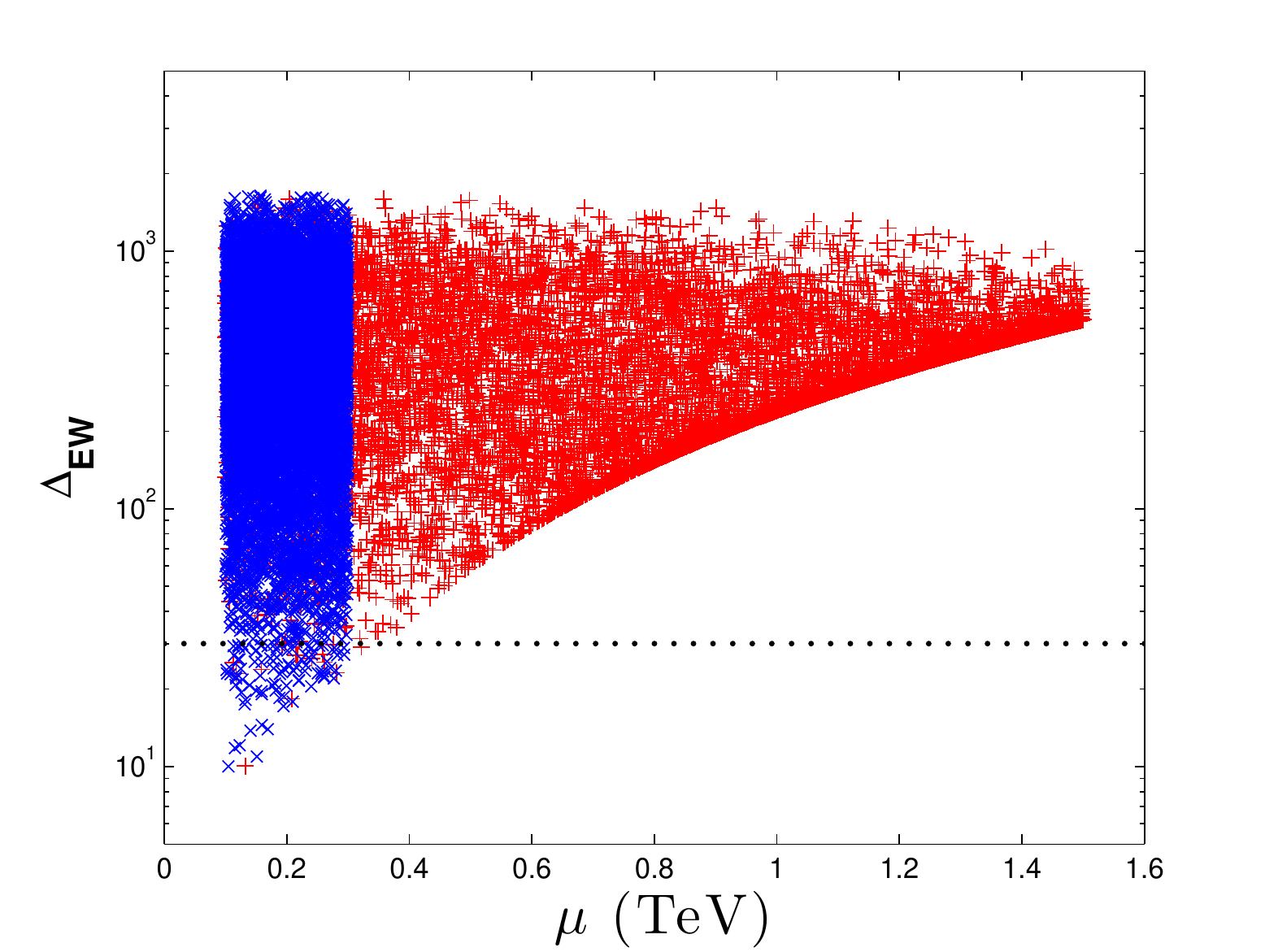}
  \includegraphics[width=7cm,clip]{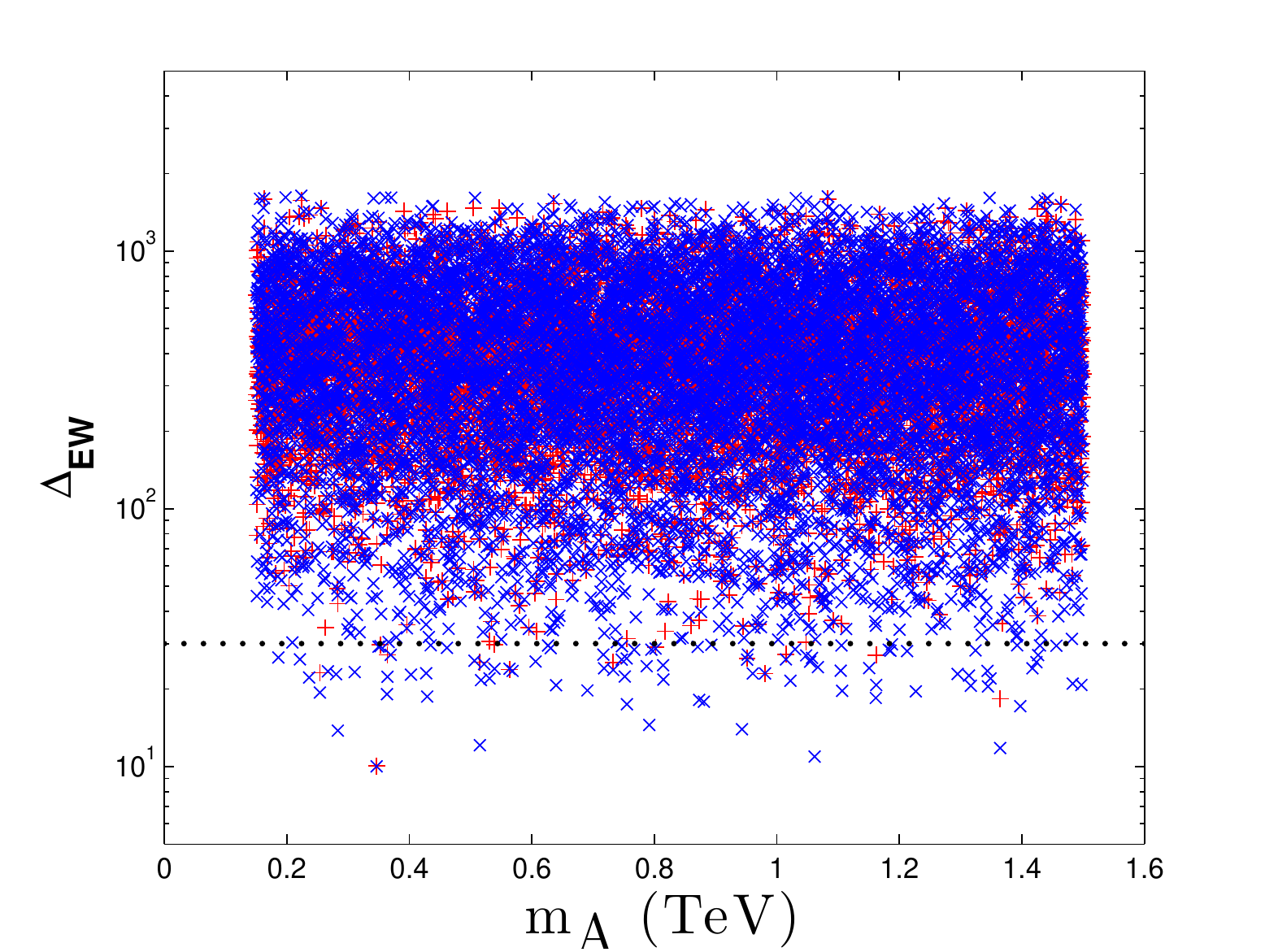}
\caption{The dependence of $\delew$ on various NUHM2 parameters
  from a scan (\ref{eq:param}) over parameter space (red pluses) and
  for the dedicated scan with 100~GeV$<\mu<$300~GeV (blue
  crosses). The line at $\delew=30$ is to guide the eye.  }
\label{fig:nuhm2}}

In frame {\it b}) of Fig.~\ref{fig:nuhm2}, we show
$\delew\ vs.\ m_{1/2}$. Here, the low values of $\delew$
span a wide range of $m_{1/2}$ values from $0.3-1.5$~TeV. Since
$m_{\tg}\sim (2.5-3)m_{1/2}$, we expect $\delew\alt 30$ for
$m_{\tg}$ values up to about 4~TeV.  For the entire parameter space
(red pluses) $\delew$ is roughly evenly distributed with
respect to the gaugino mass parameter.
In frame {\it c}), we show $\delew\ vs.\ A_0/m_0$. We see a
clear trend for low values of EWFT when $|A_0/m_0|\sim  1.5-2$.
The reason is that the hole at low magnitudes of $A_0/m_0$ and
small values of $\delew$
occurs because of the Higgs mass constraint.
Large magnitudes of GUT scale $A_0$ lead to
correspondingly large weak scale $A_t$ parameters,
which, in turn, provide large mixing in the top-squark sector.  This leads to
low EWFT and also heightened values of $m_h\sim 125$~GeV.  Frame {\it
  d}) shows $\delew\ vs.\ \tan\beta$. We see a slight preference
for low $\tan\beta \sim 10-20$ but otherwise no structure to speak of.
Frame {\it e}) shows $\delew$ versus the weak scale value of
$\mu$. The parabolic lower edge of the span of points reflects the upper
bound on $\mu$ necessary for low EWFT.  From the plot, bounds on $\mu$
can be conveniently read off: for instance, requiring $\delew\alt
30$ then requires $\mu \alt 350$~GeV. Of course, models with low
$\mu\sim 100$~GeV but multi-TeV top squarks can still be very
fine-tuned if the dominant contributions to $\delew$ 
arise from $\Sigma_u^u(\tst_i )$. In frame {\it f}), we plot
$\delew\ vs.\ m_A$.  We see that low $\delew$ can be found
over the entire range of $m_A\sim 0.15-1.5$~TeV, so this parameter is
not so relevant towards achieving low EWFT.

Next, to gain a sense of the sparticle mass ranges expected from RNS, we
plot $\delew$ versus selected sparticle masses.  First, since
$m_0\sim 2-8$~TeV for $\delew\alt 30$, we expect first and
second generation squark and slepton masses also within this range
(which is for the most part inaccessible LHC SUSY searches).  Next, in
frame {\it a}) of Fig.~\ref{fig:mass1}, we show $\delew\ vs.\ m_{\tg}$.
We find that requiring $\delew \alt 30$ requires $m_{\tg}\sim
1-4$~TeV. The lower portion of this range should be accessible to LHC14
searches, while the upper part lies beyond any LHC luminosity
upgrade\cite{lhchl}.
\FIGURE[tbh]{
\includegraphics[width=7.2cm,clip]{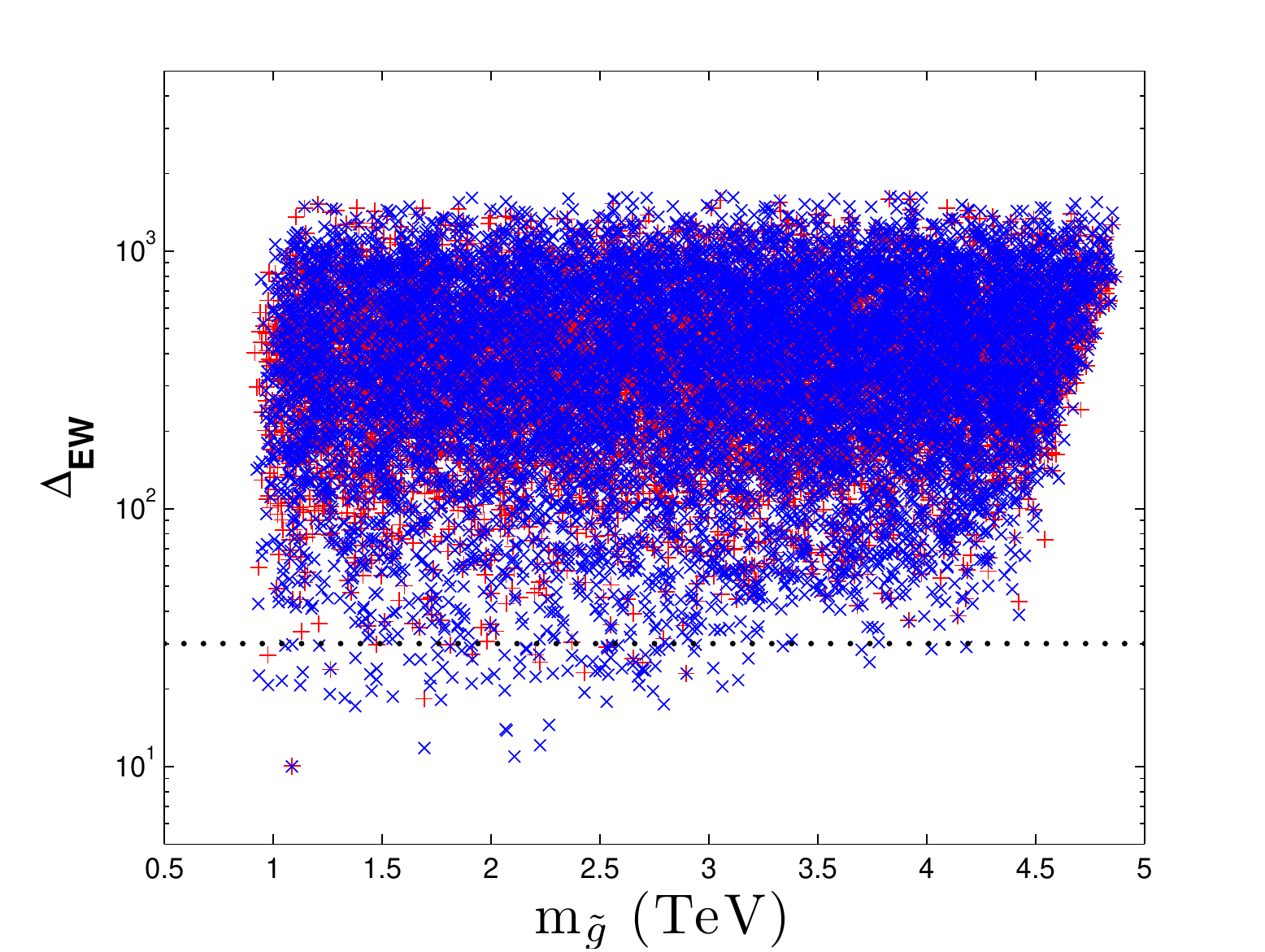}
\includegraphics[width=7.2cm,clip]{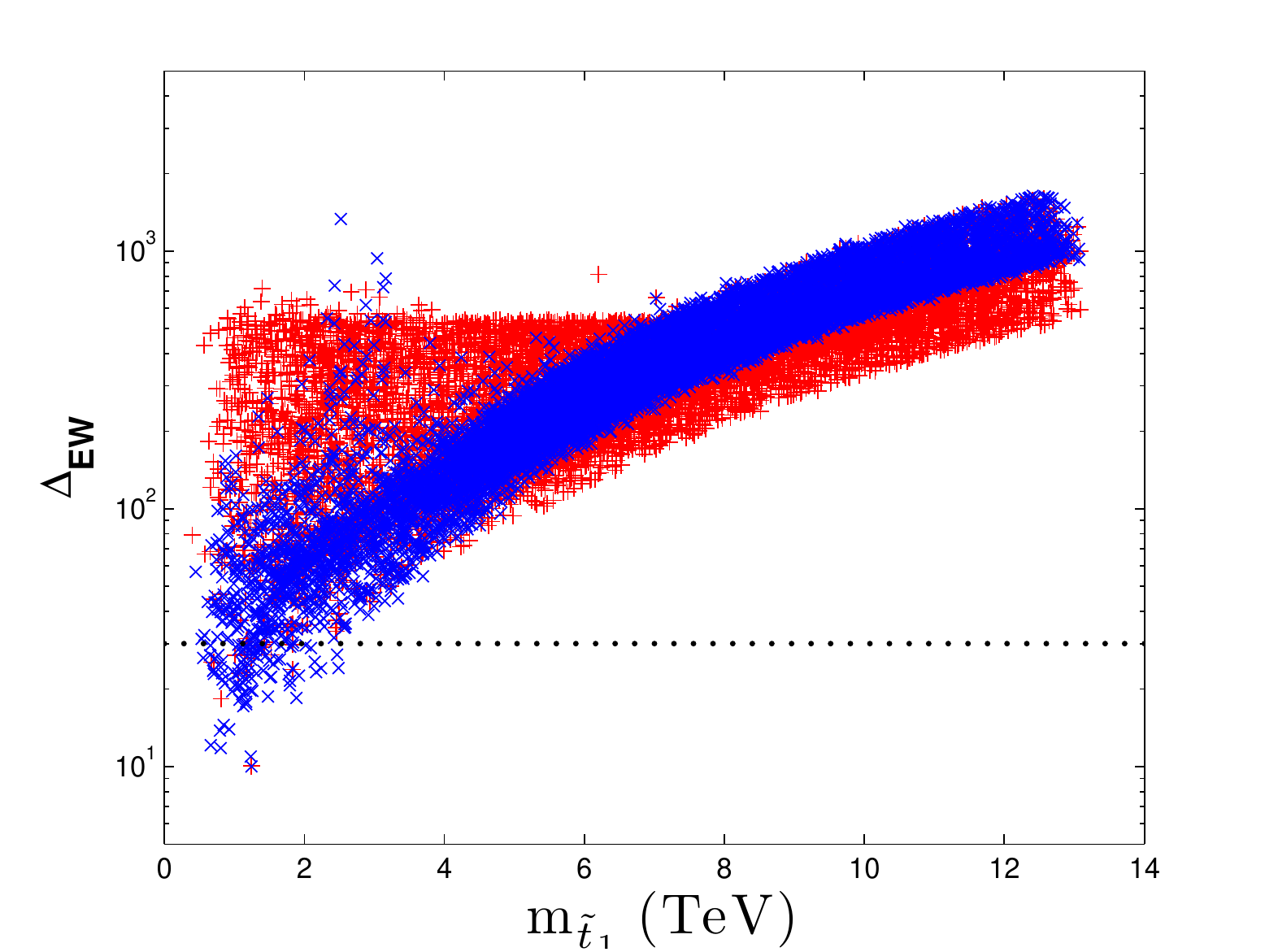}
\includegraphics[width=7.2cm,clip]{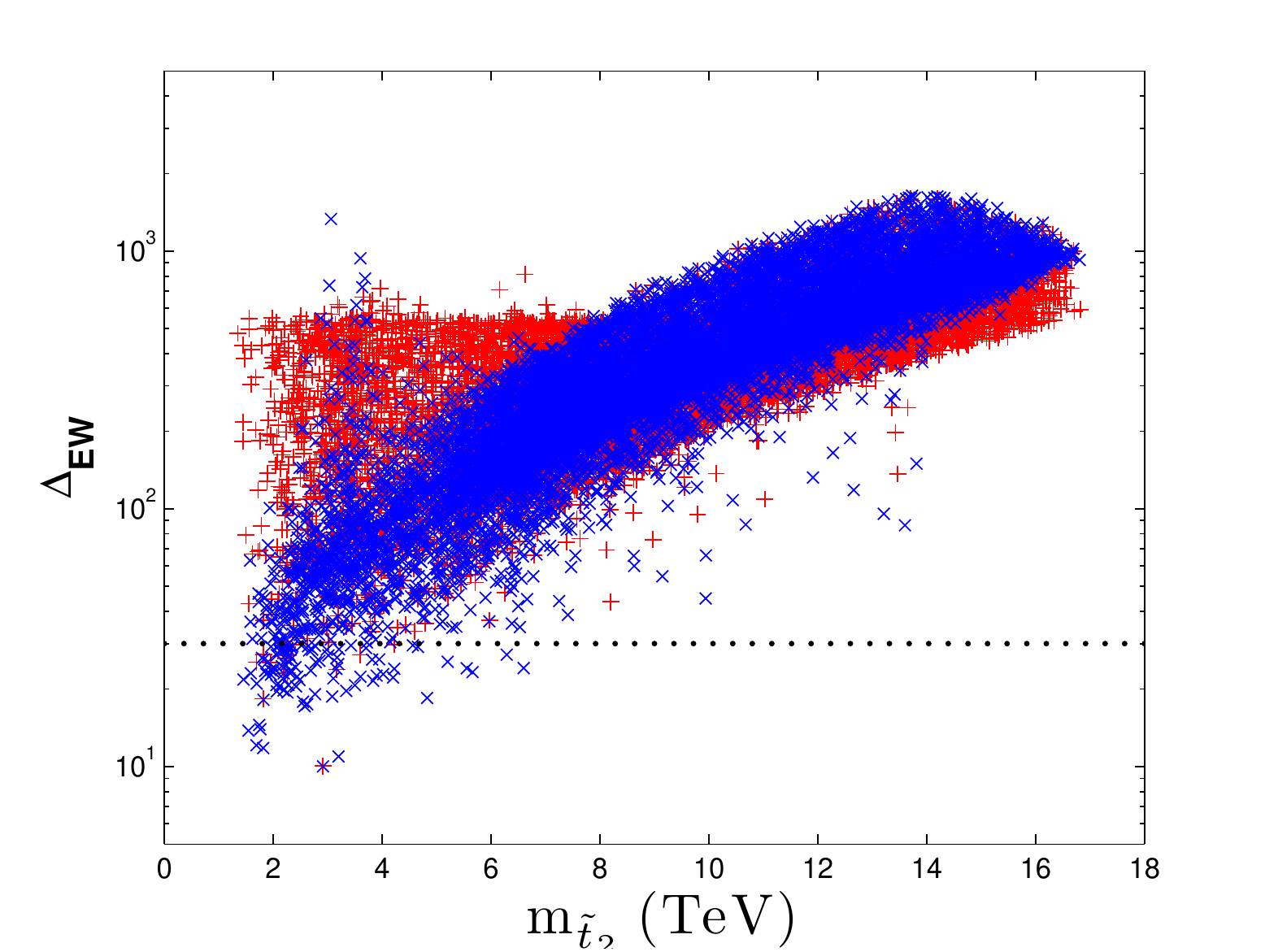}
\includegraphics[width=7.2cm,clip]{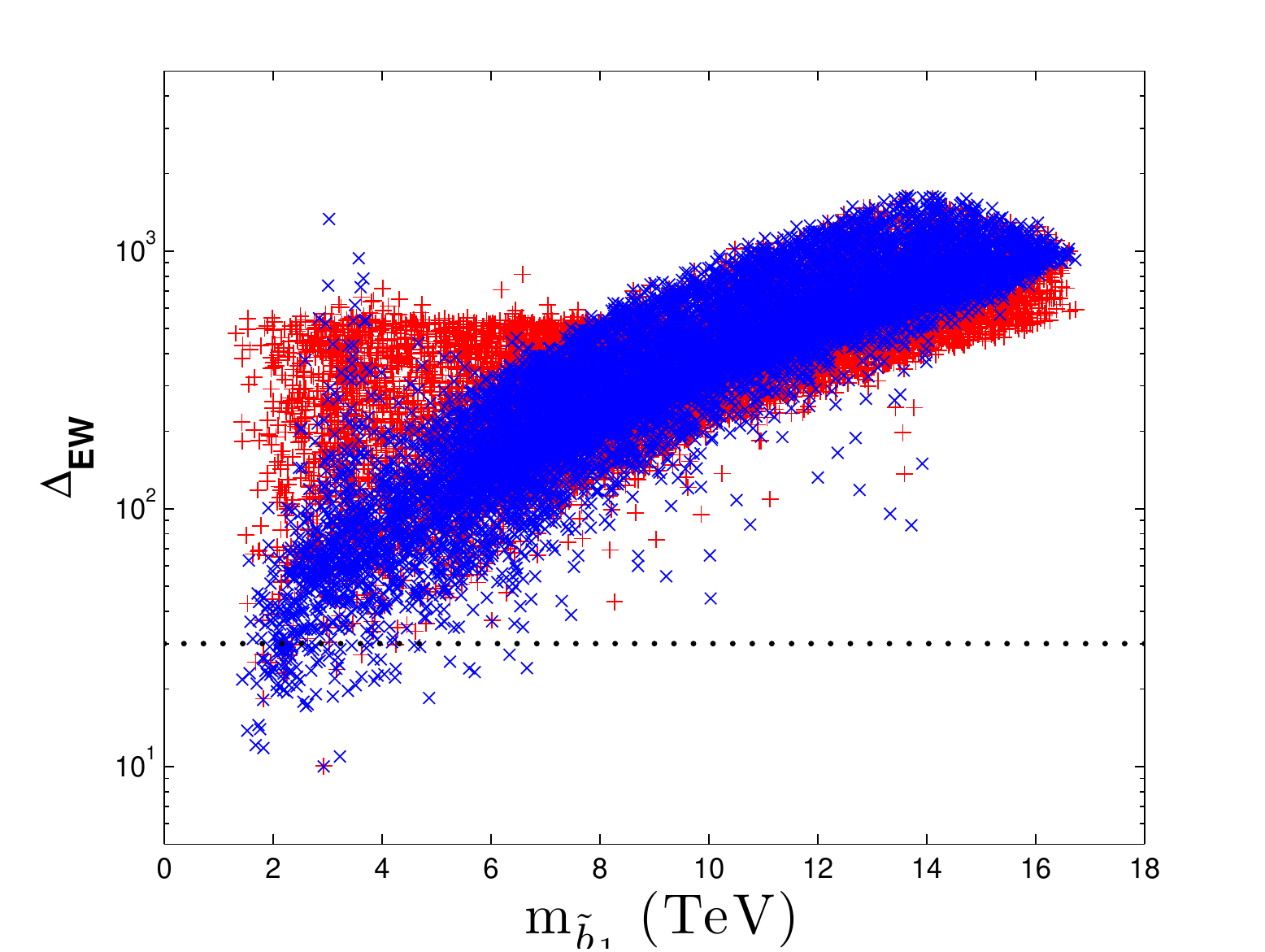}
\caption{The value of $\delew$ versus gluino and third generation
  squark masses from a scan over NUHM2 parameter space. As in
  Fig.~\ref{fig:nuhm2}, the red pluses denote the distributions from
  the complete scan, whereas the blue crosses depict the results for
  the dedicated low $\mu$ scan. The line at $\delew=30$ is to
  guide the eye. }
\label{fig:mass1}}

In frame {\it b}), we show $\delew$ versus the lighter top squark mass
$m_{\tst_1}$. Here, we see that $\delew\alt 30$ allows $m_{\tst_1}\sim
0.5-2.5$~TeV range. This is well above the range expected in generic
NS models\cite{ah,others}, where $m_{\tst_{1,2}}$ has
been advocated to lie below about 600~GeV.
In frame {\it c}), we show $\delew\ vs.\ m_{\tst_2}$. Here, we find
that $m_{\tst_2}$ can range up to $\sim 6$~TeV for $\delew\alt 30$.
Such high values of $m_{\tst_2}$ are helpful to
increase radiative corrections to the light Higgs mass $m_h$ into the
125~GeV range.  However, such heavy top squarks lie far beyond any
conceivable LHC reach.  In frame {\it d}), 
we show $\delew\ vs.\ m_{\tb_1}$. Here we see $m_{\tb_1}\sim 0.8-6$~TeV, 
which again allows for far heavier bottom squarks than previous NS models, 
where $m_{\tst_{1,2}}$ and $m_{\tb_1}$ all were suggested to be $\alt 600$~GeV.

In Fig.~\ref{fig:mass2}{\it a}), we show $\delew\ vs.\ m_{\tw_1}$.  For
RNS models, $m_{\tw_1}\simeq m_{\tz_{1,2}}\sim |\mu |$, {\it i.e.} its
mass is roughly equal to that of the two lighter neutralinos.  Since
$\tw_1$ is mainly higgsino-like near the lower edge of the envelope of
points, the distribution follows a similar pattern as for the $\delew\
vs.\ \mu$ plot in Fig.~\ref{fig:nuhm2}. We see that for $\delew\alt 20$,
$m_{\tw_1}\alt 250$~GeV. Thus, a linear collider operating with
$\sqrt{s}>2m_{\tw_1}$ will {\it directly probe} the lowest (and hence
most lucrative!) values of $\delew$ if the relatively soft visible
daughters of the chargino can be distinguished over two-photon
backgrounds\cite{krupov}. In this sense, it has been emphasized that for
models of natural SUSY, a linear $e^+e^-$ collider would be a higgsino
factory in addition to a Higgs factory\cite{bbh,bbht,ltr}!  In frame
{\it b}), we show $\delew\ vs.\ m_{\tw_2}$.  In the RNS model,
the $\tw_2$ is nearly pure wino-like and its mass can range between
$\sim 0.3-1.2$~TeV for $\delew\alt 30$.  Since RNS as presented
here includes gaugino mass unification, then typically $\tz_{1,2}$ are
higgsino-like, $\tz_3$ is bino-like and $\tz_4$ is wino-like. Since the
$SU(2)$ gauge coupling $g$ is rather large, we expect significant
rates for $\tw_2^\pm \tz_4$ production at LHC, at least for the lower
portion of the range of $m_{\tw_2}$.  In frame {\it c}), we show the
$m_{\tz_2}-m_{\tz_1}$ mass difference in RNS versus $\delew$.  For most
points with $\delew\alt 30$, we find that $m_{\tz_2}-m_{\tz_1}\alt
10-20$~GeV. Some points with $\delew\sim 30-40$ have a mass
difference as large as 100~GeV; these points arise from sampling the
lower portion of the $m_{1/2}$ range, which gives rise to gaugino masses
comparable in magnitude to $|\mu |$ so that the lighter electroweakinos
are actually gaugino-higgsino mixtures.  For the more likely small mass
gap case, the lighter neutralinos are dominantly higgsino-like and decay
via $\tz_2 \to \tz_1f\bar{f}$ (here $f$ denotes SM-fermions) through the
virtual $Z$.  Then decays into opposite-sign same-flavor (OS/SF)
isolated dileptons should occur at $\sim 3\%$ for each charged lepton
species.  The presence of low invariant mass OS/SF isolated dileptons
from boosted $\tz_2$ produced in gluino or gaugino cascade decay events
could then be a distinctive signature of RNS at the LHC.  For NUHM2
models with larger values of $\delew$, falling outside the RNS low EWFT
requirement, $m_{\tz_2}$ can be greater than $m_{\tz_1}+M_Z$ or
$m_{\tz_1}+m_h$ so that two body decays of $\tz_2$ are then allowed.
Finally, in frame {\it f}), we show $\delew\ vs.\ m_h$. Here, we see the
lower $m_h\sim 123-124$~GeV values are just slightly preferred by EWFT over
the higher range, although values of $m_h$ as high as $\sim 126.5$~GeV
occur for $\delew =30$.
\FIGURE[tbh]{
\includegraphics[width=7.2cm,clip]{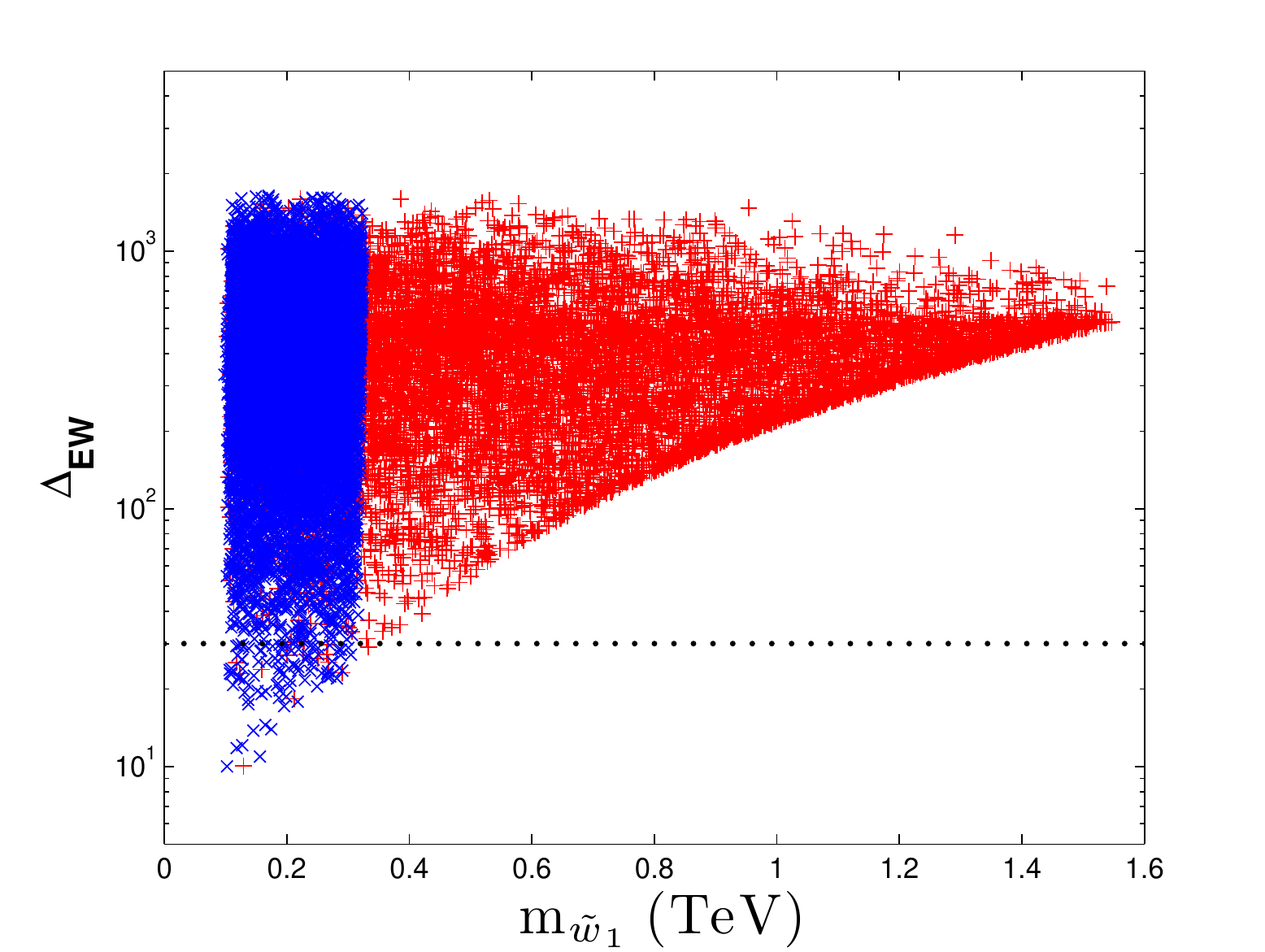}
\includegraphics[width=7.2cm,clip]{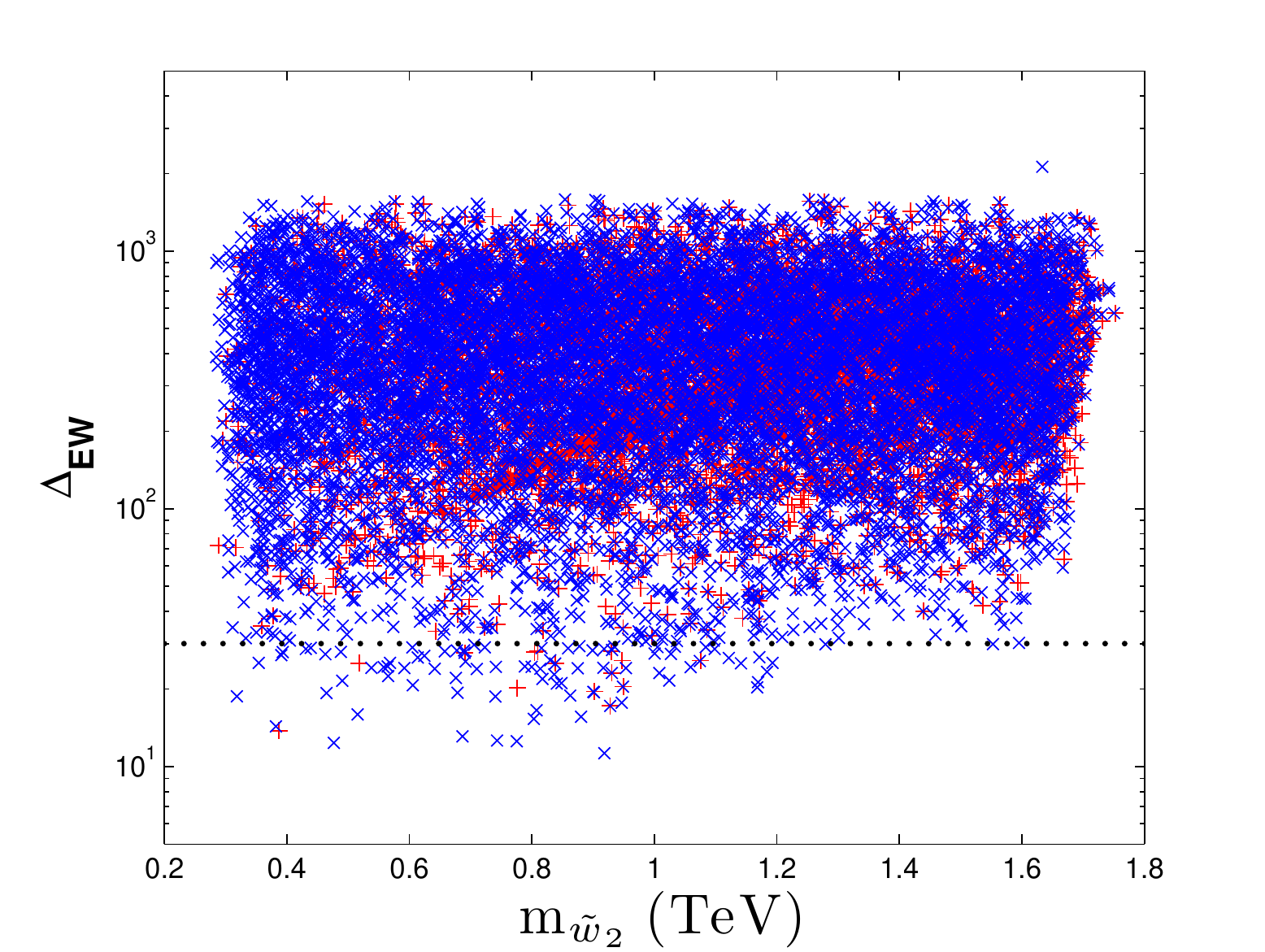}
\includegraphics[width=7.2cm,clip]{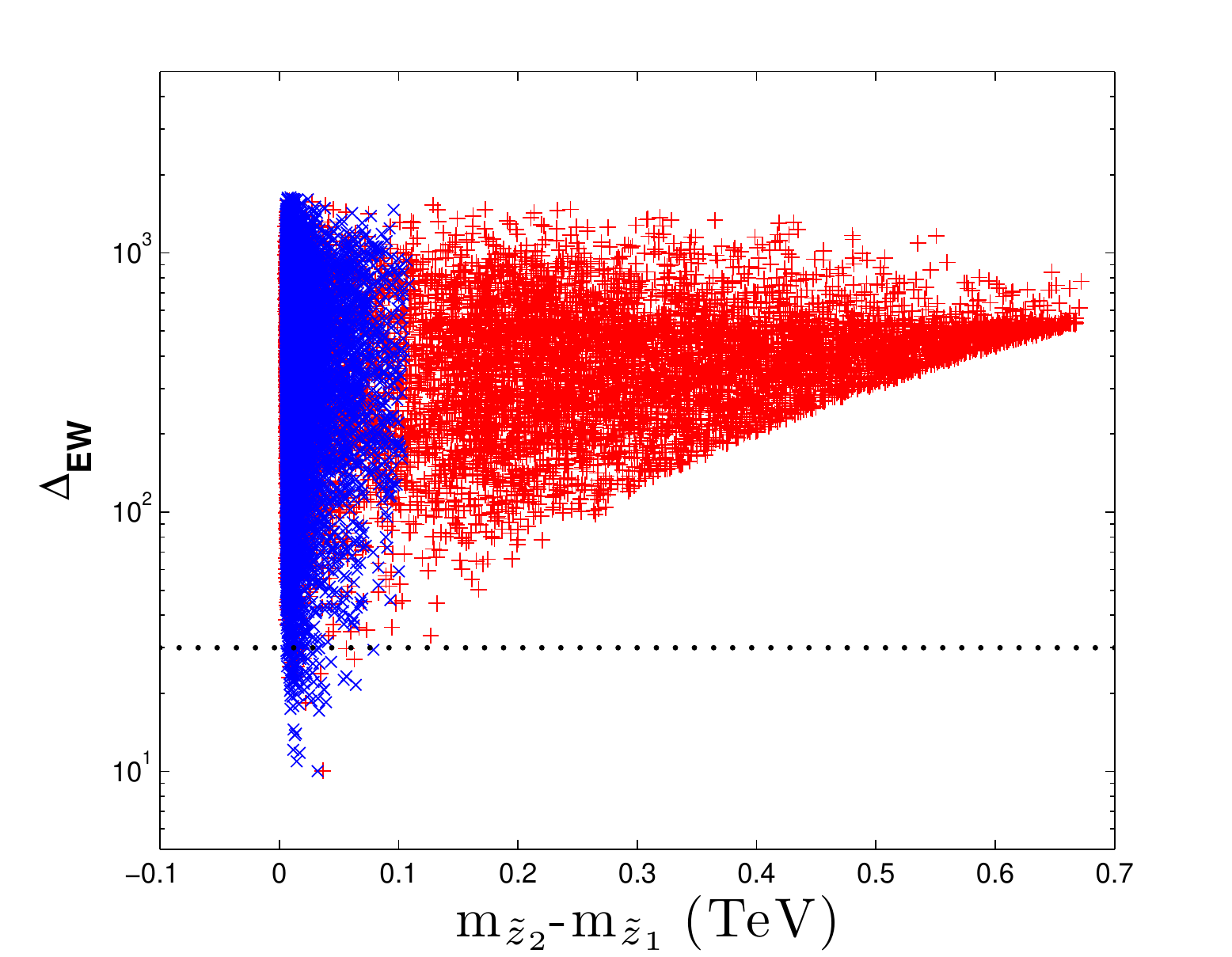}
\includegraphics[width=7.2cm,clip]{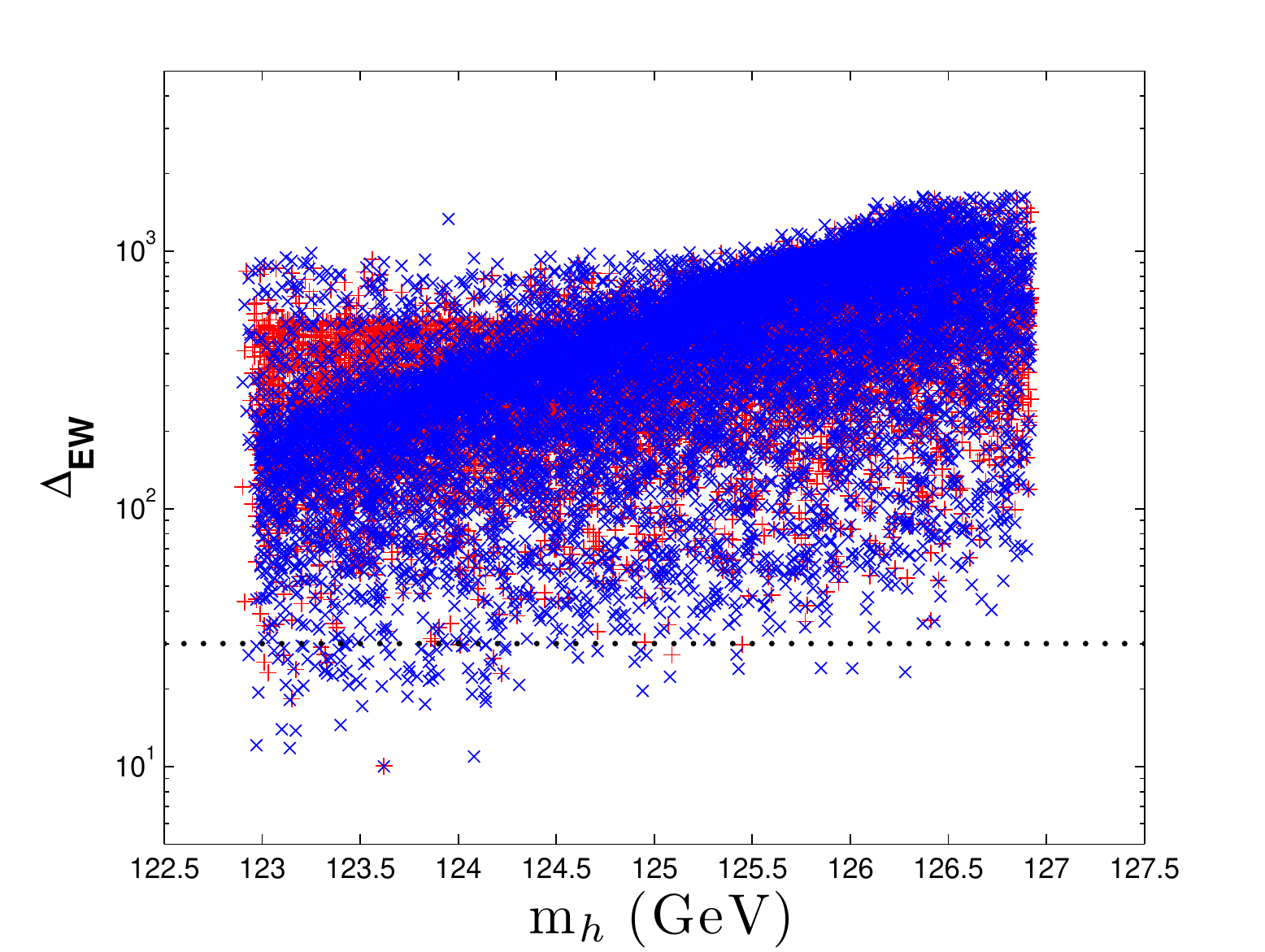}
\caption{The value of $\delew$ versus electroweak -ino
   and Higgs
  boson masses from a scan over NUHM2 parameter space. As in
  Fig.~\ref{fig:nuhm2}, the red pluses denote the distributions from
  the complete scan, whereas the blue crosses depict the results for
  the dedicated low $\mu$ scan. The line at $\delew=30$ is to
  guide the eye. }
\label{fig:mass2}}

While our methodology allows one to find a low value of $\mu^2$ for any
value of $m_0$ and $m_{1/2}$, this by itself does not guarantee a small
value of $\delew$. In addition, the GUT scale value of $m_{H_u}^2$ has
to be adjusted very precisely to obtain low EWFT, which could be viewed
as a different sort of fine-tuning: that only a very narrow range of
$m_{H_u}^2(M_{\rm GUT})$ values will yield $-m_{H_u}^2\sim M_Z^2$ at the
weak scale.\footnote{From the perspective introduced in Sec.~\ref{sec:intro}, we
would look for an underlying model where $m_{H_u}^2$ is thus
determined.}  This can be seen from Table~\ref{tab:huadj} where we plot
the value of $m_{H_u}^2(M_{\rm GUT})$ which is needed to generate small
$\mu$ solutions for two different cases of NUHM2 model parameters.
Optimistically speaking, we would view this as essentially determining
the GUT scale value of $m_{H_u}^2/m_0^2$ to be very nearly 1.65 (Case A)
or 1.71 (Case B).  It is gratifying to see that the GUT scale values of
{\it all} scalar mass parameters have no hierarchy as expected in models
of gravity-mediated SUSY breaking where all scalar masses might be
expected to be comparable at the high scale.
\begin{table}[htb]
\begin{center}
\begin{small}
\begin{tabular}{|ccc|ccc|}
 \multicolumn{3}{c|} {{\bf Case A}} & \multicolumn{3}{c}{{\bf Case B}} \cr
 $m_{H_u}^2(M_{\rm GUT})$ & $\mu$ & $\delew$
& $m_{H_u}^2(M_{\rm GUT})$ & $\mu$ & $\delew$ \cr
\hline
$1.03\times 10^7$   & 150 & 9.04  &$2.73\times 10^7$   & 150 & 15.4 \\

$1.02\times 10^7$  & 250 &18.8      & $2.72\times 10^7$  & 250 & 24.1 \\

 $1.00\times 10^7$  & 400 & 42.4  & $2.70\times 10^7$     & 400 & 49.5 \\

\end{tabular}
\end{small}
\smallskip
\caption{An illustration of the sensitivity of the EWFT fine-tuning
  measure $\delew$ to $m_{H_u}^2(M_{\rm GUT})$. For case {\bf A} the NUHM2
  parameters are $m_0=2.5$~TeV, $m_{1/2}=400$~GeV, $\tan\beta=10$,
  $m_A=1$~TeV, while for case {\bf B} we have, $m_0=4$~TeV,
  $m_{1/2}=1$~TeV, $\tan\beta=15$, $m_A=2$~TeV. For both cases, we take
$A_0=-1.6m_0$. The numbers in the Table are in~GeV units.}
\label{tab:huadj}
\end{center}
\end{table}

To be more general, we show in Fig.~\ref{fig:mHuGUT} a
scatter plot of $\delew$ versus the GUT scale ratio $m_{H_u}^2/m_0$
from our scan over NUHM2 models. We find that for points with
$\delew\alt 30$,
then $m_{H_u}(M_{\rm GUT})\sim (1-2) m_0$.
\FIGURE[tbh]{
\includegraphics[width=9cm,clip]{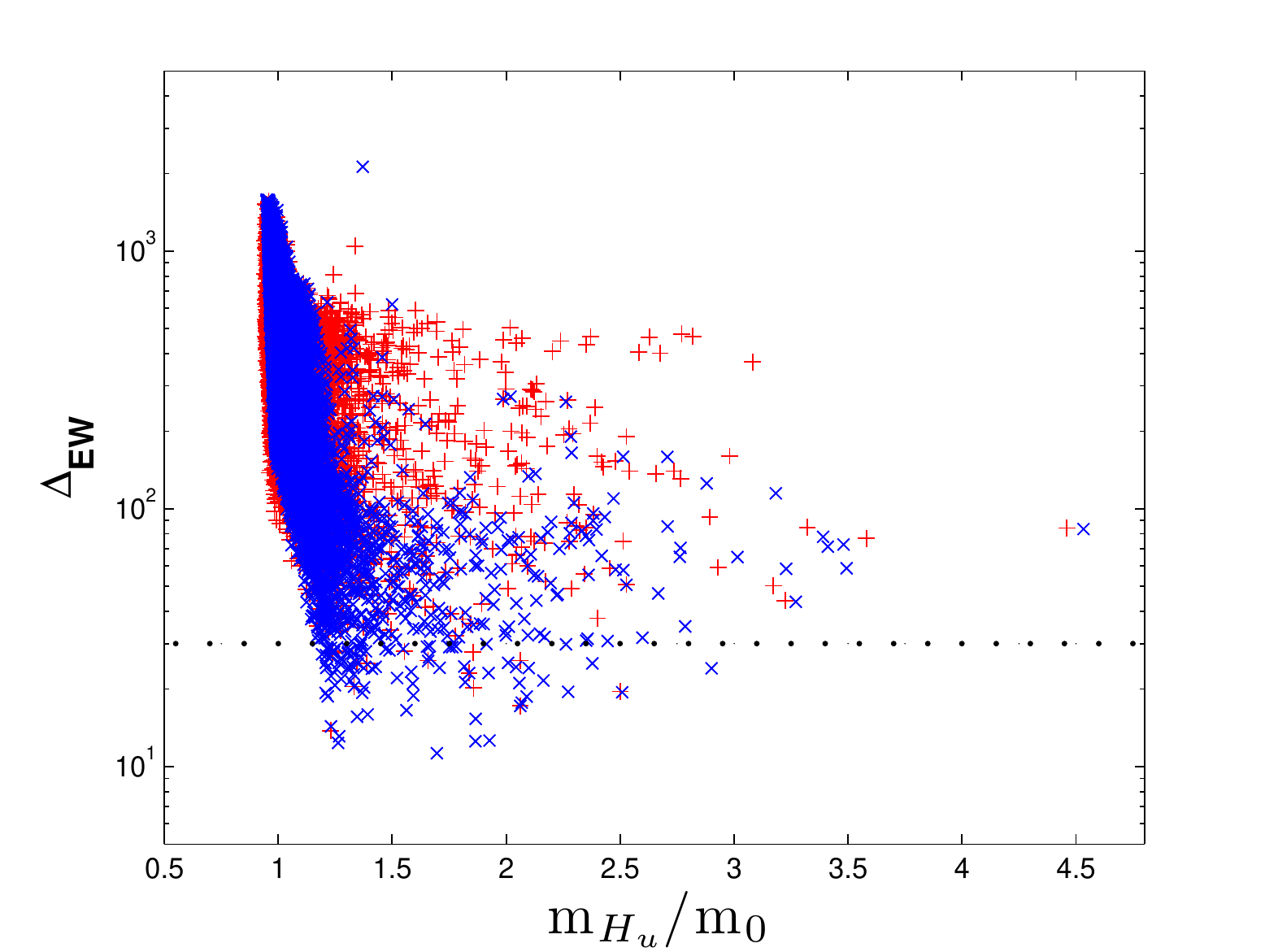}
\caption{The value of $\delew$ versus $m_{H_u}/m_0 (M_{\rm GUT})$
from the scan over the NUHM2 parameter space. As before, the red pluses
are for the scan over the entire range of $\mu$ while the blue crosses
are for the dedicated scan with $\mu$ limited to the 100-300~GeV range.
The line at $\delew=30$ is to guide the eye.}
\label{fig:mHuGUT}}

\subsection{RNS from the NUHM1 model?}

Up to this point we have focused on RNS from the NUHM2 model. However,
it is of interest to see if low EWFT is also possible within the NUHM1
\cite{nuhm2} framework in which $H_u$ and $H_d$ have equal GUT scale
mass parameters; {\it i.e.}  $m_{H_u}^2(M_{\rm GUT}) = m_{H_d}^2(M_{\rm
GUT}) \equiv m_\phi^2$. As mentioned above, $m_\phi^2$ is then adjusted
to that $m_{H_u}^2(M_{\rm SUSY}) \sim M_Z^2$. For brevity, we confine our
investigation to model lines where we fix $m_0$, $m_{1/2}$ and $A_0$ to
be the same as for the RNS2 model point, but where we vary $m_\phi$, the
common GUT scale Higgs mass parameter and $\tan\beta$. In
Fig.~\ref{fig:nuhm1} we show the value of {\it a})~$\mu$, {\it
b})~$m_A$, and {\it c})~$\delew$ versus $m_{\phi}$ for $\tan\beta =
8.85$ (the RNS2 value), 25, 40 and 50. We see that the various curves in
frame {\it a}) are quite close (except at very large values where they
dive down). This is essentially because the top Yukawa coupling that
dominantly affects $m_{H_u}^2$ (remember that $m_{H_u}^2(M_{\rm SUSY})$
determines $\mu$) hardly varies with $\tan\beta$; the small differences
arise from the (subdominant) effects of bottom Yukawa couplings.  In
contrast, the $m_A$ values in frame {\it b}) reduce considerably as
$\tan\beta$ increases. We can understand this if we remember that the
bottom Yukawa coupling-- which increases with $\tan\beta$-- drives
$m_{H_d}^2$ to low values thus reducing $m_A^2\simeq m_{H_d}^2 +
\mu^2$ for larger $\tan\beta$ values. Turning to frame {\it
c}) we see that for this model line with $\tan\beta=8.55$ (uppermost
curve), $\delew$ reduces with increasing $m_\phi$ as in the $\mu$ curve
in frame {\it a}) until the kink at which it starts increasing. We have
checked that the kink occurs when $\mu^2$ becomes so low that the
$m_{H_d}^2$ term becomes larger than all other terms in 
(\ref{eq:loopmin}). For larger values of $\tan\beta$, the $m_{H_d}^2$
contribution is suppressed, resulting in smaller values of $\delew$.
However, in none of the cases shown does $\delew$ drop below
$\sim 80$.  It may be a useful exercise to scan the NUHM1 parameter
space to see just how small the EWFT can be when all LHC constraints are satisfied.
\FIGURE[tbh]{
\includegraphics[width=7cm,clip]{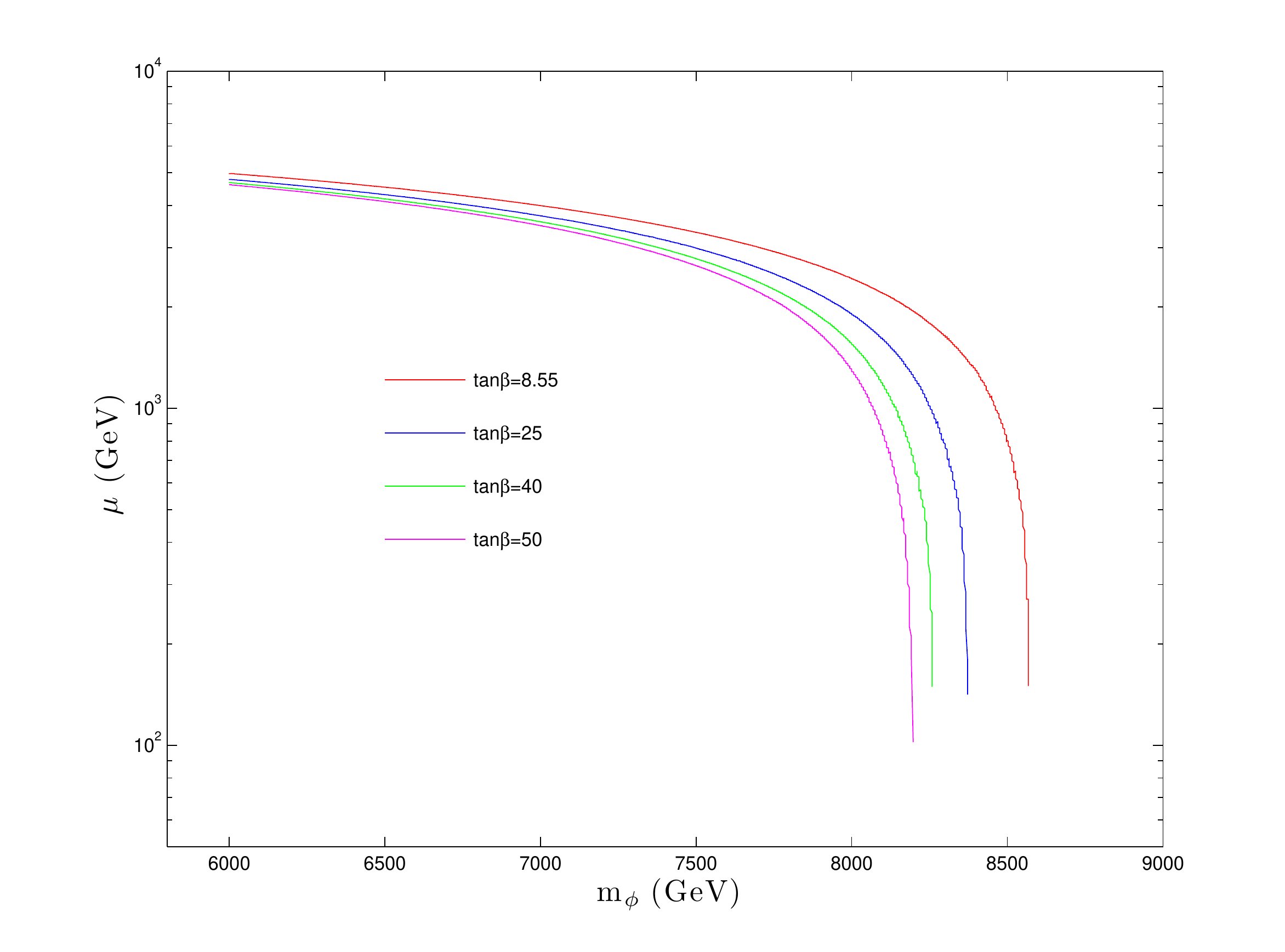}
\includegraphics[width=7cm,clip]{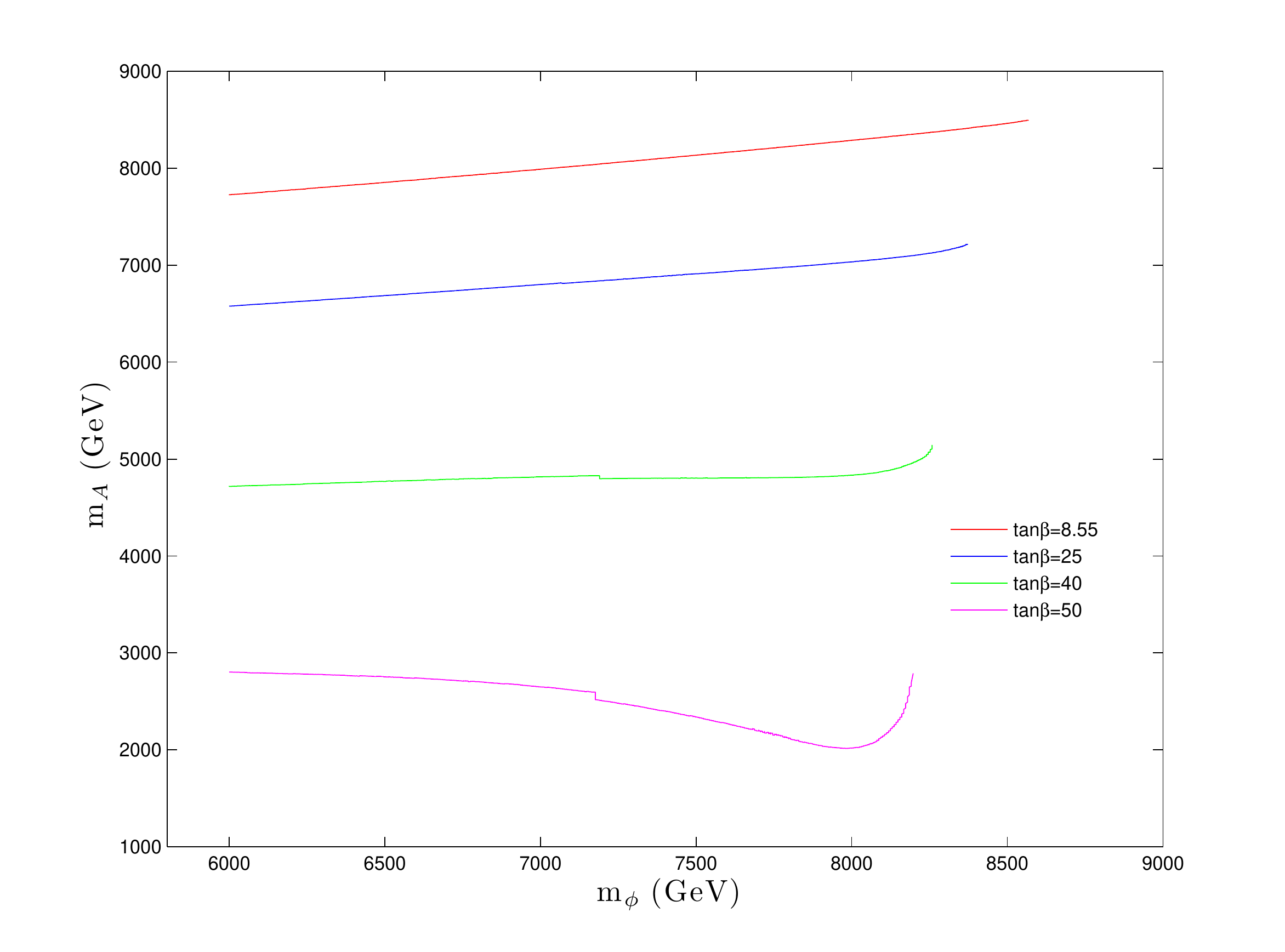}
\includegraphics[width=10cm,clip]{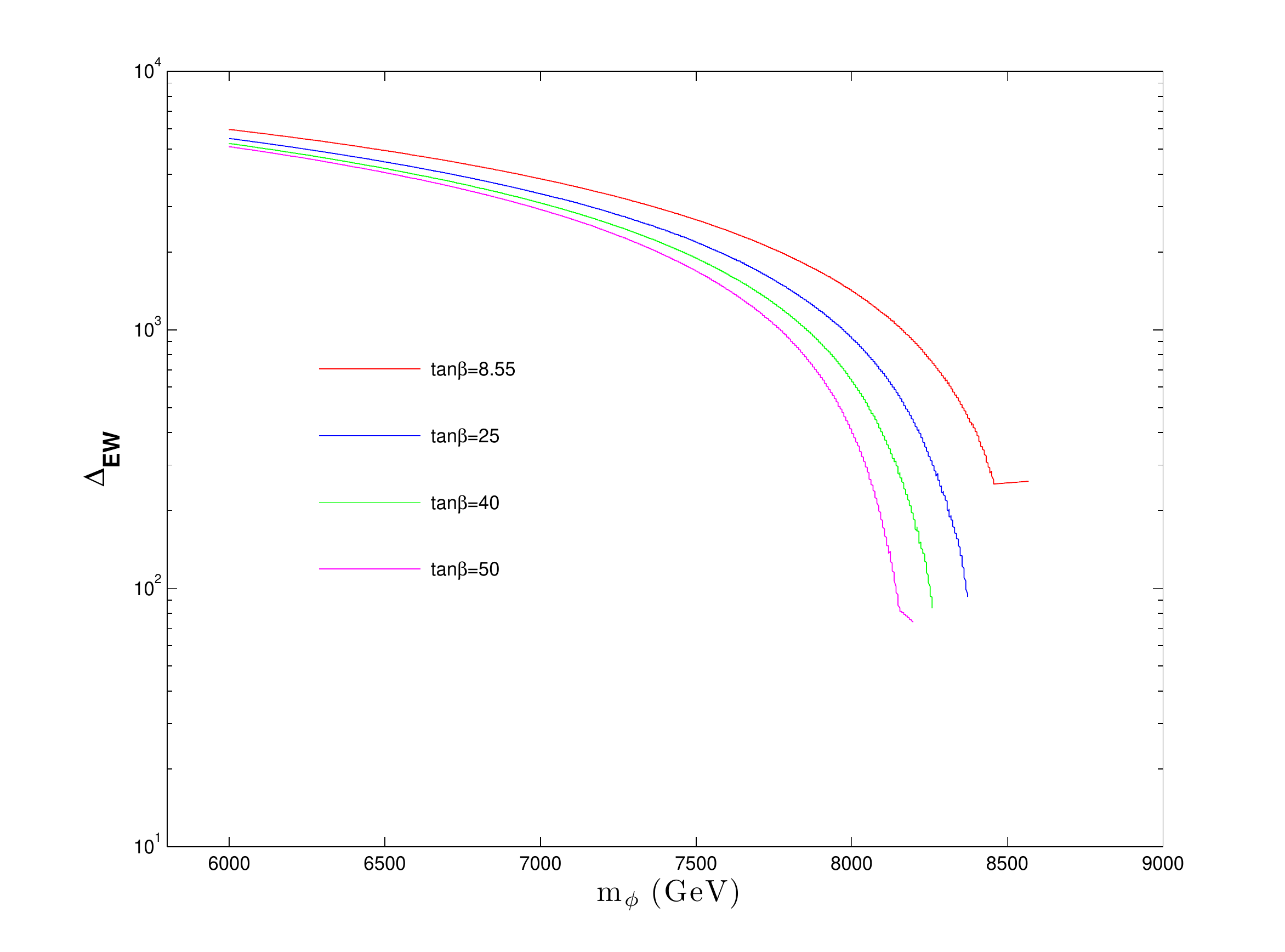}
\caption{Plot of {\it a}) $\mu$, {\it b}) $m_A$ and {\it c}) $\delew$
in the 1-parameter NUHM1 model versus $m_{\phi}$ for
RNS2 model parameters $m_0=7025$~GeV, $m_{1/2}=568.3$~GeV, $A_0=-11426.6$~GeV
and for several values of $\tan\beta$. In all the frames the order of
the lines is that of increasing $\tan\beta$, with $\tan\beta=8.55$ on
the top and $\tan\beta=50$ on the bottom.
}
\label{fig:nuhm1}}
%

\section{RNS from the NUHM3 (split generation) model}
\label{sec:nuhm3}

In this Section, we investigate if any advantage can be gained for RNS
models if we allow for a splitting between scalars of the third
generation and those of the first/second generations.  We adopt the
parameter set
\begin{equation}
m_0(1,2),\ m_0(3),\ m_{1/2},\ A_0,\ \tan\beta,\ \mu,\ m_A \qquad {\rm (NUHM3)}
\end{equation}
where $m_0(3)$ is the GUT scale third generation soft SUSY breaking mass
parameter and $m_0(1,2)$ is the corresponding (common) parameter for the
first/second generation.

We search again for RNS solutions from the split
generation parameter space by implementing a random scan over the
parameters:
\bea
m_0(3) &:&\ 0-20\ {\rm TeV},\nonumber \\
m_0(1,2) &:&\ m_0(3)-30\ {\rm TeV},\nonumber\\
m_{1/2} &:&\  0.3-2\ {\rm TeV},\nonumber \\
-3 & < & A_0/m_0\ <3, \label{eq:SGparams} \\
\mu &:&\ 0.1-1.5\ {\rm TeV}, \nonumber \\
m_A &:&\ 0.15-1.5\ {\rm TeV},\nonumber \\
\tan\beta &:&\ 3-60 .\nonumber
\eea
We implement the same LHC sparticle mass and $m_h=125\pm 2$~GeV
constraints as before.

In Fig.~\ref{fig:nuhm3}, we show $\delew$ versus $m_0(3)$ and also
versus $m_0(1,2)$. The results for $\delew$ versus other
parameters are very similar to Fig.~\ref{fig:nuhm2} so we do not repeat
them here.  From Fig.~\ref{fig:nuhm3}{\it a}), we see that RNS
solutions with $\delew\alt 30$ can be found for $m_0(3)$ values
ranging between 1-8~TeV, similar to results found in
Fig.~\ref{fig:nuhm2} for the NUHM2 model. It is interesting to note that
the smallest values of $\delew$ in the figure are no smaller
than for the NUHM2 model. 
The gap at small values of $m_0(3)$ is an artifact of the upper limit on
$m_{1/2}$ in our scan: for small values of $m_0(3)$ the lighter
$t$-squark is often driven to tachyonic masses by two-loop contributions
of heavy first/second generation squarks. We have checked that with
larger values of $m_{1/2}$ in the scan, solutions fill in the entire gap.
Again, even though the GUT
scale value of $m_0(3)$ is in the multi-TeV regime, the $\tst_2$ and
especially $\tst_1$ physical masses are considerably lower -- in the few
TeV regime -- due to radiative effects from RGE running and also large
mixing.

The key advantage of the NUHM3 model is seen in Fig.~\ref{fig:nuhm3}{\it
b}), where we plot $\delew$ versus $m_0(1,2)$. In this case, we
see that GUT scale first/second generation scalar masses can easily
range between $1-30$~TeV while still maintaining low $\delew$.
The solutions with $m_0(1,2)$ in the multi-TeV region will also produce
first/second generation squark and slepton masses which are comparable
to $m_0(1,2)$.  This allows for a much more robust solution to the SUSY
flavor/$CP$ problems. It also ameliorates the cosmological gravitino
problem if $m_{3/2}\sim m_0(1,2)$ as is expected in simple models of
gravity-mediation.
\FIGURE[tbh]{ \includegraphics[width=7cm,clip]{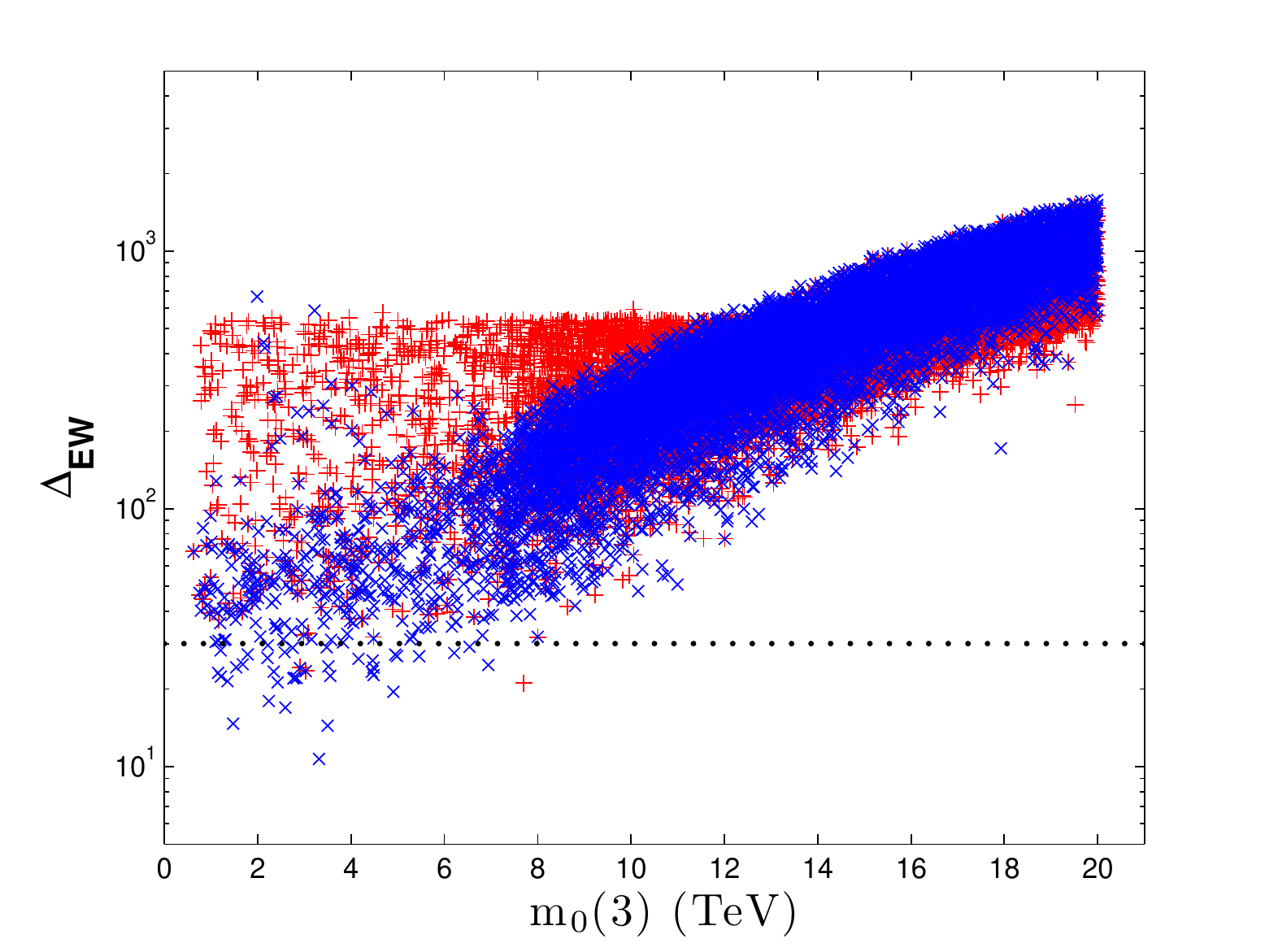}
\includegraphics[width=7cm,clip]{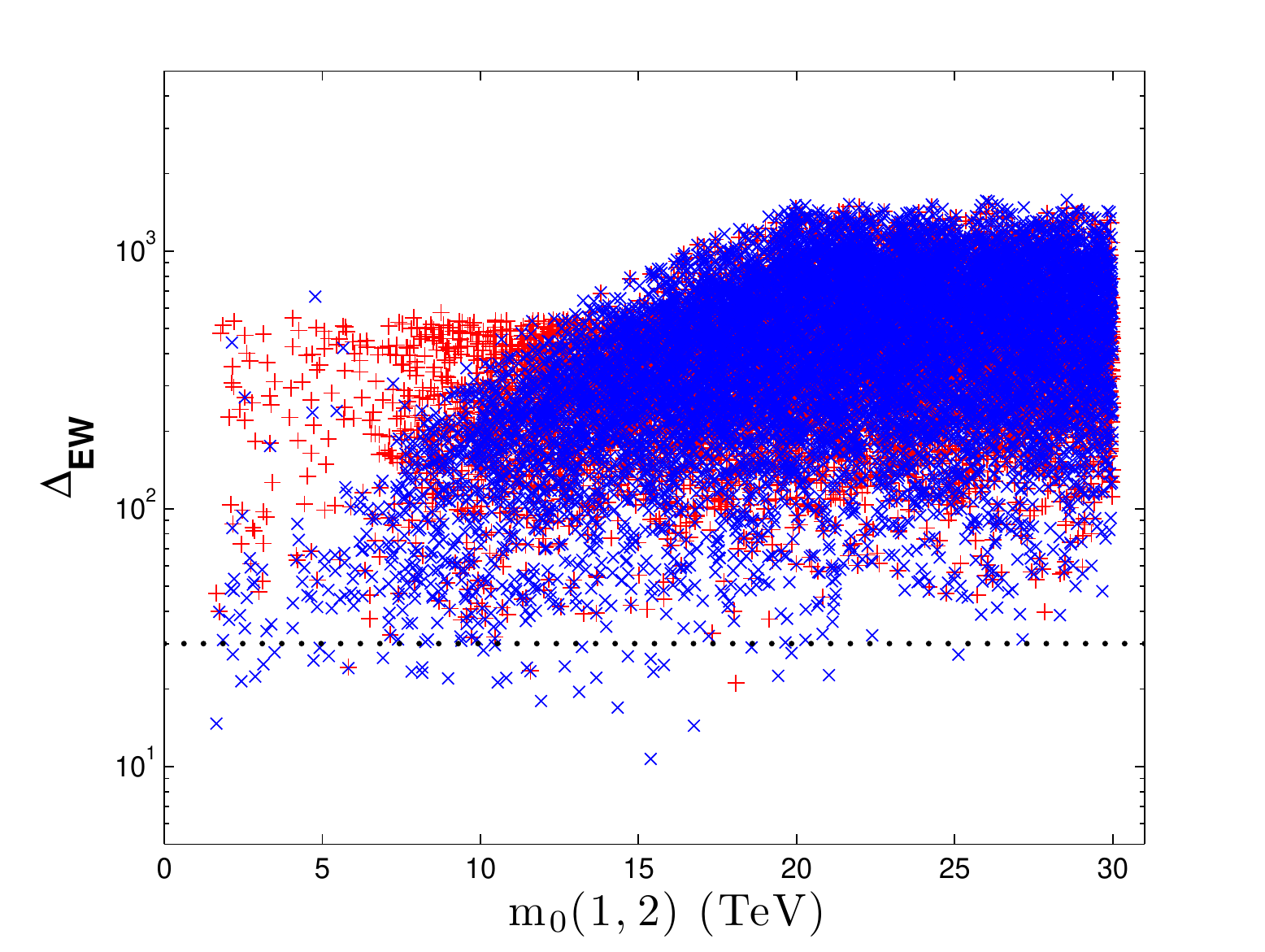}
\caption{The value of $\delew$ versus $m_0(3)$ and $m_0(1,2)$ from
a scan over NUHM3 model with split first/second and third generations. As in
  Fig.~\ref{fig:nuhm2}, the red pluses denote the distributions from
  the complete scan, whereas the blue crosses depict the results for
  the dedicated low $\mu$ scan. The line at $\delew=30$ is to
  guide the eye.}
\label{fig:nuhm3}}

We do not show plots of $\delew$ versus sparticle masses since
these are very similar to results shown in Fig.~\ref{fig:mass1} and
Fig.~\ref{fig:mass2} except for the fact that NUHM3 scans allow for much
heavier first/second generation squark and slepton masses in the 10-30
TeV range, whereas in the NUHM2 model the squarks and sleptons are
typically constrained to be below 8~TeV due to the imposed relation
$m_0(3)=m_0(1,2)$.

%
%

\section{Rare $B$ decay constraints on RNS}
\label{sec:b}

\subsection{BF($b\to s\gamma$)}
\label{ssec:bsg}

The combination of several measurements of the $b\to s\gamma $ decay rate finds that
$BF(b\to s\gamma )=(3.55\pm 0.26)\times 10^{-4}$~\cite{Asner:2010qj}.
This is slightly higher than the SM prediction\cite{Misiak:2006zs} of
$BF^{SM}(b\to s\gamma )=(3.15\pm 0.23)\times 10^{-4}$. SUSY contributions to the
$b\to s\gamma$ decay rate come mainly from chargino-stop loops and
the W-charged Higgs loops, and so are large when these particles are light
and when $\tan\beta$ is large\cite{Baer:1996kv}. Thus, in generic natural SUSY where
$m_{\tst_{1,2},\tb_1}\alt 500$~GeV, one finds generally large deviations from the SM value for
$BF(b\to s\gamma )$, as shown in Ref.~\cite{bbht}. In contrast, in radiative natural SUSY
where third generation squarks are in the~TeV range, SUSY contributions to
$BF(b\to s\gamma )$ are more suppressed. The situation is shown in Fig.~\ref{fig:BF}{\it a})
along with the measured central value (green solid line) and errors.
The red points all have $\delew<30$ and qualify as RNS points. We see the bulk of RNS points
are consistent with the measured $BF(b\to s\gamma )$, although there are outliers.

\FIGURE[tbh]{
\includegraphics[width=7cm,clip]{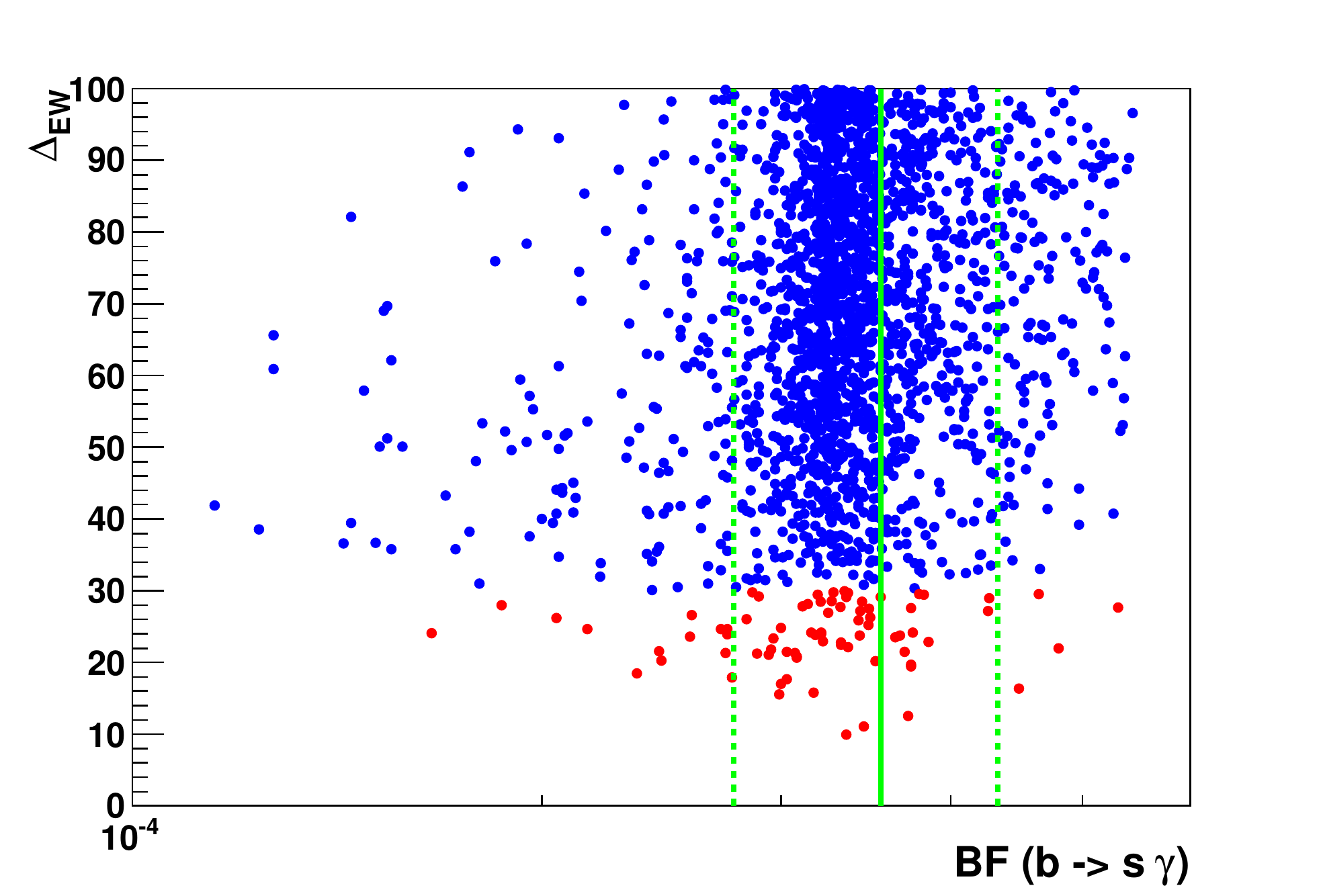}
\includegraphics[width=7cm,clip]{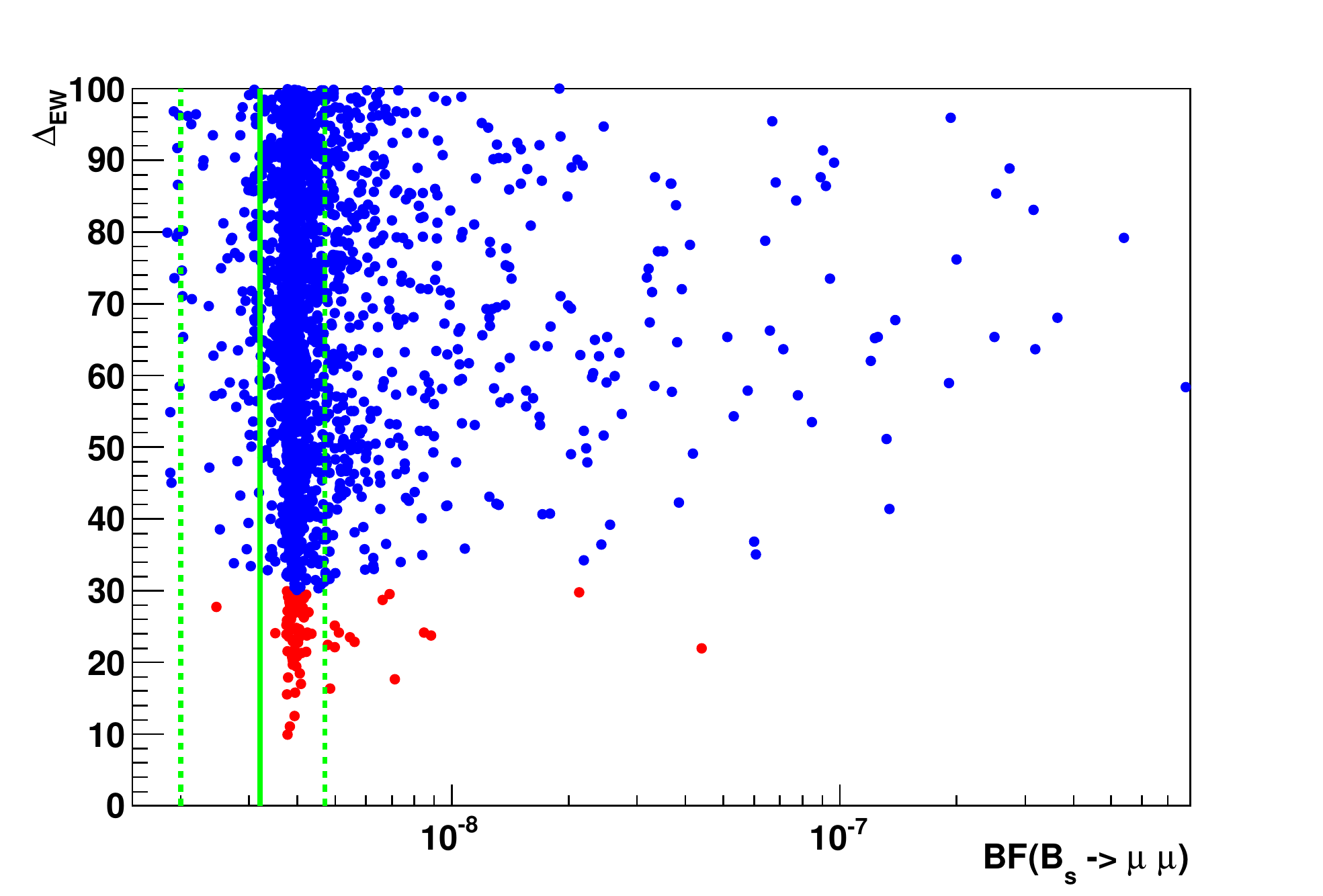}
\caption{The values of $\delew$ versus {\it a}) $BF(b\to s\gamma )$
and {\it b}) $BF(B_s\to\mu^+\mu^- )$. 
The vertical lines represent the experimental measurements with uncertainties.
}
\label{fig:BF}}

\subsection{$B_s\to \mu^+\mu^-$}

Recently, the LHCb collaboration has discovered an excess over the
background for the decay $B_s\to\mu^+\mu^-$\cite{lhcb}!  They find a
branching fraction of $BF(B_s\to\mu^+\mu^- )=3.2^{+1.5}_{-1.2}\times
10^{-9}$ in accord with the SM prediction of $(3.2\pm 0.2)\times
10^{-9}$\cite{bmm_sm}.  In supersymmetric models, this flavor-changing decay occurs
through pseudoscalar Higgs $A$ exchange\cite{Babu:1999hn}, and the
contribution to the branching fraction from SUSY is proportional to
$\frac{(\tan\beta)^6}{m_A^4}$.  We show the value of
$BF(B_s\to\mu^+\mu^- )$ from RNS in Fig.~\ref{fig:BF}{\it b}).  The
decay is most constraining at large $\tan\beta\sim 50$ as occurs in
Yukawa-unified models\cite{bkk} and low $m_A$. In the case of RNS with
lower $\tan\beta$ and heavier $m_A$, the constraint is less important.
The bulk of the RNS points in Fig.~\ref{fig:BF}{\it b}) fall well within
the newly measured error bands although there are some outlier red
points, mainly at larger values of the branching fraction.

\subsection{$(g-2)_\mu$}

In addition, the well-known $(g-2)_\mu$ anomaly has been reported as a
roughly $3\sigma$ deviation from the SM value: $\Delta a_\mu=(28.7\pm
8.0)\times 10^{-10}$\cite{gm2_ex}.  In RNS, since the
$\tmu_{1,2}$ and $\tnu_\mu$ masses are expected to be in the multi-TeV range,
only a tiny non-standard contribution to the $(g-2)_\mu$ anomaly is expected,
and alternative explanations for this anomaly would have to be sought.

\section{RNS at LHC}
\label{sec:lhc}

Here we list a few of the possibilities for an LHC search for radiative
natural SUSY.  A thorough study of signal and background simulations
will be presented in an upcoming study~\cite{rnslhc}.

The hallmark feature of radiative natural SUSY models is the presence of
light higgsino states $\tw_1$ and $\tz_{1,2}$ with masses $\sim
|\mu|\sim 100-300$~GeV, and usually a small mass gap
$m_{\tw_1}-m_{\tz_1}$ and $m_{\tz_2}-m_{\tz_1}$ of order 10-30~GeV with
a possible exception of low $m_{1/2}$ and larger $\mu$ where there can
be substantial gaugino-higgsino mixing.

One possibility for RNS at LHC is to search for clean trilepton events
from $\tw_1\tz_2$ production followed by $\tw_1\to\tz_1 \ell\nu_\ell$
and $\tz_2\to\tz_1\ell^+\ell^-$ decays where $\ell = e$ or $\mu$. This
signal has been investigated in Ref.~\cite{bbh}.  There, the $p_T(\ell
)$ values were typically found to be quite low in the 5-15~GeV range
making detection difficult. The small mass difference between the parent
and daughter neutralino will also mean that the invariant mass of the
opposite-sign same-flavor dilepton pair will be small, making it more
challenging to separate it from SM origins. Nevertheless, this reaction
certainly motivates an LHC search for clean trilepton states with very
soft lepton $p_T$ values, as low as is experimentally feasible.

While first/second generation squarks are expected to be in the
multi-TeV range, the value of $m_{\tg}$ is expected to be $\sim 1-5$
TeV. The lower portion of this mass range $m_{\tg}\sim 1-2$~TeV should
be accessible to LHC searches for gluino pair production
$pp\to\tg\tg$. For RNS models, since $m_{\tst_{1,2}}\ll m_{\tq}$, then
gluino three-body decays to third generation particles typically
dominate: $\tg\to tb\tw_1$ or $t\bar{t}\tz_i$. Thus, we would expect
$\tg\tg$ events to contain up to four $b$-jets, and 2-4 reconstructable
top quarks. A small fraction of events would contain
$\tz_2\to\tz_1\ell^+\ell^-$ where $m(\ell^+\ell^- )$ is bounded by
$m_{\tz_2}-m_{\tz_1}\sim 10-20$~GeV.  Normally the leptons from $\tz_2$
decay would be rather soft, but in the case of large boosts from the
gluino cascade decay, the opposite-sign/same-flavor pair would be highly
collimated in opening angle. We expect LHC14 with 100 fb$^{-1}$ to be able to probe
$m_{\tg}\sim 1-2$~TeV via $\tg$ cascade decays using analyses similar to those
used for gluino searches in  mSUGRA when $m_0$ is very large.~\cite{lhchl}.

A novel search for RNS at LHC is to look for pair production of the {\it
heavier} gaugino states $\tw_2$ and $\tz_3$ and $\tz_4$.  Wino pair
production occurs via the large $SU(2)$ gauge couplings and leads to
large rates for $\tw_2\tz_4$ and $\tw_2\tw_2$ processes. The decays
$\tw_2 \to W\tz_{1,2}, Z\tw_1$ and $\tz_4\to W\tw_1,Z\tz_{1,2}$ occur
with significant branching fractions and yield a variety of diboson
final states that include spectacular $W^\pm W^\pm$ and
$WZ$~\cite{andrewz} plus $\eslt$ events with soft debris from the decays
of higgsinos. Events with light Higgs bosons instead of gauge bosons in
the final state are also possible~\cite{andrewh}.

As with the mSUGRA model, a wide range of RNS signatures for LHC can
be found by exploring the $m_0\ vs.\ m_{1/2}$ plane. This plane will
look quite different from the mSUGRA case since now we will require
small $\mu\sim 100-200$~GeV in accord with EWFT and also $A_0\sim -1.6
m_0$ in accord with $m_h=125$~GeV and low EWFT.  We show the plane in
Fig.~\ref{fig:m0mhf} for $\mu =150$~GeV with $A_0=-1.6 m_0$,
$\tan\beta =10$ and $m_A=1$~TeV. Here, we plot contours $m_h=123$ and
125~GeV and also contours of $\delew=6$, 10, 15 and 50. Almost the
entire plane has low $\delew<50$, with $5.5 \leq \delew\alt 10$ in the
lower left portion.  In addition, the right-hand portion of the plane
has $m_h\agt 123-125$~GeV.  The purple-shaded region marked LEP2 has
$m_{\tw_1}<103.5$~GeV in violation of LEP2 limits on chargino pair
production. We also show recent LHC constraints from gluino/squark
searches within the mSUGRA model as the black
contour.\footnote{Strictly speaking, these are the constraints obtained
  in the mSUGRA model for $A_0=0$ from the non-observation of signals
  from gluino and first generation squark production. Since the masses
  of these sparticles depend mostly on $m_0$ and $m_{1/2}$, it is
  reasonable to suppose these also apply to the plane in
  Fig.~\ref{fig:m0mhf}.}  We extrapolate those constraints to much
higher $m_0$ values via the dashed black contour. Over the entire
plane, $m_{\tw_1}\sim m_{\tz_{1,2}}\sim \mu =150$~GeV so there would
always be light higgsino pair production at LHC.  The region with
$m_{1/2}\alt 0.6$~TeV yields $m_{\tg}\alt 2$~TeV and should be
accessible to future gluino pair production searches. Signals from
wino pair production may also be observable at the LHC, and perhaps
even at LHC8, if the heavier chargino is sufficiently light.
\FIGURE[tbh]{
\includegraphics[width=10cm,clip]{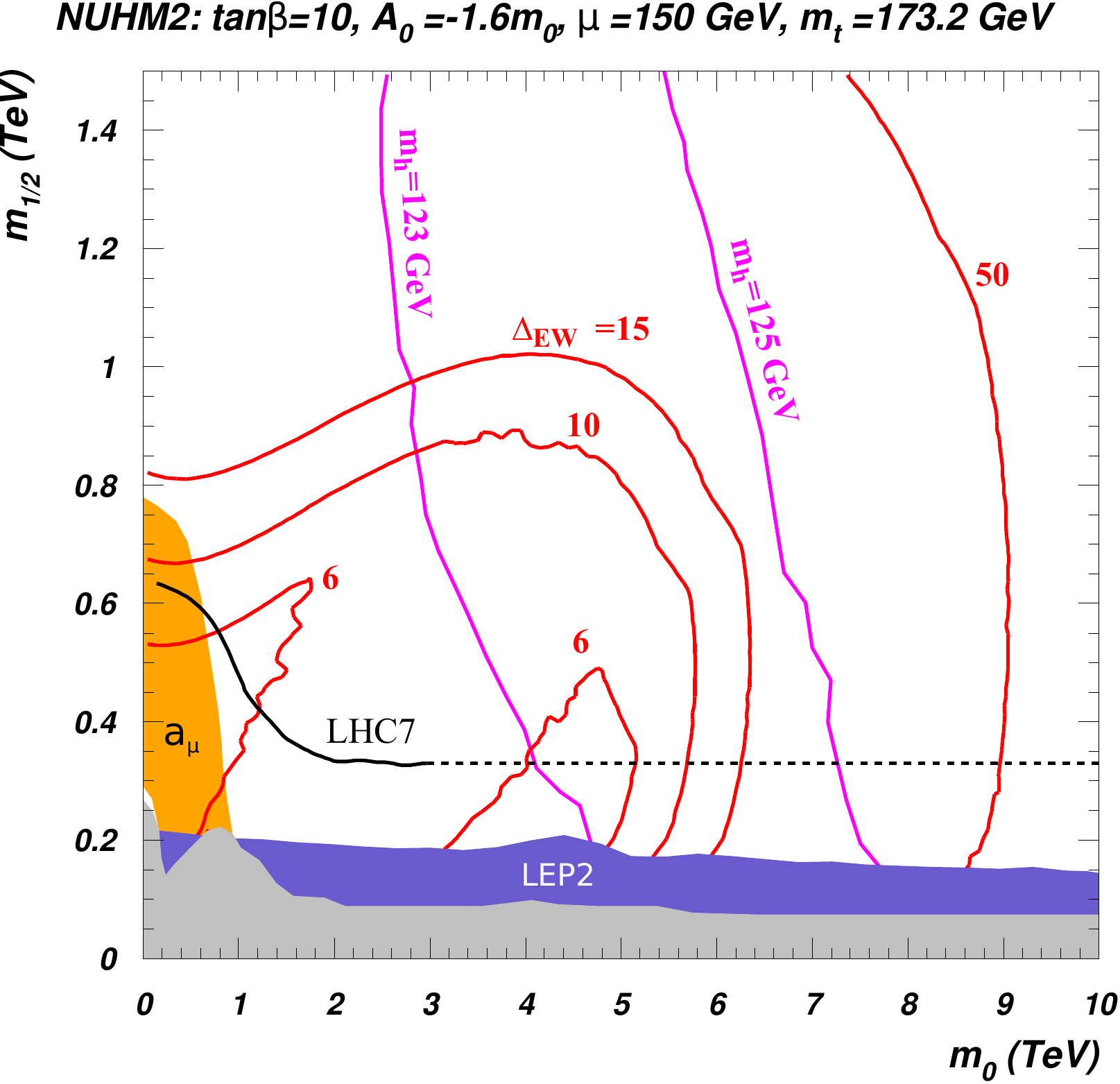}
\caption{Contours of $\delew$ (red curves) and $m_h$ (purple
curves) in the $m_0$ versus $m_{1/2}$ plane of RNS model with
$A_0=-1.6m_0$, $\tan\beta =10$, $\mu =150$~GeV and $m_A=1$~TeV.  }
\label{fig:m0mhf}}
%

\section{RNS at ILC}
\label{sec:ilc}

Since the main feature of RNS models is the presence of light higgsinos
$\tw_1$, $\tz_1$ and $\tz_2$, we expect excellent prospects for testing
RNS at a linear $e^+e^-$ collider.  Pair production for charged
higgsinos via the reaction $e^+e^-\to\tw_1^+\tw_1^-$ would yield soft
but observable decay products from $\tw_1\to\tz_1 f\bar{f}^\prime$
decay, where $f$ and $f^\prime$ are SM fermions. These decay products
should be easily detectable in the clean environment of $e^+e^-$
colliders, and moreover should be acollinear in the transverse plane as
opposed to two-photon backgrounds $\gamma\gamma\to f\bar{f}$ where
visible decay products tend to come out back-to-back. The entire $m_0\
vs.\ m_{1/2}$ plane shown in Fig.~\ref{fig:m0mhf} will be accessible to
an ILC with $\sqrt{s}\agt 2m_{\tw_1}\simeq 2|\mu|$.

Moreover, the cross section for the reaction $e^+e^-\to\tz_1\tz_2$
should also be large and provide corroborative evidence. These cross
sections will have a distinctive shape versus beam polarization as shown
in Ref.~\cite{bbh} which should be indicative of higgsino pair
production. We note here that since $E_{CM}\agt 2m_{\tw_1}\sim 2|\mu|$,
and $\delew\sim \mu^2/(M_Z^2/2)$, then an $e^+e^-$ collider
with a CM energy $E_{CM}$ directly probes 
\be 
\delew\sim
E_{CM}^2/(2M_Z^2)\, , 
\ee 
so that even a low energy $e^+e^-$ linear
collider would probe the most lucrative regions of RNS parameter space
(that portion with lowest $\delew$) and as $E_{CM}$ increases,
would discover natural SUSY or increasingly exclude it.

\section{Search for higgsino-like WIMPs from RNS}
\label{sec:dm}

One of the distinctive features of natural SUSY models is that the
lightest MSSM particle is a higgsino-like neutralino $\tz_1$.  If
$R$-parity is conserved, then the $\tz_1$ may make up all or at least a
portion of the dark matter in the universe.  Higgsinos with mass
$m_{\tz_1}>m_W,\ M_Z$ have high annihilation rates into vector boson
pairs.  Thus, if they are present in thermal equilibrium in the early
universe, then the higgsino relic density may be computed approximately
as 
\be 
\Omega_{\tz_1}^{\rm th}h^2=\frac{s_0}{\rho_c/h^2}\left(\frac{45}{\pi
g_*}\right)^{1/2}\frac{x_f}{m_{Pl}} \frac{1}{\langle\sigma v\rangle} 
\ee
where $s_0$ is the entropy density of the universe at the present time,
$\rho_c$ is the critical closure density, $h$ is the scaled Hubble
constant, $g_*$ is the number of relativistic degrees of freedom at
freeze-out, $x_f\sim25$ is the scaled freeze-out temperature, $m_{Pl}$
is the Planck mass and $\langle\sigma v\rangle$ is the thermally
averaged neutralino annihilation cross section times relative velocity.
Higgsino-like WIMPs couple with gauge strength to vector bosons so that
$\langle\sigma v\rangle$ is large and the relic density is suppressed.

We evaluate the relic density of higgsinos using Isatools\cite{isared}
from our scan over NUHM2 parameters as in Sec.~\ref{sec:nuhm2} and show
$\delew\ vs.\ \Omega_{\tz_1}h^2$ in frame {\it a}) of
Fig.~\ref{fig:DM}.  Points with $\delew<30$ indicative of RNS
are shown in red while the more highly fine-tuned points are in blue.  The
vertical green line shows the WMAP-measured value of the dark matter
density.  We see that the majority of points with $\delew <30$
have $\Omega_{\tz_1}h^2\sim 0.005-0.05$, {\it i.e.} well-below the
measured abundance. Several points have $\Omega_{\tz_1}h^2>0.12$; these
points arise from cases where $\mu\sim M_1$ where the neutralino is of
mixed bino-higgsino variety.

There exist a variety of non-standard cosmologies with features which make them more
attractive than the standard WIMP-only dark matter scenario. For instance, in stringy models with
moduli fields at the 10-100~TeV scale, the moduli may decay after BBN into SM particles, thus diluting
all relics present. Alternatively, if moduli decay to SUSY particles which cascade into the LSP, then the
neutralino abundance may be enhanced\cite{mr}.

Another possibility arises from SUSY models where the strong $CP$
problem is solved by the Peccei-Quinn mechanism\cite{pqww} with its
concomitant axion $a$. In the SUSY case, the axion superfield $\hat{a}$
also contains an $R$-even spin-0 saxion $s$ and an $R$-odd
spin-$\frac{1}{2}$ axino $\ta$.  In the case where $\tz_1$ is LSP, then
the dark matter would be a mixture of two particles: the axion and the
neutralino. Axinos which are produced thermally at high re-heat
temperature $T_R$ in the early universe would cascade decay to
neutralinos at a decay temperature $T_D<T_f$, causing a neutralino {\it
re-annihilation} which provides a much higher abundance of WIMPs than
expected in a WIMP-only picture. In addition, saxions can be produced
both thermally and via coherent oscillations, and may decay to both SUSY
and  SM particles: the former case enhances the neutralino
abundance while the latter case dilutes any relics present at the time
of decay. Calculations of the neutralino abundance in the PQ-augmented
MSSM depend on the various PQ parameters along with $T_R$ and the SUSY
particle spectrum and have been presented in
Ref.~\cite{ckls,blrs,bls}. In the case of models with a standard
under-abundance of neutralinos, the neutralino abundance is almost always
{\it enhanced} beyond its standard value $\Omega_{\tz_1}h^2$. If this
scenario is applied to the case of RNS, then we may most likely expect
an enhanced higssino-like WIMP abundance beyond its standard value. In
this scenario, axions will also be produced via coherent oscillations at
temperature around the QCD phase transition.  Thus, the higgsinos could
make up either a small or a large fraction of the relic dark matter,
with axions comprising the remainder. The important point here is that
it is very difficult to suppress the higgsino abundance below its
standard thermal value which is shown in Fig.~\ref{fig:DM}{\it a}). 
Thus, we would expect relic higgsinos to
be present in the universe today, but with an abundance which is
suppressed by between $1-15$ from the measured value.  This opens up the
opportunity to detect relic higgsinos, albeit while these would only
constitute a fraction of the measured dark matter abundance. At the same
time, there is also the possibility to detect relic axions.
\FIGURE[tbh]{
\includegraphics[width=7cm,clip]{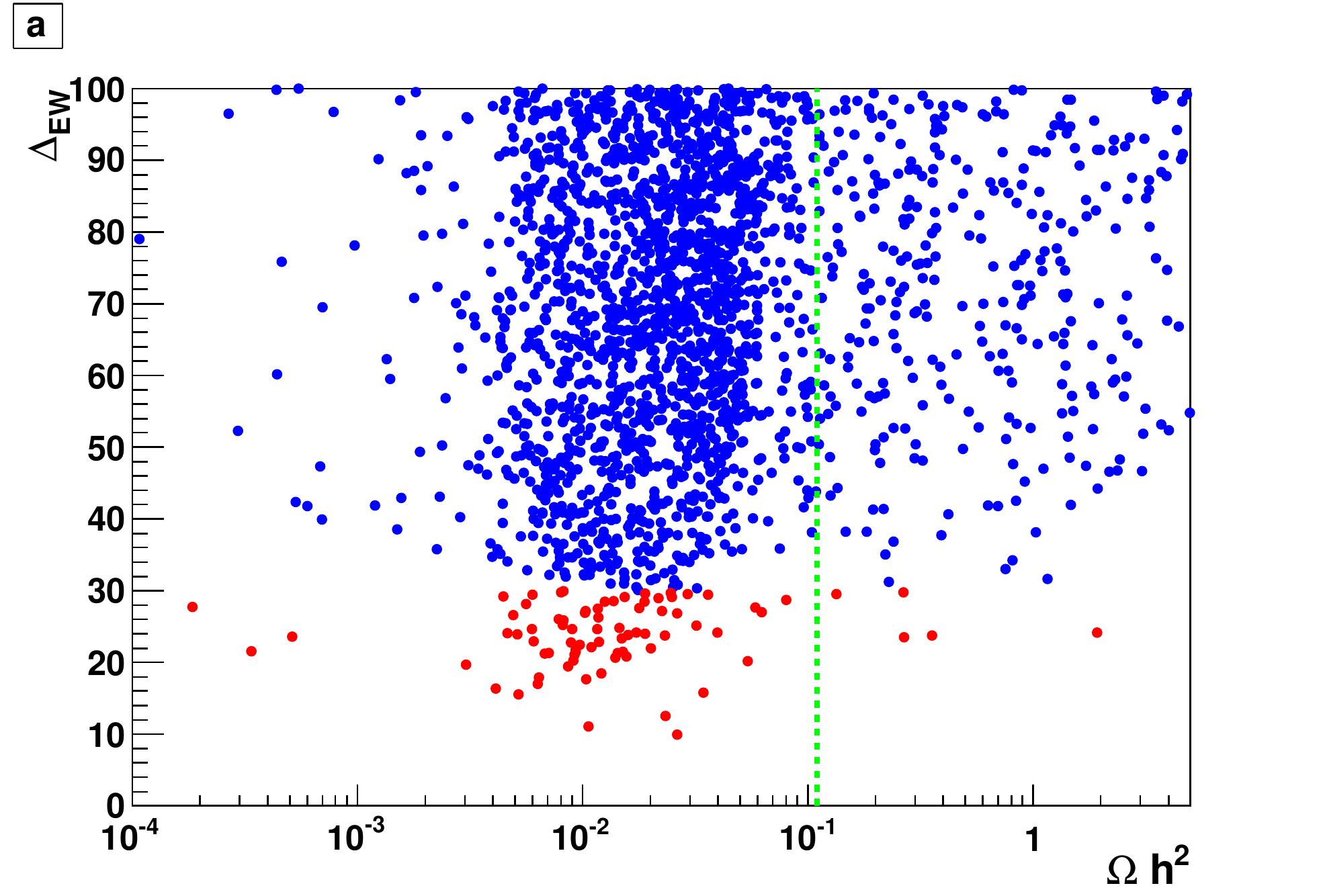}
\includegraphics[width=7cm,clip]{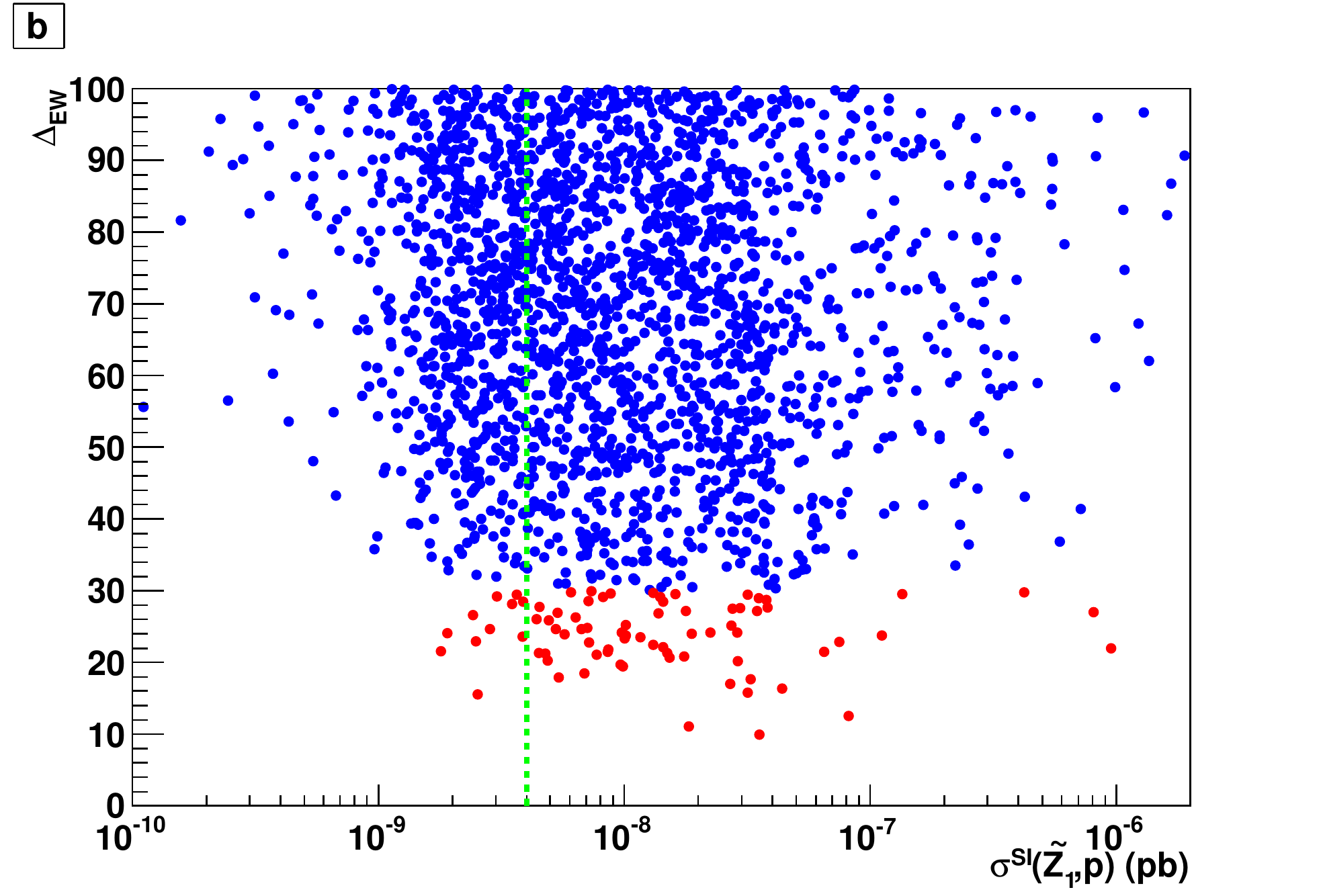}
\includegraphics[width=7cm,clip]{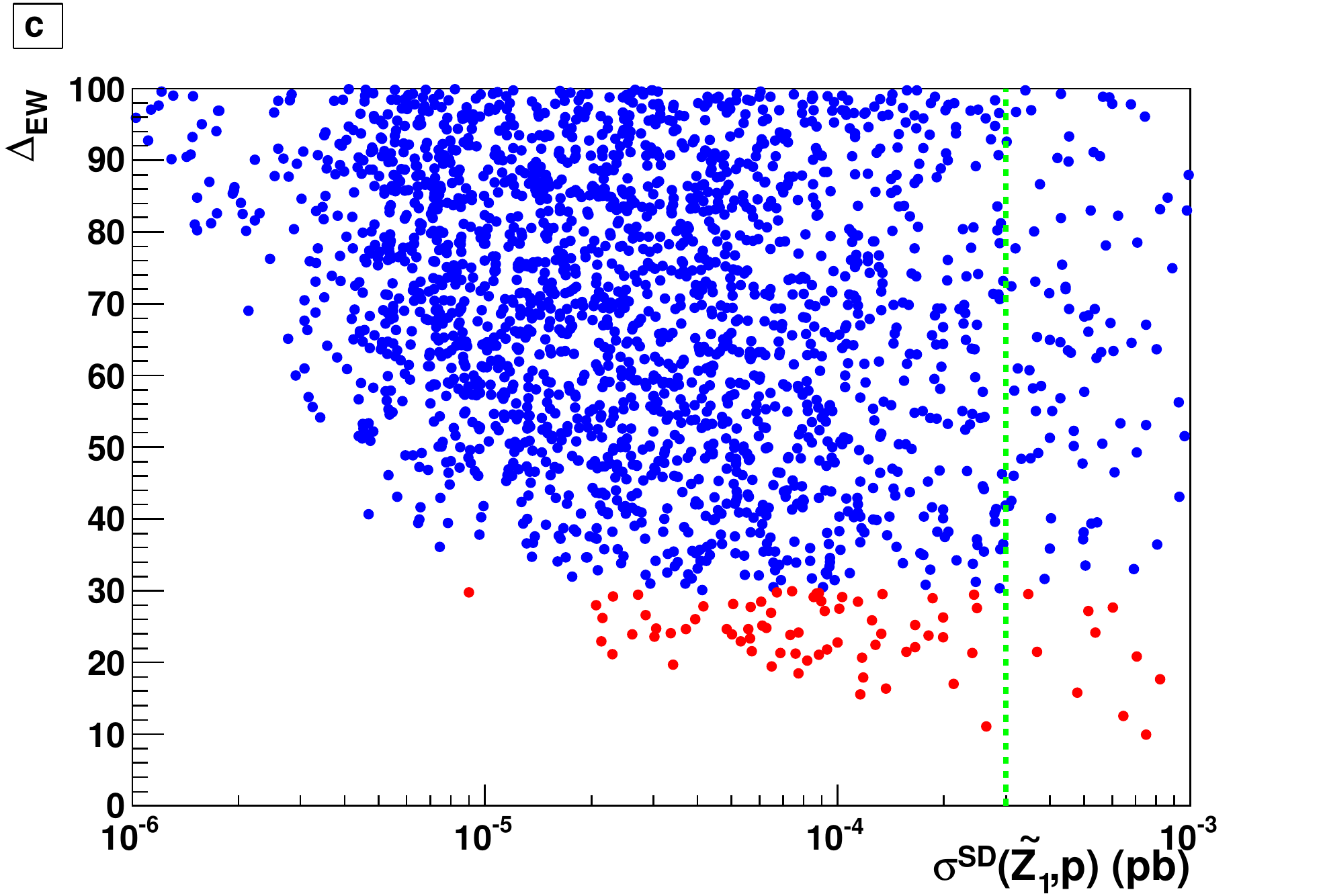}
\includegraphics[width=7cm,clip]{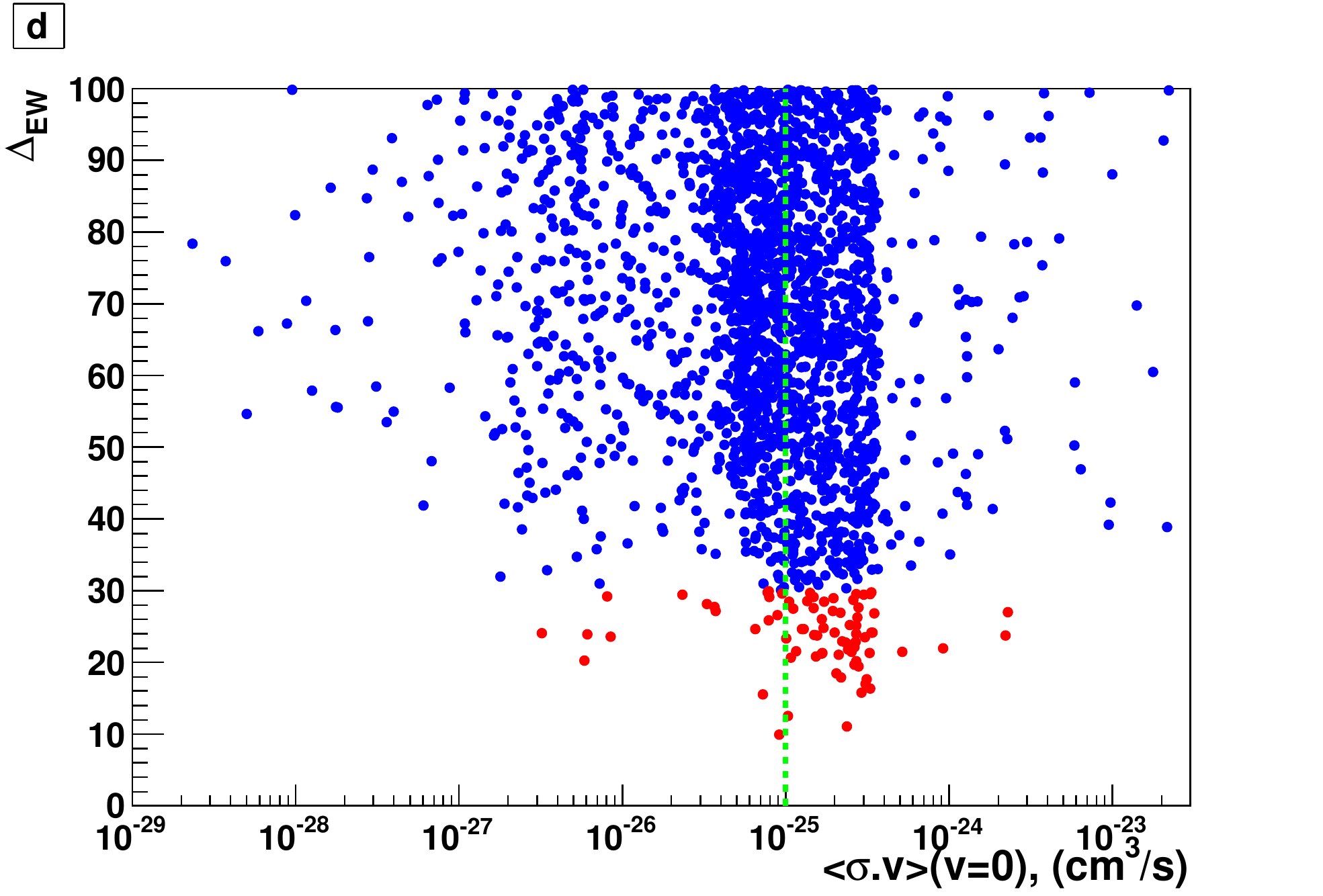}
\caption{The value of $\delew$ versus neutralino relic density
and direct and indirect WIMP detection rates. Vertical green lines
denote upper experimental limits obtained assuming that the WIMP
saturates the observed density of dark matter. 
The predictions in the last three frames need to be re-scaled
by a factor $\Omega_{\tz_1}h^2/0.11$ if the neutralinos make up only 
part of the dark matter. }
\label{fig:DM}}

With a view towards detecting relic higgsinos from RNS, we show in
Fig.~\ref{fig:DM}{\it b}) the value of $\delew$ versus the
spin-independent neutralino-proton scattering cross section $\sigma^{SI}
(\tz_1 p)$ in pb from Isatools\cite{bbbo}. The red points with
$\delew<30$ occur with $\sigma^{SI} (\tz_1 p)\sim
10^{-9}-10^{-7}$~pb. For comparison, we show via the green vertical line
the Xe-100 limit from 225 live days\cite{xe100} for $m_{\tz_1}\sim 150$
GeV.  Naively, many of the RNS points would be excluded if higgsinos
comprised the entire dark matter density.  However, in the mixed
axion/higgsino dark matter scenario, the expected local abundance of
WIMPs can be scaled down by factors of $1-15$ typically. Even with this
rescaling of the expected local abundance, we still expect relic
higgsinos to be within detection range of near-future WIMP detectors.

In Fig.~\ref{fig:DM}{\it c}), we show the spin-dependent
neutralino-proton cross section $\sigma^{SD}(\tz_1 p)$ in pb. The bulk
of RNS points with $\delew<30$ populate the region with
$\sigma^{SD}(\tz_1 p)\sim 2\times 10^{-5}-10^{-3}$ pb.  The IceCube
neutrino detector at the South Pole is sensitive to detection of
neutrinos arising from higgsino annihilation in the core of the sun. The
expected detection rate depends on the sun's ability to sweep up
neutralinos via $\tz_1 p$ collisions which depends mainly on $\sigma^{SD}(\tz_1 p)$. 
For reference, we also show the
current IceCube WW limit~\cite{icecube} at $\sigma^{SD}(\tz_1 p)\sim
3\times 10^{-4}$~pb. This limit depends on the assumption that WIMPs
comprise the entire DM abundance, and would need to be rescaled for a
mixed axion/higgsino cosmology.

Fig.~\ref{fig:DM}{\it d}) shows the thermally averaged neutralino annihilation
cross section times relative velocity, evaluated as $v\to 0$. This
quantity enters linearly into indirect searches for neutralino
annihilation in the cosmos into $\gamma$s or $e^+$, $\bar{p}$ or
$\bar{D}$. For the case of RNS, the bulk of points with
$\delew<30$ inhabit the region around
$\langle\sigma v\rangle |_{v\to 0}\sim  10^{-25}$~cm$^3$/sec.  The
vertical green line shows the upper limit on the annihilation cross
section times velocity for very non-relativistic dark matter in dwarf
spheroidal satellite galaxies of the Milky Way annihilating to $W$ boson
pairs obtained by the Fermi collaboration\cite{fermi}, assuming a $\sim
150$~GeV WIMP. Models with a larger annihilation cross section would
have led to a flux of gamma rays not detected by the experiment,
assuming a Navarro-Frenk-White profile\cite{nfw} for each dwarf galaxy in the
analysis.  We see that the Fermi bound might exclude the bulk of points
assuming higgsinos saturate the DM density.
This bound changes rather slowly with the WIMP mass,
being just a factor of 2 weaker for a WIMP mass of 300~GeV.  Further
searches and improvements by the Fermi-LAT Collaboration and/or the
impending AMS results should provide more stringent probes of the RNS
model.

\section{Summary and conclusions}
\label{sec:conclude}

Models of natural supersymmetry reconcile the lack of a SUSY signal at
LHC with the principle of electroweak naturalness.  Natural SUSY models
are characterized by light higgsinos of mass $\sim 100-300$~GeV, three
light third generation squarks with mass less than about 500~GeV and
gluinos of mass less than about 1.5~TeV. First/second generation squarks
may be much heavier -- in the multi-TeV regime -- thus avoiding LHC
searches and providing at least a partial decoupling solution to the
SUSY flavor and $CP$ problems. Attractive as they are, generic NS models based
on the MSSM are at odds with the recent discovery of a light Higgs
scalar at $\sim 125$~GeV which requires~TeV-scale top squarks along with
large top-squark mixing.

We presented here an improved natural SUSY model dubbed {\it radiative
natural SUSY}~\cite{ltr}, or RNS.  RNS is a SUSY model based on the MSSM
which may be valid all the way up to the GUT scale. Thus, it maintains
the desirable features of gauge coupling unification and radiative
electroweak symmetry breaking while avoiding the introduction of extra
possibly destabilizing gauge singlets or other forms of exotic matter.
The main features of the RNS model include 1)~a low value of
superpotential higgsino mass $|\mu |\sim 100-300$~GeV, and 2)~a weak
scale value of $-m_{H_u}^2\sim M_Z^2$: both these qualities are required
to fulfill electroweak naturalness at the tree level. The term
$m_{H_u}^2$ is driven to low values radiatively by the same mechanism
leading to REWSB and depends on a large top quark Yukawa coupling.  We
proceed further by evaluating EWFT at the 1-loop level.  In this case,
top squark masses enter the computation of $\delew$ and are
also driven radiatively to few-TeV values.  By allowing for large
top-squark mixing ($|A_0|\sim (1-2) m_0$), top-squark contributions to
EWFT are suppressed at the same time as the light Higgs boson mass is
uplifted: thus, the model reconciles electroweak fine-tuning with
$m_h\simeq 125$~GeV all in the context of the MSSM valid up to the GUT
scale.

RNS may be realized in the two-parameter non-universal Higgs models
NUHM2. In this case, low EWFT with $\delew\alt 30$ can be
attained for model parameters which lead to a distinctive mass spectrum: 
\bi
\item light higgsino-like $\tw_1$ and $\tz_{1,2}$ with mass $\sim 100-300$~GeV,
\item gluinos with mass $m_{\tg}\sim 1-4$~TeV,
\item heavier top squarks than generic NS models: $m_{\tst_1}\sim 1-2$~TeV 
      and $m_{\tst_2}\sim 2-5$~TeV,
\item first/second generation squarks and sleptons with mass
$m_{\tq,\tell}\sim 1-8$~TeV. The $m_{\tell}$ range can be pushed up to
20-30~TeV if non-universality of generations with $m_0(1,2)> m_0(3)$ is
allowed.  
\ei
The RNS model with the above spectra also fulfills limits from rare
$B$-decay measurements, which can be an Achilles heel for generic NS
models with much lighter third generation squarks.

The RNS model can be tested at LHC for $m_{1/2}$ in the lower portion of
its range whereupon gluino pair production and/or gaugino pair
production may lead to observable signals.  Many of the associated SUSY
events will contain light higgsinos arising from cascade decays with
soft decay products which may be observable. The case of OS/SF dileptons
with mass $\alt 10-20$~GeV would signal the presence of
$\tz_2\to\tz_1\ell^+\ell^-$ decay.

Linear $e^+e^-$ colliders would likely provide the definitive test of
RNS models since pair production of charged higgsinos should be easily
observable and the lowest energy machines will scrutinize the most
lucrative parameter choices with the lowest values of $\delew$.
For RNS, an ILC type machine would be a higgsino factory in addition to
a Higgs factory.

In RNS, we also expect the presence of higgsino-like WIMPs which have
large rates for direct and indirect WIMP detection.  Since higgsinos are
thermally underproduced, we expect them to constitute only a portion of
the measured dark matter abundance, with perhaps axions comprising the
remainder. Detectability via WIMP searches will depend on the higgsino
fraction of the dark matter. 

The many elegant features presented above impel us to regard RNS as the
possible new paradigm SUSY model.  Its consequences for detection at
colliders and at dark matter detectors merits a high level of scrutiny.

\acknowledgments

We thank Javier Ferrandis for various calculations concerning the
effective potential.  This work was supported in part by the
U.S. Department of Energy.


\appendix

\section{Radiative corrections to the Higgs potential minimization conditions}

The Higgs portion of the scalar potential in the MSSM is given by
\be
V_{Higgs}=V_{\rm tree}+\Delta V,
\ee
where the tree level portion for the neutral Higgs sector is given by
\bea
V_{\rm tree}=(m_{H_u}^2+\mu^2)|h_u^0|^2 +(m_{H_d}^2+\mu^2)|h_d^0|^2 \nonumber \\
-B\mu (h_u^0h_d^0+h.c.)+{1\over 8}(g^2+g^{\prime 2})
(|h_u^0|^2-|h_d^0|^2)^2
\eea
and the radiative corrections (in the effective potential approximation, and using the
$\overline{DR}$ regularization scheme, as appropriate for SUSY models) by
\be
\Delta V=\sum_{i}\frac{(-1)^{2s_i}}{64\pi^2} (2s_i+1)c_i m_i^4\left[
\log\left(\frac{m_i^2}{Q^2}\right)-\frac{3}{2}\right] ,
\ee
where the sum over $i$ runs over all fields that couple to Higgs fields,
$m_i^2$ are the {\it Higgs field dependent} mass squared values, and
$c_i=c_{col}c_{cha}$, with $c_{col}=3\ (1)$ for colored (uncolored) particles and
$c_{cha}= 2\ (1)$ for charged (neutral) particles and $s_i$ is their spin quantum number.

Minimization of the scalar potential  allows one to compute the
gauge boson masses in terms of the Higgs field vacuum expectation values $v_u$ and $v_d$, and
leads to the well-known conditions that
\bea
B\mu v_d&=&\left( m_{H_u}^2+\mu^2-g_Z^2(v_d^2-v_u^2)\right)v_u+\Sigma_u  \\
B\mu v_u&=&\left( m_{H_d}^2+\mu^2+g_Z^2(v_d^2-v_u^2)\right)v_d+\Sigma_d \, ,
\label{eq:min}
\eea
where
\be
\Sigma_{u,d}=\frac{\partial\Delta V}{\partial h_{u,d}} \bigg|_{min}
\ee
and $h_{u,d}^0=(h_{u,dR}^0+ih_{u,dI}^0)/\sqrt{2}$, $g_Z^2=(g^2+g'^2)/8$.
By $SU(2)$ invariance, the scalar potential $V$ depends on the scalar
fields as \cite{wss}
$V(h_u^{\dagger}h_u,h_d^{\dagger}h_d,h_u h_d+{\rm c.c.})$, then we have
\bea
\Sigma_u&=&\Sigma_u^uv_u+\Sigma_u^dv_d\, ,\\
\Sigma_d&=&\Sigma_d^uv_u+\Sigma_d^dv_d\ \ {\rm and}\\
\Sigma_d^u &=&\Sigma_u^d
\eea
where
\bea
\Sigma_u^u&=&\frac{\partial\Delta V}{\partial | h_u|^2}\bigg|_{min},\\
\Sigma_d^d&=&\frac{\partial\Delta V}{\partial | h_d|^2}\bigg|_{min}\ \ \ {\rm and}\\
\Sigma_u^d&=&\frac{\partial\Delta V}{\partial (h_u h_d+{\rm c.c.})}\bigg|_{min}.
\eea
In this case, the minimization conditions may be expressed as
\bea
M_Z^2/2 &=& \frac{(m_{H_d}^2+\Sigma_d^d)-(m_{H_u}^2+\Sigma_u^u)\tan^2\beta}{\tan^2\beta -1}-\mu^2 ,
\label{app:mzs}\\
B\mu &=&\left( (m_{H_u}^2+\mu^2+\Sigma_u^u)+(m_{H_d}^2+\mu^2+\Sigma_d^d)\right)\sin\beta\cos\beta +\Sigma_u^d .
\label{app:bmu}
\eea 
The advantage of writing the minimization conditions in terms of
$\Sigma_u^u$ and $\Sigma_d^d$ and $\Sigma_{ud}$ is that the corrections
to $m_{H_u}^2$, $m_{H_d}^2$ and $B\mu$ are neatly separated so 
that $\Sigma_u^d$ terms do not
appear in Eq.~(\ref{app:mzs}), and so do not contribute to
the fine-tuning calculation.

The contributions of the various $\Sigma$s can be written as:
\bea
\Sigma_u^u &=& \sum_i
\frac{1}{32\pi^2}(-1)^{2s_i}(2s_i+1)c_i\frac{\partial m_i^2}{\partial
  |h_u|^2}\bigg |_{min} F(m_i^2)\;,
\label{eq:siguu} \nonumber \\
\Sigma_d^d &=& \sum_i
\frac{1}{32\pi^2}(-1)^{2s_i}(2s_i+1)c_i\frac{\partial m_i^2}{\partial
  |h_d|^2}\bigg |_{min} F(m_i^2)\;,
\label{eq:sigdd}  \\
\Sigma_u^d &=& \sum_i
\frac{1}{32\pi^2}(-1)^{2s_i}(2s_i+1)c_i\frac{\partial m_i^2}{\partial
  (h_u h_d + {\rm c.c.})}\bigg |_{min} F(m_i^2)
=\Sigma_d^u .\nonumber
\label{eq:sigud}
\eea
where
\be
F(m^2)= m^2\left(\log\frac{m^2}{Q^2}-1\right) .
\ee
with the optimized scale choice $Q^2 =m_{\tst_1}m_{\tst_2}$.

For the top squark contributions, we find
\bea
\Sigma_u^u (\tst_{1,2})&=& \frac{3}{16\pi^2}F(m_{\tst_{1,2}}^2)
\left[ f_t^2-g_Z^2\mp \frac{f_t^2 A_t^2-8g_Z^2
(\frac{1}{4}-\frac{2}{3}x_W)\Delta_t}{m_{\tst_2}^2-m_{\tst_1}^2}\right]
\label{eq:sigtuu} \\
\Sigma_d^d (\tst_{1,2})&=& \frac{3}{16\pi^2}F(m_{\tst_{1,2}}^2)
\left[ g_Z^2\mp \frac{f_t^2\mu^2+8 g_Z^2 (\frac{1}{4}-\frac{2}{3}x_W)\Delta_t}{m_{\tst_2}^2-m_{\tst_1}^2}\right]
\label{eq:sigtdd}
\eea
where $\Delta_t=(m_{\tst_L}^2-m_{\tst_R}^2)/2+M_Z^2\cos 2\beta (\frac{1}{4}-\frac{2}{3}x_W)$ and $x_W\equiv\sin^2\theta_W$.
In the denominator of (\ref{eq:sigtuu}) and (\ref{eq:sigtdd}),
the tree level expressions of $m_{\tst_{1,2}}^2$ should be used.

For $b$-squark contributions, we have
\bea
\Sigma_u^u (\tb_{1,2})&=& \frac{3}{16\pi^2}F(m_{\tb_{1,2}}^2)
\left[ g_Z^2\mp \frac{f_b^2\mu^2-8g_Z^2
(\frac{1}{4}-\frac{1}{3}x_W)\Delta_b}{m_{\tb_2}^2-m_{\tb_1}^2}\right]
\label{eq:sigbuu} \cr
\Sigma_d^d (\tb_{1,2})&=& \frac{3}{16\pi^2}F(m_{\tb_{1,2}}^2)
\left[ f_b^2-g_Z^2\mp \frac{f_b^2 A_b^2-8 g_Z^2 (\frac{1}{4}-\frac{1}{3}x_W)\Delta_b}{m_{\tb_2}^2-m_{\tb_1}^2}\right]
\label{eq:sigbdd}
\eea
where $\Delta_b=(m_{\tb_L}^2-m_{\tb_R}^2)/2-M_Z^2\cos 2\beta (\frac{1}{4}-\frac{1}{3}x_W)$.
The expressions for $\Sigma_u^u(\ttau_{1,2} )$ and $\Sigma_d^d(\ttau_{1,2})$ are similar to
$\Sigma_u^u(\tb_{1,2})$ and $\Sigma_d^d(\tb_{1,2})$ but with $b\to \tau$, $c_{col}=1$ and
$(\frac{1}{4}-\frac{1}{3}x_W )\to (\frac{1}{4}-x_W )$.

For first/second generation sfermion contributions, we find
\bea
\Sigma_{u,d}^{u,d}(\tf_{L,R}) &=& \frac{c_{col}}{16\pi^2}F(m_{\tf_{L,R}}^2)\left(4g_Z^2(T_3-Qx_W\right)
\eea
where $T_3$ is the weak isospin and $Q$ is the electric charge assignment (taking care to flip
the sign of $Q$ for $R$-sfermions). For instance,
\bea
\Sigma_u^u (\tu_L) &=& \frac{3}{16\pi^2}F(m_{\tu_L}^2)\left(f_u^2-4g_Z^2 (\frac{1}{2}-\frac{2}{3}x_W)\right)\\
\Sigma_u^u (\tu_R) &=& \frac{3}{16\pi^2}F(m_{\tu_R}^2)\left(f_u^2-4g_Z^2 (\frac{2}{3}x_W)\right)\\
\Sigma_d^d (\tu_L) &=& \frac{3}{16\pi^2}F(m_{\tu_L}^2)\left(4g_Z^2 (\frac{1}{2}-\frac{2}{3}x_W)\right)\\
\Sigma_d^d (\tu_R) &=& \frac{3}{16\pi^2}F(m_{\tu_R}^2)\left(4g_Z^2 (\frac{2}{3}x_W)\right) .
\eea
These contributions, arising from electroweak $D$-term contributions to masses, cancel out
separately for squarks and sleptons in the limit of mass degeneracy, due to the fact that
weak isospins and electric charges (or weak hypercharges) sum to zero in each generation.
For this reason, we sum these contributions before taking the maximum contribution to
the fine-tuning measure $\delew$.

For chargino contributions, we find
\bea
\Sigma_u^u(\tw_{1,2}^\pm )&=& \frac{-g^2}{16\pi^2} F(m_{\tw_{1,2}}^2)\left(1\mp
\frac{M_2^2+\mu^2-2m_W^2\cos 2\beta}{m_{\tw_2}^2-m_{\tw_1}^2}\right) \\
\Sigma_d^d(\tw_{1,2}^\pm )&=& \frac{-g^2}{16\pi^2} F(m_{\tw_{1,2}}^2)\left(1\mp
\frac{M_2^2+\mu^2+2m_W^2\cos 2\beta}{m_{\tw_2}^2-m_{\tw_1}^2}\right) .
\eea

For contributions from neutralinos, we find\footnote{Unlike the case of
  other contributions where it is easy to explicitly find the
  eigenvalues of the Higgs-field-dependent squared mass matrices, this
  is not possible for the neutralino. To evaluate the derivatives of the
  eigenvalues of the {\it squared} neutralino mass matrix that appear in
  (\ref{eq:sigdd}), we use the technique introduced in Ref.~\cite{an} and
  elaborated further in Ref.~\cite{in}.}

\bea \Sigma_u^u(\tz_i)&=& \frac{1}{16\pi^2}
  \frac{F(m_{\tz_i}^2)}{D(\tz_i)} \left[K(\tz_i)- 2(g^2+g'^2)\mu^2
  M_Z^2\cos^2\beta (m_{\tz_i}^2-m_{\pino}^2) \right] \, ,\\
  \Sigma_d^d(\tz_i)&=& \frac{1}{16\pi^2} \frac{F(m_{\tz_i}^2)}{D(\tz_i)}
  \left[K(\tz_i)- 2(g^2+g'^2)\mu^2 M_Z^2\sin^2\beta
  (m_{\tz_i}^2-m_{\pino}^2) \right]\, , \eea 
where \bea K(\tz_i)&=&
  -m_{\tz_i}^6 (g^2+g'^2) \nonumber\\ && +m_{\tz_i}^4 \left[ g^2
  (M_1^2+\mu^2)+g'^2 (M_2^2+\mu^2)+(g^2+g'^2)M_Z^2\right] \nonumber\\
  &&-m_{\tz_i}^2 \left[\mu^2 (g^2 M_1^2+g'^2 M_2^2)+(g^2+g'^2)M_Z^2
  m_{\pino}^2 \right], \eea 
$D(\tz_i)=\prod_{j\neq i}
  (m^2_{\tz_i}-m^2_{\tz_j})$, and $m_{\pino} = M_1 \cos^2\theta_W +M_2
  \sin^2\theta_W$.  
Our neutralino corrections differ in form as well as
numerically from those in the literature where these were calculated
using the neutralino mass (not mass squared) matrix~\cite{an,deboer}.

For weak bosons, we find
\bea
\Sigma_u^u(W^\pm )&=& \Sigma_d^d(W^\pm)=\frac{3g^2}{32\pi^2} F(m_W^2)\\
\Sigma_u^u(Z^0 )&=& \Sigma_d^d(Z^0)=\frac{3g^2}{64\pi^2\cos^2\theta_W} F(M_Z^2) .
\eea

For Higgs bosons, we find
\bea
\Sigma_u^u(h,H)&=& \frac{g_Z^2}{16\pi^2} F(m_{h,H}^2)\left(
1\mp \frac{M_Z^2+m_A^2 (1+4\cos 2\beta +2\cos^2 2\beta )}{m_{H}^2-m_{h}^2}\right),\\
\Sigma_d^d(h,H)&=& \frac{g_Z^2}{16\pi^2} F(m_{h,H}^2)\left(
1\mp \frac{M_Z^2+m_A^2 (1-4\cos 2\beta +2\cos^2 2\beta )}{m_{H}^2-m_{h}^2}\right)
\eea
and

\bea
\Sigma_u^u(H^\pm )&=& \Sigma_d^d(H^\pm)=\frac{g^2}{32\pi^2} F(m_{H^\pm}^2) .
\eea

For SM fermions $t$, $b$ and $\tau$, we find
\bea
\Sigma_u^u(t)&=& -\frac{3f_t^2}{8\pi^2} F(m_t^2),\\
\Sigma_d^d(t)&=& 0
\eea

\bea
\Sigma_u^u(b)&=& 0,\\
\Sigma_d^d(b)&=& -\frac{3f_b^2}{8\pi^2} F(m_b^2)
\eea

\bea
\Sigma_u^u(\tau)&=& 0,\\
\Sigma_d^d(\tau)&=& -\frac{f_{\tau}^2}{8\pi^2} F(m_{\tau}^2) .
\eea

%
%

\end{document}